\documentclass[nofootinbib,amsmath,amssymb,preprint,11pt]{revtex4}
\usepackage{graphicx,epsfig,psfrag,bm,amssymb}
\usepackage{dcolumn}
\usepackage{bm}
\usepackage{color}
\usepackage{mathrsfs,amsfonts,color}
\usepackage[utf8]{inputenc}
\usepackage{hyperref}
\usepackage{relsize}

\newcommand{\be}{\begin{equation}}
\newcommand{\ee}{\end{equation}}

\newcommand{\veck}{\mathbf{k}}
\newcommand{\vecr}{\mathbf{r}}
\newcommand{\vecx}{\mathbf{x}}
\newcommand{\vecv}{\mathbf{v}}
\newcommand{\vecp}{\mathbf{p}}
\newcommand{\vecq}{\mathbf{q}}
\newcommand{\vecu}{\mathbf{u}}
\newcommand{\vecS}{\mathbf{S}}
\newcommand{\vecR}{\mathbf{R}}
\newcommand{\vecG}{\mathbf{G}}
\newcommand{\vesc}{v_{\rm esc}}
\newcommand{\vmin}{v_{\rm min}}
\newcommand{\Zeff}{Z_{\rm eff}}
\newcommand{\TeV}{{\rm TeV}}
\newcommand{\GeV}{{\rm GeV}}
\newcommand{\MeV}{{\rm MeV}}
\newcommand{\keV}{{\rm keV}}
\newcommand{\eV}{{\rm eV}}
\newcommand{\sigmabar}{\overline{\sigma}}

\newcommand{\He}{$^4$He}

\definecolor{blueish}{rgb}{0., 0.42,0.9}

\begin{document}
\title{Searches for light dark matter using condensed matter systems}
\author{Yonatan Kahn}
\email{yfkahn@illinois.edu}
\affiliation{Department of Physics, University of Illinois at Urbana-Champaign, Urbana, IL 61801, USA}
\affiliation{Illinois Center for Advanced Studies of the Universe, University of Illinois at Urbana-Champaign, Urbana, IL 61801, USA}
\author{Tongyan Lin}
\email{tongyan@physics.ucsd.edu}
\affiliation{Department of Physics, University of California, San Diego, CA 92093, USA}

\date{\today}
\begin{abstract} 
Identifying the nature of dark matter (DM) has long been a pressing question for particle physics. In the face of ever-more-powerful exclusions and null results from large-exposure searches for TeV-scale DM interacting with nuclei, a significant amount of attention has shifted to lighter (sub-GeV) DM candidates. Direct detection of the light dark matter in our galaxy by observing DM scattering off a target system requires new approaches compared to prior searches. Lighter DM particles have less available kinetic energy, and achieving a kinematic match between DM and the target mandates the proper treatment of collective excitations in condensed matter systems, such as charged quasiparticles or phonons. In this context, the condensed matter physics of the target material is crucial, necessitating an interdisciplinary approach. In this review, we provide a self-contained introduction to direct detection of keV--GeV DM with condensed matter systems. We give a brief survey of dark matter models and basics of condensed matter, while the bulk of the review deals with the theoretical treatment of DM-nucleon and DM-electron interactions. We also review recent experimental developments in detector technology, and conclude with an outlook for the field of sub-GeV DM detection over the next decade.

 \end{abstract}

\maketitle

\tableofcontents

\section{Introduction}
\label{sec:Intro}

The past several decades have featured an immense accumulation of gravitational evidence for dark matter (DM): to our knowledge, 23\% of our universe feels the force of gravity but not the strong nuclear force or electromagnetism to any measurable extent. This fact explains myriad astronomical and cosmological observations across widely varying distance and time scales. The early observations of Rubin \cite{1970ApJ...159..379R}, who noted that stars on the outskirts of galaxies rotated faster than would be inferred based on the Newtonian gravitational potential of the visible stars and gas, imply that galaxies host ``halos'' of DM which extend far beyond the visible matter. This DM comprises the majority of the mass of the galaxy. The beautiful precision of fits to data on the fluctuations of the cosmic microwave background (CMB) \cite{Aghanim:2018eyx}, photons which take a snapshot of the universe 380,000 years after the Big Bang, requires a ``dark'' component of the universe which gravitates but does not interact strongly with photons \cite{Boehm:2001hm,Chen:2002yh}. Furthermore, numerical simulations show that DM provides the ``gravitational scaffolding'' for galaxies to form~\cite{2015ARA&A..53...51S} -- without DM, we might not be here at all! 

Given that DM exists in the universe, how do we find it in the laboratory? The range of possible DM masses is extremely broad. At the high end, DM could be as heavy as the Planck scale ($10^{19} \ \GeV/c^2$) if it is an elementary particle, or even heavier if DM were a composite particle or comprised of small black holes formed shortly after the big bang. At the low end, DM could be as light as $\sim 10^{-22} \ \eV/c^2$, the scale at which the de Broglie wavelength of DM exceeds the sizes of the smallest dwarf galaxies; the uncertainty principle implies that lighter DM cannot be meaningfully considered as bound to the galaxy. At both extremes of mass, and for much of the intervening 50 orders of magnitude, it is possible that DM has \emph{no} appreciable non-gravitational interactions. While there exist several creative proposals for detecting the gravitational signatures of such DM, from pulsar-timing array probes of ultra-light DM \cite{Porayko:2018sfa} to networked quantum sensors for Planck-scale DM \cite{Carney:2019pza}, the experimental and observational prospects of this ``nightmare scenario'' can be overall quite grim.

However, early-universe thermodynamics provides a hopeful clue: the fundamental interactions of particles in the Standard Model (SM) of particle physics -- electrons, protons, neutrinos, and so on -- in the fractions of a second after the Big Bang can predict with great accuracy the abundances of light elements billions of years later \cite{PhysRev.73.803}. The spectacular success of this paradigm suggests that a plausible scenario is one where dark matter is a new fundamental particle, which we will denote $\chi$, with \emph{some} non-gravitational interactions between the DM and the SM. If these interactions let DM establish thermal contact with the SM at some point during the evolution of the universe, equilibrium thermodynamics could easily explain the fact that the ratio of the DM and ordinary matter abundance is an order-1 number today. Indeed, in many models of purely gravitationally-coupled DM, the problem is an overabundance of DM which would have driven the curvature of the universe positive and resulted in a Big Crunch \cite{MOROI1993289}. The \emph{hypothesis of thermal contact} provides an elegant mechanism for safely depleting the primordial DM abundance through annihilations into a thermal plasma, and moreover, provides sharp correlations among the DM-SM coupling, the DM mass, and the observed late-time DM abundance.

The hypothesis of thermal contact restricts the allowed DM mass range considerably: DM cannot be too light or it would have been too fast to clump and form structures, and it cannot be too weakly-coupled or it could never have made thermal contact. The allowed parameter space is
\be
m_\chi c^2 \in \ \sim [\keV, 100 \ \TeV]
\ee
where the upper end of the mass range is a constraint from quantum-mechanical unitarity on the DM annihilation amplitude \cite{Griest:1989wd}. For DM masses ranging from the upper limit down to about 1 GeV, a well-motivated candidate with connections to other fundamental physics such as supersymmetry has existed for decades: the WIMP, or ``weakly-interacting massive particle.'' A vigorous international experimental program has searched for WIMP DM in the laboratory, but so far to no avail: for more than 20 years, all searches have turned up null.\footnote{The only persistent positive claim, from the DAMA collaboration \cite{Bernabei:2010mq}, was recently conclusively refuted by another experiment using identical detectors \cite{Amare:2021yyu}.} 

To see where DM might be hiding, consider a typical search strategy for \emph{direct detection}: an experiment looks for kinetic energy deposited by DM scattering on atomic nuclei. The source of DM in such an experiment is the DM that pervades our galaxy, where the DM mass density and velocity in our Solar neighborhood can be inferred from gravitational measurements:\footnote{For the remainder of this review we will use the natural unit conventions common in particle physics and set $\hbar = c = 1$.}
\be
\rho_\chi \equiv n_\chi m_\chi \simeq 0.3-0.5 \ \GeV/{\rm cm}^3, \qquad v_\chi \simeq 10^{-3}.
\ee
Assuming a DM mass of $m_\chi = 100 \ \GeV$ and a dark matter-nucleus interaction cross section of $\sigma_{\chi N} =  A^2 \alpha_W^2/m_\chi^2$ where $\alpha_W \simeq 0.03$ is the coupling constant of the weak nuclear force and $A$ is atomic mass number, we may estimate the scattering rate per nucleus as
\be
R_\chi = n_\chi \sigma_{\chi N} v_\chi \simeq 3 \times 10^{-26}/{\rm s}
\ee
for a heavy nucleus of $A=100$.
Immediately we see the need for condensed matter detectors: Avogadro's number of scattering targets must be present in an experimentally-manageable volume to have any hope of seeing a statistically-significant number of events.  Even so, there may only be a handful events in a year, and the targets must also be highly radiopure and well-shielded to search for extremely rare DM interactions.  This is the approach taken by collaborations such as XENON \cite{Aprile:2017aty} and LZ \cite{Akerib:2019fml}, which are building larger and larger detectors containing multiple tons of liquid xenon, in order to probe DM that may be hiding with a smaller-than-expected cross section.   

Despite the need for condensed matter targets, for DM with mass of $m_\chi = 100 \ \GeV$ the scattering can be modeled as elastic scattering off free nuclear targets. To see why, note that since the DM is non-relativistic, its typical momentum and kinetic energy are
\be
p_\chi \simeq m_\chi v_\chi \simeq 100 \ \MeV \left(\frac{m_\chi}{100 \ \GeV}\right), \qquad E_\chi \simeq \frac{1}{2}m_\chi v_\chi^2 \simeq 50 \ \keV \left(\frac{m_\chi}{100 \ \GeV}\right),
\ee
which are much larger than any of the scales where many-body effects become important. Furthermore, DM of this mass does not have enough kinetic energy to excite internal nuclear states, so the scattering kinematics are those of classical elastic scattering. Typical nuclear recoil detectors like XENON and LZ exploit the ionization and scintillation signals created by a fast struck nucleus, resulting in a detector energy threshold at the few hundred eV scale.

\begin{table}[t!]
\begin{center}
\begin{tabular}{|c|c|c|}
\hline
DM mass & DM energy or momentum & CM scale \\
\hline
\hline
50 MeV & $p_\chi \sim 50 \ \keV$ & zero-point ion momentum in lattice \\
\hline
20 MeV & $E_\chi \sim 10 \ \eV$ & atomic ionization energy \\
\hline
2 MeV & $E_\chi \sim 1 \ \eV$ & semiconductor band gap \\
\hline
100 keV & $E_\chi \sim 50 \ {\rm meV}$ & optical phonon energy \\
\hline
\end{tabular}
\end{center}
\caption{Energy and momentum scales relevant for DM scattering in CM systems}
\label{tab:scales}
\end{table}%

That said, DM may also be hiding ``in plain sight'' with a large cross section at the low end of the ``thermal contact'' mass range, below 1--10 GeV \cite{Boehm:2002yz,Boehm:2003hm,Fayet:2004bw}. It is this \emph{light dark matter} (``light'' here referring to mass, not any kind of electromagnetic interactions) on which we focus in this review.  Because of the low kinetic energy, \emph{sub-GeV DM} may be invisible to ton-scale nuclear recoil detectors, no matter how strong its interactions \cite{Essig:2011nj}. However, as $m_\chi$ decreases, $n_\chi$ increases, so even experiments with relatively small targets (gram-scale rather than ton-scale) can have comparable discovery prospects for the same thermally-motivated cross sections, if the energy threshold can be reduced. Furthermore, from the point of view of maximizing the DM signal, it is optimal to have systems with available excitations that match the low energies and momentum transfers associated with DM masses in the keV--GeV range. Since the DM mass is much lower than a nucleus mass in this regime, nuclear recoils are a poor kinematic match, but the wide range of available excitations in condensed matter systems offers a promising way forward. In particular, we will see the relevance of the following degrees of freedom:
\begin{itemize}
\item DM-nuclear scattering $\leftrightarrow$ phonons
\item DM-electron scattering $\leftrightarrow$ electron quasiparticles and plasmons
\end{itemize}
Bringing atoms closer together generically lowers the excitation energy, so solid-state systems are also beneficial from both energy threshold and target density considerations compared to atomic or molecular targets. Importantly, for sub-GeV DM, a full condensed matter (CM) treatment of any solid or liquid target is mandatory, because the DM momentum and energy scales are no longer the largest scales in the problem and the targets (electrons or nuclei) may not be approximated as free particles. See Tab.~\ref{tab:scales} for a comparison of DM and CM scales. The myriad tools of CM, along with the plethora of novel materials with unusual or exotic properties, may then be brought to bear on the problem of DM detection, and indeed such pursuits have already engendered a fruitful and creative cross-disciplinary collaboration over the past decade.

This review endeavors to provide a self-contained introduction to searches for light (sub-GeV) dark matter using condensed matter systems. In particular, no background in quantum field theory, condensed matter physics, or particle physics will be assumed: the fortuitous fact that dark matter is non-relativistic means that the main results of the subject can be understood completely at a technical level using only quantum mechanics. This review is structured as follows. In Sec.~\ref{sec:DMReview}, we lay out the essential properties of sub-GeV DM (namely, its kinematics and dynamics) which govern its interactions with generic detectors. In Sec.~\ref{sec:CMReview}, we survey the key objects and tools of condensed matter physics which describe the behavior of quasiparticles and collective modes in solid-state systems. We hope that Sec.~\ref{sec:DMReview} can provide a lightning introduction to DM to students or researchers unfamiliar with particle physics, and likewise for Sec.~\ref{sec:CMReview} for students or researchers unfamiliar with condensed matter physics; both should be accessible to beginning graduate students. Sec.~\ref{sec:DMN} provides the theoretical backbone for DM-nuclear scattering, including the transition from scattering off single nuclei to excitation of collective modes like phonons; Sec.~\ref{sec:DMe} provides the analogous material for DM-electron scattering, moving from single-electron scattering in atoms to a many-body treatment using the dielectric function relevant for solid-state detectors. In Sec.~\ref{sec:Migdal} we discuss the Migdal effect, where DM-nuclear scattering can lead to electronic excitations in atoms or solids, the calculation of which combines the tools developed in the previous sections. In Sec.~\ref{sec:Detection}, we provide our theorists' perspective on the experimental techniques used to detect sub-GeV DM. We conclude in Sec.~\ref{sec:Conclusion} with an outlook on the next decade in the field.

\section{Introduction to Light DM}
\label{sec:DMReview}

In an arbitrary detector of volume $V$  and density $\rho_T$, Fermi's Golden Rule gives the scattering rate for DM per unit target mass \cite{Trickle:2019nya}: 
\be
R_\chi =  \frac{1}{\rho_T}\frac{\rho_\chi}{m_\chi} \int  \, d^3 \vecv  f_\chi(\vecv) \frac{V\, d^3 \vecp'_\chi}{(2\pi)^3} \sum_f |\langle f, \vecp_\chi' | \Delta H_{\chi T} | i, \vecp_\chi \rangle|^2 2 \pi \delta(E_f - E_i + E'_{\chi} - E_\chi).
\label{eq:Rgeneral}
\ee
where $f_\chi(\vecv)$ is the lab-frame DM velocity distribution, $\Delta H_{\chi T}$ is the non-relativistic Hamiltonian governing the interactions between DM and the target constituents, and  $|i \rangle$, $|f \rangle$ are the initial and final detector states with energies $E_i$ and $E_f$ respectively. At this point, the only assumption we have made about the target system is that it can be treated with non-relativistic quantum mechanics, and we make no assumptions about the DM spin. We also generally work with systems in the ground state at zero temperature, so that we do not have to sum over an ensemble of initial states, and we will use $|i \rangle$ and $|0 \rangle$ interchangeably to refer to the initial (ground) state. 

We assume that the DM interactions with the target $\Delta H_{\chi T}$ may be treated as a perturbation on the free-particle DM Hamiltonian, such that unperturbed eigenstates are plane waves $|\vecp \rangle$, and that there is no entanglement between the DM and the target so that $|i, \vecp_\chi\rangle \equiv |i\rangle \otimes |\vecp_\chi \rangle$ and similarly for $|f, \vecp_\chi'\rangle$. To simplify the expression further, we assume a single operator dominates in $\Delta H_{\chi T}$  such that the matrix element factorizes into Fourier components $\vecq$ as:
\begin{align}
\langle f, \vecp_\chi' | \Delta H_{\chi T} | i, \vecp_\chi \rangle &\equiv \! \int  \! \frac{d^3 \vecq}{(2\pi)^3} \, \langle \vecp_\chi' | \mathcal{O}_{\chi}(\vecq) | \vecp_\chi \rangle  \times \langle f | \mathcal{O}_{T}(\vecq) |i\rangle \\
	& = \frac{1}{V} \sqrt{ \frac{\pi \bar \sigma(q)}{\mu_\chi^2} } \langle f | \mathcal{O}_{T}(\vecq) |i\rangle
	\label{eq:factorized_matrixelement}
\end{align}
where the $\mathcal{O}_{\chi}$ and $\mathcal{O}_{T}$ operators only act on the DM and target system states, respectively. In the second line, we have inserted plane wave states for the DM, {\emph e.g.}, $ e^{i \vecp_{\chi} \cdot \vecr}/\sqrt{V}$, and used the fact that the matrix element $\langle \vecp_\chi' | \mathcal{O}_{\chi}(\vecq) | \vecp_\chi \rangle$ will lead to momentum conservation with $\vecq \equiv \vecp_\chi - \vecp_\chi'$. The quantity $(\pi \bar \sigma(q)/\mu_\chi^2)^{1/2}$ (where $q \equiv |\vecq|$) corresponds to the strength of the interaction potential  in terms of a cross section $\bar \sigma(q)$ and mass parameter $\mu_{\chi}$, and we will give examples later for particular models. With this convention, $\mathcal{O}_{T}$ is a dimensionless operator and while it only acts on the target system,  it could still depend on the DM model, such as the strength of DM coupling to the electron, proton, and neutron constituents of the system.

To continue our factorization of the DM and target system portions of the above rate, we can introduce an auxiliary variable $\omega$ and integrate over $\omega$ with a delta function $\delta( \omega + E'_{\chi} - E_\chi)$. This gives the rate as
\be
R_\chi =  \frac{1}{\rho_T}\frac{\rho_\chi}{m_\chi} \int  \, d^3 \vecv  f_\chi(\vecv) \int \frac{d^3 \vecq}{(2\pi)^3} \,  d\omega\, \delta( \omega + E'_{\chi} - E_\chi) \, \frac{\pi \bar \sigma(q)}{\mu_\chi^2}    \times \underbrace{ \frac{2\pi}{V} \sum_f |\langle f | \mathcal{O}_{T}(\vecq) |i\rangle|^2 \delta(E_f - E_i -\omega) }_{\mathlarger{S(\vecq, \omega)}} .
\label{eq:Rfactorized}
\ee
Note that we can swap between $\vecq$ and $\vecp'_{\chi}$ using momentum conservation for the DM, but that we have \emph{not} assumed the target eigenstates have definite momentum. In fact, for the majority of the examples discussed here, the relevant final states in the target will not be momentum eigenstates. Eq.~(\ref{eq:Rfactorized}) gives a factorized form of the rate, where all of the dynamics of the target system are contained in the final terms of the expression. This target response piece is called the {\emph{dynamic structure factor}}, and denoted $S(\vecq, \omega)$. The factor of $1/V$ is included in the normalization to indicate that we are dealing with an intrinsic quantity (since the sum over final states also scales as $V$) rather than an extrinsic quantity.  As noted above, the target response does still depend on details of the DM model. To obtain the rate, the target response is weighted by the DM potential strength, and integrated over the phase space in terms of momentum transferred $\vecq$ and energy deposited $\omega$ by the DM, as well as the DM velocity distribution.   We will use the form of the rate in Eq.~(\ref{eq:Rfactorized}) throughout the review.

We will next explore the kinematics of DM scattering, specified by $f_\chi(\vecv)$ and $E_\chi - E_\chi'$, and then the dynamics that give rise to the interaction strengths $\bar \sigma(q)$. The treatment of the target-dependent piece will be taken up in Sec.~\ref{sec:CMReview}.

\subsection{Kinematics}

\begin{figure}[t!]
\begin{center}
\includegraphics[width=0.55\textwidth]{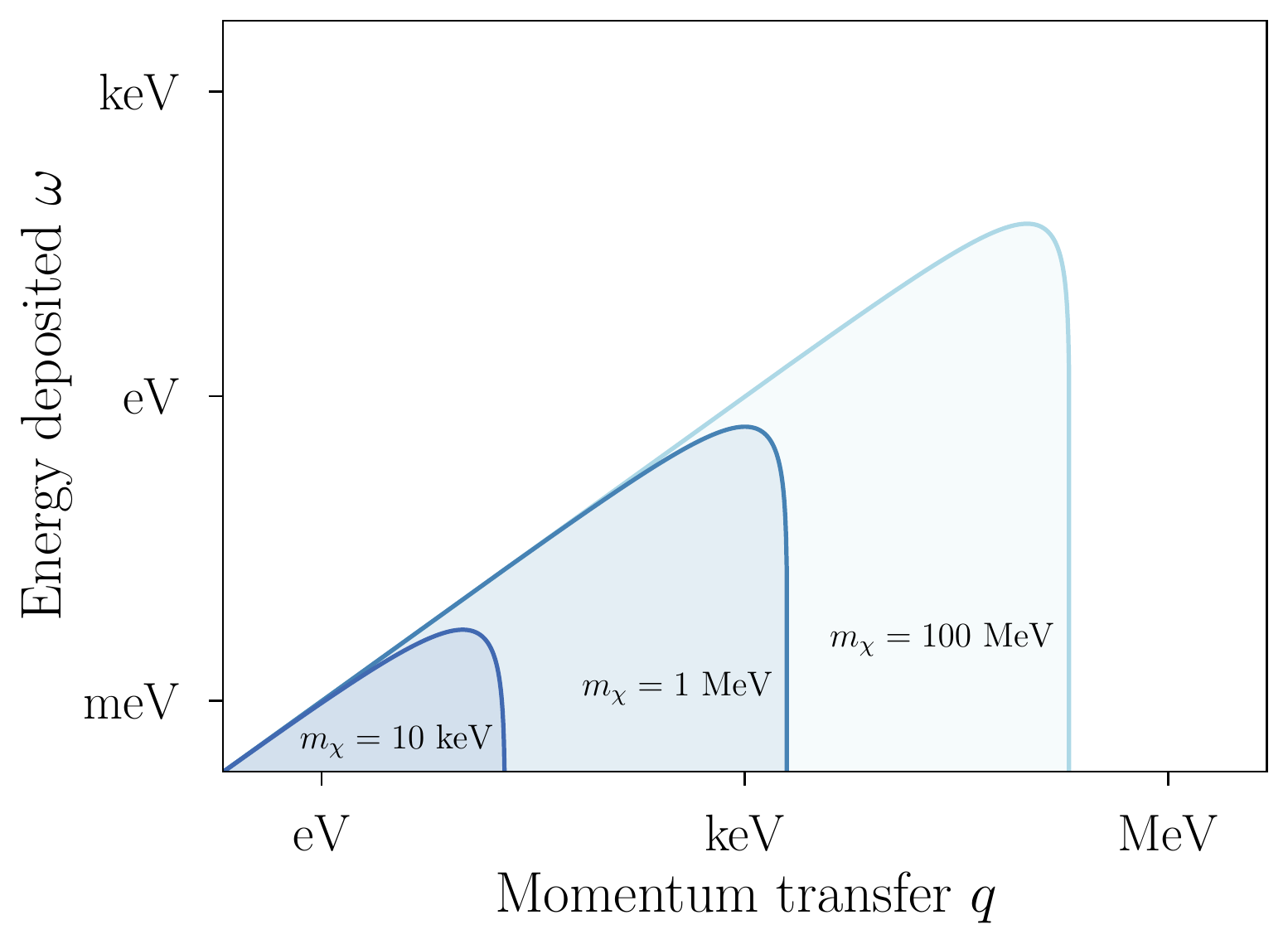}~~
\caption{ \label{fig:parabola}
Kinematically-allowed regions for DM scattering with $v = 10^{-3}$ for various DM masses, adapted from Ref.~\cite{Trickle:2019nya}.
}
\end{center}
\end{figure}

Suppose incoming DM with momentum $\vecp_\chi = m_\chi \vecv$ scatters off a detector target and exits with momentum $\vecp_\chi'$. Using that for nonrelativistic DM, the energy eigenstates of the DM Hamiltonian are $E_\chi = p_\chi^2/2m_\chi$ and $E_\chi' = p_\chi'^2/2m_\chi$, we may write the energy deposited in the target in terms of the momentum transfer $\vecq$:
\be
\omega_\vecq = E_\chi - E_\chi' = \frac{1}{2}m_\chi v^2 - \frac{(m_\chi \vecv - \vecq)^2}{2m_\chi} = \vecq \cdot \vecv - \frac{q^2}{2m_\chi}.
\label{eq:omegaqdef}
\ee
Eq.~(\ref{eq:omegaqdef}) defines the kinematically-allowed region in $\omega, \vecq$ for DM scattering as a function of DM mass and velocity.\footnote{For bosonic DM, there is the additional possibility of \emph{absorption}, where the entire mass-energy of the DM is transferred to the target, yielding $\vecq = m_\chi \vecv$ and $\omega \approx m_\chi$. Condensed matter systems then provide sensitivity to eV-mass DM and below. While we focus exclusively on the scattering process in this review, see Refs.~\cite{An:2013yua,Hochberg:2016ajh,Bloch:2016sjj,Hochberg:2016sqx,SuperCDMS:2019jxx,Arvanitaki:2017nhi,DAMIC:2016qck,Chigusa:2020gfs,Mitridate:2020kly,Mitridate:2021ctr} for a dedicated treatment of absorption in various targets.} As shown in Fig.~\ref{fig:parabola}, for fixed $\vecv$, this region is bounded by an inverted parabola in the $\omega-q$ plane; as $v$ increases, the parabola moves up in $\omega$ since the DM has more kinetic energy. The upper boundary of the parabola corresponds to forward scattering, $\vecq \cdot \vecv = q v$, which gives the largest possible energy deposit $\omega$ for a given $q$. The apex of the parabola corresponds to $q = m_{\chi} v$ and $\omega = \frac{1}{2}m_\chi v^2$, where the target absorbs all of the kinetic energy of the incoming DM and $\vecp_\chi' = 0$. The right boundary of the parabola corresponds to maximum momentum transfer for a given energy deposit, which reduces to elastic ``brick-wall'' scattering when $\vecp'_\chi = -\vecp_\chi$ and $\omega_\vecq \to 0$.

Of course, incoming DM does not have a single velocity $\vecv$, but a range of velocities given by the probability distribution $f_\chi(\vecv)$, the \emph{DM velocity distribution}, which is normalized by $\int d^3 \vecv \, f_\chi(\vecv) = 1$. Eq.~(\ref{eq:omegaqdef}) implies that for a given $\omega$, $q$ in the scattering phase space, there is a minimum DM initial velocity required:
\be
v_{\rm min}(q,\omega) = \frac{\omega_\vecq}{q} + \frac{q}{2 m_{\chi}}
\label{eq:vmingen}.
\ee
We can see this restriction explicitly in the rate by taking an isotropic approximation, in which we assume the target-dependent piece of Eq.~(\ref{eq:omegaqdef}) depends only on $q$ and not on $\vecq$. (Including the full $\vecq$ dependence can be important, however, for anisotropic target systems.) Using the delta function $\delta(\omega - \omega_\vecq)$ to integrate Eq.~(\ref{eq:Rfactorized}) over the angle $\hat \vecq \cdot \hat \vecv$, we obtain the isotropic rate:
\be
	R_{\chi}^{\rm iso} = \frac{1}{\rho_T}\frac{\rho_\chi}{m_\chi} \int \frac{q\, dq}{(2\pi)^{2}} \,  d\omega \, \eta(\vmin(q,\omega)) \times \,\frac{\pi \bar \sigma(q)}{\mu_\chi^2} \times S(q,\omega) 
\label{eq:Riso}
\ee
where we have introduced a function for the mean inverse DM speed:
\be
\eta(\vmin) =  \int_{\vmin}^\infty d^3 \vecv \, \frac{f_{\chi}(\vecv)}{v}.
\label{eq:etadef}
\ee

Accurately determining the DM velocity distribution is a challenging problem which is an active area of research; here we will be content with some simple models which set the relevant scales, and refer the reader to the recent literature on this topic for a more in-depth study. In general, models of the DM velocity distribution come from a combination of simple analytical arguments, simulations which model the many-body gravitational dynamics of forming and merging galaxies, and astrometry data on nearby stars such as the \emph{Gaia} catalogue \cite{Gaia:2018ydn}. Non-gravitational DM self-interactions can qualitatively and quantitatively change the velocity distribution, but these are model-dependent. A complementary approach, which exploits the positivity and normalization properties of $f(\vecv)$ to derive consistency conditions on observed spectra and rates in a \emph{halo-independent} fashion \cite{Fox:2010bz,Fox:2010bu}, was developed for WIMP-nuclear scattering but has recently been applied to sub-GeV DM-electron scattering \cite{Chen:2021qao}.

\begin{table}[t!]
\begin{center}
\begin{tabular}{c|c}
\hline
\hline
Local DM density $\rho_0$ & 0.55 GeV/cm$^3$ \\
\hline
Mean DM speed $v_0$ & 233 km/s \\
\hline
Galactic escape velocity $v_{\rm esc}$ & 528 km/s \\
\hline
Solar velocity in galactic frame & 246 km/s \\
\hline
Earth velocity with respect to the sun & 30 km/s \\
\hline
\end{tabular}
\end{center}
\caption{Approximate parameters for the DM phase space distribution, from \cite{Evans:2018bqy}.}
\label{tab:DMHaloParams}
\end{table}

A good starting point for the DM velocity distribution is the \emph{Standard Halo Model} (SHM) \cite{Drukier:1986tm}, a Maxwellian in the rest frame of the Milky Way:
\be
f_{\rm R}(\vecv) = \frac{1}{(2\pi \sigma_v^2)^{3/2} N_{\rm R, esc}}e^{-\frac{|\vecv|^2}{2\sigma_v^2}} \Theta(\vesc - |\vecv|),
\ee
where $\sigma_v$ is the velocity dispersion and $\vesc$ is the Galactic escape velocity. Note that the distribution has been truncated with a hard cutoff at the escape velocity, leading to a normalization constant
\be
N_{\rm R, esc} = {\rm erf}\left(\frac{\vesc}{\sqrt{2} \sigma_v}\right)- \sqrt{\frac{2}{\pi}} \frac{\vesc}{\sigma_v} {-\frac{\vesc^2}{2\sigma_v^2}}.
\ee
A Maxwellian velocity distribution with no cutoff is a self-consistent solution to the collisionless Boltzmann equation for a spherical isotropic DM density distribution $\rho(r) \propto 1/r^2$ (known in the literature as the ``isothermal sphere'' model), which yields the observed flat rotation curves; the escape velocity cutoff renders the total DM mass finite within the galaxy. The mean DM speed is $v_0 = \sqrt{2} \sigma_v$. By the virial theorem, any test particle at the location of the solar radius $r_\odot$ ($\simeq 8$ kpc from the center of the Milky Way) which has had sufficient time to gravitationally equilibrate will have average velocity equal to $\sqrt{GM/r_\odot}$, where $M$ is the mass enclosed within a radius $r_\odot$. Indeed, the Sun itself is one such test particle, so one may infer the mean DM speed (and hence the velocity dispersion) by setting $v_0$ equal to the ``Local Standard of Rest,'' which is the circular velocity of the Sun around the center of the Milky Way, corrected for the small random ``peculiar velocity'' of the Sun with respect to its neighboring stars. Similarly, the escape velocity may be bounded from below by the maximum velocity of the fastest stars in the galaxy. Some representative values for $v_0$ and $\vesc$ are given in Tab.~\ref{tab:DMHaloParams}; see \cite{Evans:2018bqy,Baxter:2021pqo} for discussions of the associated uncertainties.

In the arguments above, the Milky Way dark matter halo is idealized as a self-gravitating spherical distribution in equilibrium, but in our cosmological history, the Milky Way formed through a rich history of mergers of smaller halos. A wealth of data, both from simulations and observations, have persuasively shown that the true DM velocity distribution in the Milky Way is not a perfect Maxwellian. Simulations of halo formation give access to the DM phase space distribution directly and suggest that the bulk of the distribution near $v_0$ differs from a Maxwellian at the $\sim 20\%$ level \cite{Vogelsberger:2008qb}, which is tied to the fact that the density distribution is not exactly $1/r^2$. The high-velocity tail is the most uncertain, as it relies critically on the merger history of the Milky Way, which is a stochastic process and requires careful observational reconstruction. In general, mergers lead to imprints in both the DM distribution and stellar distribution, such that observations of the stellar phase space distribution can be used to infer the Milky Way's history. In some cases, the result is small-dispersion remnants called \emph{streams}; one striking example is the Sagittarius stellar stream \cite{2003ApJ...599.1082M,2006ApJ...642L.137B}, which may have a DM component that comprises $\sim 5\%$ of the local DM density with a local velocity $v_{\rm Sag} \simeq 400$ km/s and dispersion $\sigma_{\rm Sag} \simeq 10$ km/s. Other mergers may leave more diffuse remnants, like the radially-anisotropic stellar substructure observed by \emph{Gaia} (dubbed the ``Gaia Enceladus'' or ``Gaia sausage''), implying corresponding DM substructure that may be $\sim 20\%$ of the local DM density \cite{Evans:2018bqy,Bozorgnia:2019mjk}. It is an open question to what extent the smooth component of the DM distribution may be correlated with stars, but the consequences for light DM phenomenology are largely driven by the quasi-Maxwellian bulk of the distribution, though important effects at experimental thresholds do arise from the high-velocity tail \cite{Radick:2020qip,Buch:2020xyt}.

Since direct detection experiments are done in terrestrial laboratories, what we actually need is the lab-frame DM distribution $f_\chi(\vecv)$, which may be obtained from the rest-frame distribution $f_{\rm R}(\vecv)$ by a Galilean boost along the velocity of the Earth $\vecv_\oplus$:
\be
f_\chi(\vecv; t) = f_{\rm R}(\vecv + \vecv_\oplus(t)).
\ee
Here, $\vecv_\oplus$ represents the velocity of the Earth with respect to the Galactic rest frame, which contains both the solar velocity with respect to the Galactic rest frame and the relative Earth-Sun motion. Note that even though the rest-frame distribution is stationary, the lab-frame distribution acquires a time dependence due to the yearly motion of the Earth around the Sun (\emph{annual modulation}) and the rotation of the Earth over 24 hours (\emph{daily modulation}). The velocity of the Earth with respect to the Sun is about 30 km/s, so the dominant effect of annual modulation is a $\sim 10\%$ change in both the DM flux and the high-velocity cutoff of the distribution depending on the time of year. By contrast, the linear velocity of the Earth's surface at the equator is only $\sim 0.5$ km/s, so the effect of daily modulation is an order-1 change in the \emph{direction} of the mean DM velocity, but not the speed or the flux. Experiments which are sensitive to the direction of the DM ``wind'' thus have an important handle on the DM distribution, and as we will see, light DM experiments are particularly suited to this observable. Over the course of a day, during which $|\vecv_\oplus|$ can be taken as a constant, a convenient parameterization of the direction of $\vecv_\oplus(t)$ relevant for daily modulation is
\be
\vecv_\oplus (t) = |\vecv_\oplus|
\left( \! \begin{array}{c} \sin\theta_e \sin \vartheta \\ \sin\theta_e \cos\theta_e ( \cos \vartheta - 1) \\ \cos^2 \theta_e + \sin^2 \theta_e \cos\vartheta \end{array} \! \right),
\ee
where $\vartheta = 2\pi \times \left( \tfrac{t}{24\, \text{h} } \right)$, 
$\theta_e \approx \, 42^\circ$ is the inclination of the Earth's rotation axis, and at time $t=0$ in these coordinates the $(x,y)$ plane of a dark matter detector is perpendicular to the DM wind.\footnote{Note that the time dependence of $|\vecv_\oplus|$ leads to annual modulation, which can also have important effects on light DM scattering \cite{Lee:2015qva}.} 

Accounting for the daily variation of the DM velocity distribution in the lab frame then leads to a daily modulation of the event rate. To calculate the time-dependent rate per unit mass, it is useful to absorb the energy-conserving delta function into the velocity distribution, yielding
 \be
 g(\vecq, \omega; t) \equiv \int d^3 \vecv \, f_R(\vecv + \vecv_\oplus(t)) \delta(\omega - \omega_\vecq)
 \label{eq:gdef}
 \ee
as the anisotropic analogue to $\eta(\vmin)$ with time dependence through $\vecv_\oplus(t)$. Rearranging the factors in Eq.~(\ref{eq:Rfactorized}), we obtain
\be
R_\chi(t) =  \frac{1}{\rho_T}\frac{\rho_\chi}{m_\chi}   \int \frac{d^3 \vecq}{(2\pi)^3} \,  d\omega \,  g(\vecq, \omega; t) \, \frac{\pi \bar \sigma(q)}{\mu_\chi^2} S(\vecq, \omega).
\label{eq:Rdaily}
\ee

The local DM density $\rho_\chi$ normalizes the overall rate of any DM direct detection experiment. Historically, measurements of this quantity have relied on vertical accelerations of stars outside of the plane of the Galactic disk \cite{Read:2014qva}, leading to a $\sim 50\%$ uncertainty.\footnote{The light DM community has traditionally used both $\rho_\chi = 0.3 \ \GeV/{\rm cm}^3$ or $\rho_\chi = 0.4 \ \GeV/{\rm cm}^3$ for most experimental limits, though both of these values are likely outdated; some care is therefore required to translate between limits from different experiments.} This method uses the Jeans equations and assumes the disk stars are in equilibrium, which is already in conflict with some \emph{Gaia} observations; new techniques may use angular stellar accelerations from \emph{Gaia} to determine $\rho_\chi$ directly from the Poisson equation with a minimum of assumptions \cite{Buschmann:2021izy}. The density at Earth may also acquire a time dependence through the ``gravitational focusing'' effect \cite{Lee:2013wza}, where slower DM particles are bent towards Earth when the Earth is behind the Sun, increasing the density of slow DM particles in March. This may compete with the annual modulation signal from the speed distribution, where the DM flux peaks in June. Indeed, a study of the full kinematics of DM requires treatment of the entire 6-dimensional phase space distribution $f_\chi(\vecx, \vecv; t)$.

\subsection{Dynamics}

We now turn to the dynamics of DM, namely its interactions with SM particles and with itself. For sub-GeV DM in particular, the interactions are strongly constrained by the related requirements of a consistent thermal history of DM (leading to a late-time abundance of DM, the \emph{relic abundance}, which matches CMB observations) and the suppression of any additional sources of energy injection after the time of the CMB, which could reionize the universe and distort the CMB anisotropies. Therefore, to understand the interactions of DM in the laboratory, we first briefly review DM interactions in the early universe.

\subsubsection{Early universe}

As a starting point, consider the hypothesis that DM was once in thermal equilibrium with the SM. At temperatures well above $m_\chi$, annihilation processes such as $\chi \bar{\chi} \to e^+ e^-$ occur at equal rates to the reverse reaction $e^+ e^- \to \chi \bar{\chi}$. As the temperature $T$ of the universe drops below the DM mass, the reverse reaction becomes Boltzmann-suppressed and the DM number density drops exponentially. However, all of these processes are taking place in an expanding universe, and once the annihilation rate $\Gamma_{\chi \bar{\chi}}$ drops below the expansion rate, the annihilation shuts off and the DM abundance becomes fixed. This sequence of events is known as \emph{thermal freeze-out} \cite{kolb1994early}, and directly relates the annihilation cross section to the relic abundance; the larger the cross section, the less DM left in the universe today. The existence of an annihilation channel which couples DM to the SM, for example $\chi \bar{\chi} \to e^+ e^-$, implies that there must be a related scattering process $\chi e^- \to \chi e^-$ which permits direct detection. There is an important caveat to this story, which is that at late times where $T \ll m_\chi$, the annihilation rate is never exactly zero. Even if $\Gamma_{\chi \bar{\chi}}$  is small enough to not meaningfully affect the overall DM density, residual annihilations can still inject enough ionizing particles into the CMB to distort the observed anisotropies. The upshot is that DM lighter than 10 GeV is ruled out if its thermally-averaged annihilation cross section is independent of velocity \cite{Slatyer:2015jla}. This does not rule out \emph{all} sub-GeV DM candidates, but it does place important restrictions on the spin and interactions of light DM from the freeze-out mechanism --  for example, Dirac fermion DM is ruled out but scalar DM is allowed \cite{Izaguirre:2015yja}.

It is also possible for light DM to be in thermal contact with the SM without being in thermal equilibrium. A well-studied example of this is the \emph{freeze-in mechanism} \cite{Hall:2009bx,Essig:2011nj,Chu:2011be,Dvorkin:2020xga}, where the initial abundance of DM is zero, but very weak interactions in the SM plasma populate the DM with $e^+ e^- \to \chi \bar{\chi}$. The DM abundance is always small enough that the reverse reaction does not occur with an appreciable rate, and the DM never equilibrates. The production shuts off at late times when the temperature drops below $m_\chi$, or for $m_\chi < m_e$, when the temperature drops below $m_e$ and positrons drop out of equilibrium. This model avoids the CMB energy injection constraints because DM annihilation never occurs. If we allow for the possibility of number-changing DM interactions, such as $\chi \chi \to \chi \chi \chi$, there are several other scenarios for generating the correct relic abundance, including strongly-interacting DM (SIMP) \cite{Hochberg:2014dra,Hochberg:2014kqa} and elastically decoupling relics (ELDER) \cite{Kuflik:2015isi,Kuflik:2017iqs}. The possibilities expand further if we allow a \emph{dark sector}, containing the DM and possibly other particles, that is thermalized with its own temperature $T_\chi$. For our purposes it suffices to note that there are multiple examples of viable models for light DM.

There are many other important bounds from cosmological observations on the parameter space of light DM. The most important (and least model-dependent) is the \emph{warm dark matter bound}: DM which was in thermal equilibrium with the SM must have mass greater than $\sim 1$ keV, or it would have been too relativistic to gravitationally clump. More precisely, requiring that DM not damp the observed matter power spectrum constrains $m_\chi \gtrsim 12 \ \keV$, and the additional interaction of DM with baryons would produce a drag force which would affect CMB anisotropies, strengthening the bound to $m_\chi \gtrsim 20 \ \keV$ \cite{Dvorkin:2020xga}. In fact, a similar bound applies for freeze-in DM: despite the fact that it was never in thermal equilibrium, its phase space distribution inherits some of the properties of the SM plasma. There are also upper bounds on the DM-proton and DM-electron cross sections for the massive mediator limit, though at values well above those required for thermal freeze-out or freeze-in \cite{Nadler:2019zrb,Maamari:2020aqz}. In addition, there are constraints on the DM self-interaction cross section, $\sigma_{\chi \chi}/m_\chi \lesssim 1 \ {\rm cm}^2/{\rm g}$~\cite{Tulin:2017ara}. These bounds are somewhat subtle because simulations typically assume contact interactions between DM, while exchange of a light mediator would lead to a long-range force. Even with all of these constraints, though, light DM remains a viable possibility, and indeed some of the strongest constraints on the DM-CM interaction strength in the mass range of MeV--GeV are now coming from direct detection experiments. 

\subsubsection{Laboratory interactions}

The thermal histories for sub-GeV DM described above require, at a minimum, one additional ingredient: a new force which mediates the thermal contact between the DM and SM. Indeed, DM cannot interact with the SM through the strong force (otherwise DM would not be ``dark'' with respect to baryons), and neither can it be the weak force, which has too small of an annihilation cross section to generate the correct relic abundance of sub-GeV DM. In principle, it could be the photon if DM had a small enough electric charge to be cosmologically ``dark,'' but the CMB excludes this possibility for freeze-out because such a small charge would not lead to sufficient annihilation and would yield an overabundance of DM unless other annihilation channels are introduced~\cite{McDermott:2010pa}.

A benchmark model of such a new force is a \emph{dark photon}  \cite{Fayet:1980ad,Fayet:1980rr,Holdom:1985ag,Okun:1982xi}, denoted $A'$. In this model, DM has a charge $g_{D}$ under a ``dark'' version of electromagnetism, but unlike electromagnetism, the dark photon may be massive with mass $m_{A'}$. In addition, because the quantum numbers of the dark photon are the same as the ordinary photon, the two states may mix, which is usually introduced as a \emph{kinetic mixing} parameter $\varepsilon \ll 1$. This mixing implies that particles with electromagnetic charge $Qe$ also have a dark photon charge, which is given by $\varepsilon Qe$. Combined, the dark photon couplings with the dark matter and charged particles allows for thermal contact between the dark matter and SM. In certain regions of parameter space, the requirement of obtaining the correct relic abundance fixes the size of the couplings \cite{Alexander:2016aln,Battaglieri:2017aum}:
\be
\alpha_D \varepsilon^2 \simeq \begin{cases}10^{-14} \frac{m_{A'}^4}{m_\chi^2 \ \MeV^2}, \qquad {\rm freeze-out \ } (m_e < m_\chi < m_{A'}) \\ 10^{-24} \frac{m_e}{m_\chi}, \qquad {\rm freeze-in \ } (m_{A'} \ll m_{\chi}, \ m_{\chi} < m_e).\end{cases}
\ee
For a given $m_\chi$ and $m_{A'}$, then, these thermal histories predict the DM scattering rate at direct detection experiments, leading to concrete \emph{thermal targets} in parameter space which are the goals of a number of experimental programs.

In the non-relativistic limit, the dark photon model yields the following interaction Hamiltonian between DM and charged particles, to leading order in the relative velocity:
\be
\Delta H_{\chi Q} =  \int \frac{d^3 \vecq}{(2\pi)^3} e^{i \vecq \cdot (\vecr_Q - \vecr_\chi)} \frac{\varepsilon Q e g_D}{q^2 + m_{A'}^2}
\label{eq:DarkPhotonHamiltonian}
\ee
where $\vecr_\chi$ is the DM position operator, $\vecr_Q$ is the position operator of a particle of electric charge $Qe$, $e = \sqrt{4\pi \alpha}$ is the electron charge with $\alpha \simeq 1/137$ the fine structure constant, and $g_D$ is the dark charge.\footnote{Note that we are using Heaviside-Lorentz conventions for the electric charge as is common in high-energy physics, where $\alpha = e^2/(4\pi)$. This differs by factors of $4\pi$ from cgs-Gaussian units where $\alpha = e^2$.} Because the potential is translation-invariant and depends only on the relative coordinate $\vecr_\chi - \vecr_Q$, the matrix element of Eq.~(\ref{eq:DarkPhotonHamiltonian}) may be evaluated between plane-wave DM states:
\be
\langle \vecp_\chi' | \Delta H_{\chi Q} | \vecp_\chi \rangle = \int  \frac{d^3 \vecq}{(2\pi)^3} \frac{d^3 \vecr_\chi}{V} \, e^{i (\vecp_\chi - \vecp'_\chi) \cdot \vecr_\chi} e^{i \vecq \cdot (\vecr_Q - \vecr_\chi)} \frac{\varepsilon Q e g_D}{q^2 + m_{A'}^2} =  \frac{1}{V} \frac{\varepsilon Q e g_D}{q^2 + m_{A'}^2} e^{i \vecq \cdot \vecr_Q}
\label{eq:DMMatrixElement}
\ee
where in the last equality the integration over the DM coordinate enforces momentum conservation, $\vecq = \vecp_\chi - \vecp_\chi'$. The matrix element of $\Delta H_{\chi Q}$  thus has the factorized form of Eq.~(\ref{eq:factorized_matrixelement}), with
\be
	\langle f,\vecp_\chi' | \Delta H_{\chi Q} | i, \vecp_\chi \rangle = \frac{1}{V} \frac{\varepsilon Q e g_D}{q^2 + m_{A'}^2} \langle f | e^{i \vecq \cdot \vecr_Q} | i \rangle \equiv \frac{1}{V} \mathcal{V}(q)  \langle f | e^{i \vecq \cdot \vecr_Q} | i \rangle
\ee  
where we identify the cross section $\bar \sigma(q)$ as proportional to the scattering potential $\mathcal{V}(q)$, 
\be
	\bar \sigma(q) = \frac{\mu_{T\chi}^2}{\pi} \left (\frac{\varepsilon Q e g_D}{q^2 + m_{A'}^2}\right)^2 \equiv  \frac{\mu_{T\chi}^2}{\pi} (\mathcal{V}(q))^2
\ee
and $\mu_{T\chi} = \frac{m_T m_\chi}{m_T + m_\chi}$ is the DM-target reduced mass; for a target proton or electron, for instance, $m_T = m_p$ or $m_e$ respectively.

It is common in the DM literature to rewrite $\bar \sigma(q) = \bar \sigma_{T} F_{\rm DM}^{2}(q)$, where $\bar \sigma_{T}$ is a fiducial cross section at fixed momentum $q_{0}$,
\be
\bar{\sigma}_T = \frac{\mu_{T\chi}^2}{\pi} \left (\frac{\varepsilon Q e g_D}{q_0^2 + m_{A'}^2}\right)^2
\ee
 and $F_{\rm DM}(q)$ is a momentum-dependent  \emph{DM form factor}
\be
F_{\rm DM}(q) \equiv \frac{q_0^2 + m_{A'}^2}{q^2 + m_{A'}^2}.
\label{eq:FDMDef}
\ee
which parameterizes the momentum dependence of the scattering potential. For $T = e$, $\bar \sigma_{T}$ can be interpreted as a cross section for DM scattering off a free electron at a reference momentum $q_0$, which is typically taken to be the inverse Bohr radius, $q_0 = 1/a_0 = \alpha m_e \simeq 3.7 \ \keV$. For $T = p$, $\bar{\sigma}_p$ is the DM-proton cross section and $q_0$ is an arbitrary reference momentum which is often taken to be $q_{0} = m_\chi v_{0}$. The two limits $F_{\rm DM} \to 1$ and $F_{\rm DM} \to (q_0/q)^2$ correspond to a \emph{heavy mediator}, $m_{A'} \to \infty$, or \emph{light mediator}, $m_{A'} \to 0$, respectively. Since the mass of the dark photon is unknown, these two limiting cases span the range of possibilities for the scattering amplitude. In position space, the heavy mediator limit corresponds to a contact interaction, $\mathcal{V}(\vecr_\chi - \vecr_Q) \propto \delta^{(3)}(\vecr_\chi - \vecr_Q)$. 

Plugging in some numerical values, we find that for the freeze-out scenario with $m_{A'} > m_\chi$, the typical electron cross section is
\be
\bar{\sigma}_e  \simeq 3 \times 10^{-39} {\rm cm}^2 \left(\frac{10 \ \MeV}{m_\chi}\right)^2,
\ee
independent of $m_{A'}$, $g_D$, and $\varepsilon$. This is a feature, not a bug, because the same DM-SM interaction (with the same $m_{A'}$ dependence) fixed the relic abundance in the early universe. Assuming a typical electron density of $n_e = 10^{24}$/cm$^3$ in a generic material, the mean free path of a 10 MeV DM particle in a generic detector is
\be
\lambda = (n_e \bar{\sigma}_e)^{-1} \simeq 4 \times 10^{12} \ {\rm m}.
\ee
Unlike ordinary Coulomb scattering between charged SM particles, then, there is no possibility of multiple scattering in any detector (or even of correlating scattering events between two nearby detectors on an event-by-event basis); thermal relic DM experiments are truly rare-event searches.

The dark photon model illustrates the ``top-down'' approach, where we began with a particular model of DM dynamics in the early universe to derive DM interactions in the laboratory. That approach predicted a particular coupling strength of DM to electron and proton number density in the nonrelativistic limit. Another approach one might take is to start with a general scalar or vector mediator coupling to electron, proton, or neutron number density. Motivated by the search for WIMP-nucleus scattering, the other case we will consider in this review is DM that couples to protons and neutrons only, mediated by a Yukawa interaction. The DM-nucleon Hamiltonian is given by
\be
\Delta H_{\chi n} =  \int \frac{ d^{3} \vecq}{(2\pi)^{3}} \ e^{i \vecq \cdot (\vecr_{n} - \vecr_{\chi})}  \frac{y_n y_{D}}{q^{2} + M^{2}}
\label{eq:nucleonHamiltonian}
\ee
for a mediator of mass $M$, where $n$ denotes either a proton or a neutron. The coupling $y_{n}$ now plays the role of the charge of a nucleon with respect to this new mediator, and $y_{D}$ is the DM coupling. We will assume equal proton and neutron coupling for simplicity. As before, we can define a DM-nucleon fiducial cross section
\be
	\bar \sigma_{n}  = \frac{\mu_{\chi n}^2}{\pi} \left (\frac{y_n y_D}{q_0^2 + M^2}\right)^2
\ee
where $q_{0} = m_{\chi} \sigma_{v}$ as before. Again, there is also a DM form factor $F_{\rm DM}(q)$, which is identical to Eq.~(\ref{eq:FDMDef}) but with the replacement $m_{A'} \to M$.

For this benchmark, the cosmology is quite different from the previous dark photon model. For sub-GeV DM, the annihilation process $\chi \bar\chi \to n \bar n$ is clearly not possible when the temperature of the universe is well below $T \approx$  GeV, while at higher temperatures, one needs to specify a microscopic coupling of DM to quarks or gluons, which can be model-dependent. In addition, there are strong constraints on mediators coupling to quarks or gluons from observations of rare meson decays. The upshot is that thermal freeze-out scenarios with sharp benchmark values of $y_{n} y_{D}$ are excluded for sub-GeV dark matter~\cite{Krnjaic:2015mbs}. One can also consider enlarging the dark sector, which does lead to viable thermal relic possibilities. Combining cosmological and laboratory constraints then lead to upper bounds on $y_{n} y_{D}$ as a function of $M$, and therefore on $\bar \sigma_{n}$. For MeV-GeV mass dark matter and the massless mediator limit $(M \ll q_{0})$, the bounds on $\bar \sigma_{n}$ are the weakest, with the potential for large signals in direct detection experiments. However, there are more stringent limits on sub-MeV DM and the massive mediator limit~\cite{Knapen:2017xzo,Krnjaic:2015mbs,Green:2017ybv}. Despite these caveats, for the sake of uniformity we will primarily consider this type of interaction throughout our discussion of DM-nucleus interactions in Sec.~\ref{sec:DMN}.

Finally, to generalize the interactions even further, we can take a completely  ``bottom-up'' approach where all possible DM-SM interactions consistent with Galilean and translation invariance are enumerated. As discussed in Refs.~\cite{Fitzpatrick:2012ix,Catena:2019gfa}, in the non-relativistic limit there are 14 operators associated with exchange of a new bosonic mediator of mass $M$, all of which contribute to the Hamiltonian as
\be
\Delta H_{\chi T} = \int \frac{d^3 \vecq}{(2\pi)^3} e^{-i \vecq \cdot \vecr_\chi} F_{\chi}(\vecq, \vecp_\chi, \vecS_\chi) \times \frac{1}{q^2 + M^2} \times e^{i \vecq \cdot \vecr_\psi} F_{T}(\vecq,\vecp_\psi, \vecS_\psi)
\label{eq:BottomUpHamiltonian}
\ee
where $\psi = e, p, n$ is a Standard Model fermion.  $F_{\chi}$ is a function only of momentum transfer $\vecq$ and properties of $\chi$ such as $\vecp_\chi$ and $\vecS_\chi$, while $F_{T}$ similarly depends only on $\vecq, \vecp_\psi$ and $\vecS_{\psi}$.  This again gives the same factorized matrix elements as in Eq.~(\ref{eq:factorized_matrixelement}). $M$ is the mass of the mediator which may be a scalar, pseudoscalar, vector, or axial vector. In all cases, the potential still only depends on the relative coordinate between $\chi$ and $\psi$. As before, the DM part of the matrix element may be factored out, with the remaining piece defining a target response function. Without a top-down model to rely on, there are no a priori target values for the coefficients of these operators (though non-renormalizable operators should be suppressed by a sufficiently high energy scale that they would not have already been probed in high-energy collider experients), but some combinations arise from top-down models in the non-relativistic limit. For example, in addition to the $e^{i \vecq \cdot \vecr_Q}$ operator in Eq.~(\ref{eq:DarkPhotonHamiltonian}), a light dark photon mediator will let the DM source a dark magnetic field proportional to the DM velocity, which will couple to the electron spin analogous to the spin-orbit coupling which contributes to the fine structure of hydrogen. As illustrated by this example, spin-dependent interactions which arise from top-down models are typically suppressed compared to spin-independent interactions by the small DM velocity, $v_\chi \ll 1$, and/or involve low-mass parity-odd mediators such as axial vectors which are highly constrained by cosmological, astrophysical, or low-energy observables (see for example Ref.~\cite{Kahn:2016vjr}). For these reasons, we focus on the spin-independent interaction as a benchmark model in this review.

\section{Introduction to Condensed Matter Systems}
\label{sec:CMReview}

 In contrast to the panoply of possible DM masses and interactions, the non-relativistic Hamiltonian of any condensed matter system is universal:
\be
H_{\rm CM} = \sum_k \frac{\vecp_k^2}{2m_e} + \sum_I \frac{\vecp_I^2}{2M_I} - \alpha \sum_{I, k} \frac{Z_I}{ |\vecr_k - \vecr_I|} + \frac{\alpha}{2} \sum_{k \neq l} \frac{1}{|\vecr_k - \vecr_l|} + \frac{\alpha}{2} \sum_{I \neq J} \frac{Z_I Z_J}{|\vecr_I - \vecr_J|}.
\label{eq:CMham}
\ee
Here, lowercase letters index electrons and uppercase letters index ions of charge $Z_I$ with masses $M_I$, and the sums run over all the electrons and ions in the material. Despite the fact that the only interactions are through the Coulomb potential (and its relativistic generalization, which includes for example spin-orbit coupling), it is obviously impossible to exactly solve the associated many-body Schr\"{o}dinger equation when there are Avogadro's number of terms. 

In the first part of this section, which largely follows Ref.~\cite{kaxiras2019quantum}, we review some of the techniques to determine two of the elementary excitations that appear in any solid state system -- electrons and phonons -- and give a basic description of their properties. This provides the groundwork, with which we can focus on the problem of most interest for dark matter detection: computing the condensed matter part of the matrix element which appears in the DM scattering rate, Eq.~(\ref{eq:Rgeneral}):
\be
S(\vecq, \omega) \equiv \frac{2\pi}{V} \sum_f |\langle f | \left(\sum_k f_{e} e^{i \vecq \cdot \vecr_k} + \sum_I f_I e^{i \vecq \cdot \vecr_I}\right) | i \rangle|^2 \delta(E_f - E_i - \omega)
\label{eq:Sqw}
\ee
where $f_{e}$ is a (normalized) DM coupling with electrons and $f_{I}$ is the DM coupling to ions, and we have summed over all constituents of the system. Here we take $f_e$ and $f_I$ to be $\vecq$-independent constants because we have factored out all of the $q$ dependence into the DM cross section in Eq.~(\ref{eq:factorized_matrixelement}). 

The function $S(\vecq, \omega)$ is known as the \emph{dynamic structure factor}. Note that different conventions exist in the literature for the overall normalization of $S(\vecq, \omega)$ in terms of factors of $2\pi$ and volume, and the couplings $f_{e,I}$ are typically normalized relative to an overall interaction strength. For example, for the dark photon model, we would be interested in the operator
\be
\mathcal{O}(\vecq) = \frac{\varepsilon e g_D}{q^2 + m_{A'}^2}\left(-\sum_k e^{i \vecq \cdot \vecr_k} + \sum_I Z_I e^{i \vecq \cdot \vecr_I}\right).
\label{eq:ODarkPhoton}
\ee
Typically the prefactor in front is absorbed into a fiducial DM cross section and form factor, leading to the natural definition $f_{e} = -1$ and $f_{I} = Z_{I}$ in this work. Note also that we will focus on the particular choice of structure factor above, which depends only on the position operators for electrons $\vecr_k$ and ions $\vecr_I$, as this is the structure factor relevant for the most commonly-studied models in the literature. In other models, the leading nonrelativistic coupling could have additional dependence on the target momenta and spins, and requires defining additional structure factors.

In Eq.~(\ref{eq:Sqw}), it is important to note that the initial and final states $ | i \rangle, | f \rangle$ are states of the interacting many-body system, not states for noninteracting particles. This is obvious for the case where the final state consists of phonons, which are collective excitations of the ions, but it is also true for the case where DM couples to electrons. The systems we focus on in this review are those for which the internal interactions among microscopic constituents may be strong, but where the response at low energy and momentum transfer may be described by long-lived elementary excitations, such as phonons and electron quasiparticles. A key quantity which determines the importance of including these many-body states is the momentum transfer $\vecq$ from the DM to the CM system. For sufficiently large $\vecq$, the cross terms between different electrons or nuclei will have large relative phases and average out to zero, and the scattering can be effectively treated as if DM interacted incoherently with an individual electron or nucleus. As a very rough estimate, the scale at which many-body effects become crucial is
\be
q_{\rm coh} \lesssim \frac{2\pi \hbar}{a} \simeq 5 \ {\rm keV}/c,
\label{eq:qcoh}
\ee
where $a$ is the nearest-neighbor spacingof lattice sites in typical solids and we have temporarily restored $\hbar$ and $c$ for clarity. Of course, this is only an estimate and for accurate rates, many-body effects also need to be accounted for at larger momentum transfers. This is particularly true if scattering is restricted to low energy transfers $\omega$, comparable to the energy thresholds of the elementary excitations. In any case, the point is that this kinematic regime can be accessed by DM in non-mutually-exclusive ways:
\begin{itemize}
\item Sub-MeV DM carries a maximum momentum of $m_\chi v \lesssim 1 \ \keV$, so $q < 2 m_\chi v < q_{\rm coh}$ for any interaction. Thus, many-body effects are crucial for the lightest DM candidates. 
\item DM of any mass interacting through a light mediator. As $m_{A'} \to 0$ in Eq.~(\ref{eq:ODarkPhoton}), the prefactor scales as $1/q^4$ and thus the rate integral is weighted toward the smallest kinematically-allowed momentum transfers.
\end{itemize}
In contrast, for GeV-scale WIMPs, the momentum transfers which lead to detectable energy deposits are large enough that the DM scattering can be treated as single-particle scattering.

In the remainder of this section, we will give an overview of theoretical tools used to describe many-body excitations in Sec.~\ref{sec:CMexcitations} and then elaborate on how these tools are used to compute the dynamic structure factor in Sec.~\ref{sec:DMstructurefactor}.
We will then apply this to calculate DM scattering rates in Secs.~\ref{sec:DMN}, \ref{sec:DMe}, and \ref{sec:Migdal}. Aside from this theoretical treatment, one can also use direct measurements of the dynamic structure factor, as determined by SM probes of the material. In the case of DM-electron scattering, the appropriate dynamic structure factor is related to the \emph{complex dielectric response}, which measures the linear response of a given material to spatially- and temporally-varying electromagnetic perturbations. In the case of DM-nucleus scattering, a similar dynamic structure factor also governs neutron scattering. We also briefly comment on the possibility of using experimental measurements to determine $S(\vecq, \omega)$ in this section.

\subsection{Elementary excitations: band structures and quasiparticle excitations \label{sec:CMexcitations} }

Given that Eq.~(\ref{eq:CMham}) is impossible to solve exactly, the key to condensed matter theory is that we can describe systems with {\emph{emergent}} weakly-interacting many-body states, also known as quasiparticles in the case of electrons. Electron quasiparticles have the same charge and spin as electrons, but are truly many-body states and can have a different spectrum of excitations. Nevertheless, we will see a useful approximation is to treat the electron quasiparticles with a ``single-particle'' wavefunction. As is common, we will also simply refer to these excitations as electrons. In the case of the ions, the emergent modes are the phonons, which can be directly obtained by a change of basis into collective coordinates; as these modes do not have a direct particle counterpart, they are more often called ``collective excitations.'' The interactions of the electrons also lead to a collective excitation, called the plasmon, which we will touch on in Sec.~\ref{sec:DMstructurefactor}. There are also many other emergent modes beyond what we will discuss here, for instance spin waves, which are called \emph{magnons}.

For the majority of this review, we will consider crystalline solid-state systems, where the constituent atoms are arranged in periodic lattices. (Superfluid helium, discussed in Sec.~\ref{sec:helium}, will require a different treatment.)
The minimal repeating structure in such a lattice is called the \emph{primitive unit cell}, and in 3 dimensions can be defined by three basis vectors called the primitive lattice vectors, ${\bf a}_{1}, {\bf a}_{2}, {\bf a}_{3}$. Any point in the crystal may be reached by translating a point in the unit cell centered at the origin by a lattice vector $\vecR$ which is a linear combination of the primitive lattice vectors with integer coefficients. The discrete translational invariance imposed by the lattice yields a discrete Fourier transform: any function periodic on the lattice, $f(\vecr) = f(\vecr + \vecR)$, may be written as
\be
f(\vecr) = \sum_{\vecG} e^{i \vecG \cdot \vecr} f(\vecG); \qquad e^{i \vecG \cdot \vecR} = 1,
\label{eq:GExpand}
\ee
where the vectors $\vecG$ are \emph{reciprocal lattice vectors}. In Fourier space (conventionally referred to as \emph{reciprocal space}), the analogue to the unit cell are the reciprocal space primitive lattice vectors ${\bf b}_{1}, {\bf b}_{2}, {\bf b}_{3}$ where ${\bf a}_{i} \cdot {\bf b}_{j} = 2 \pi \delta_{ij}$, and $\vecG$ is any linear combination of ${\bf b}_{i}$ with integer coefficients. 
The reciprocal space primitive lattice vectors can be defined as the inverse of the matrix of lattice vectors, with an explicit expression
\begin{align}
	{\bf b}_{i} = 2 \pi  \sum_{j,k} \frac{\epsilon_{ijk} \, {\bf a}_{j} \times {\bf a}_{k} }{ ({\bf a}_1 \times {\bf a}_2) \cdot {\bf a}_3}.
\end{align}

\begin{figure}[t!]
\begin{center}
(a) \hspace{1.2cm} \includegraphics[width=0.25\textwidth]{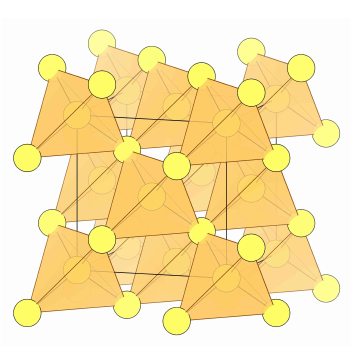} \hspace{1.5cm}
(b) \hspace{1.2cm}\includegraphics[width=0.3\textwidth]{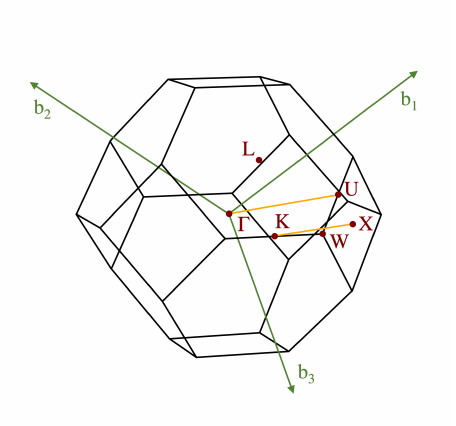}  \vspace{0.5cm}\\
(c)\includegraphics[width=0.42\textwidth]{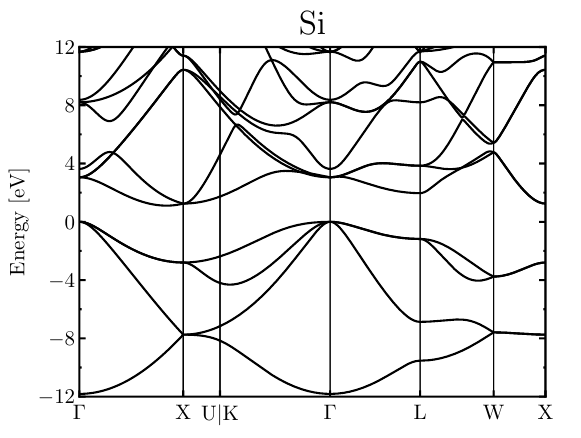} 
(d)\includegraphics[width=0.38\textwidth]{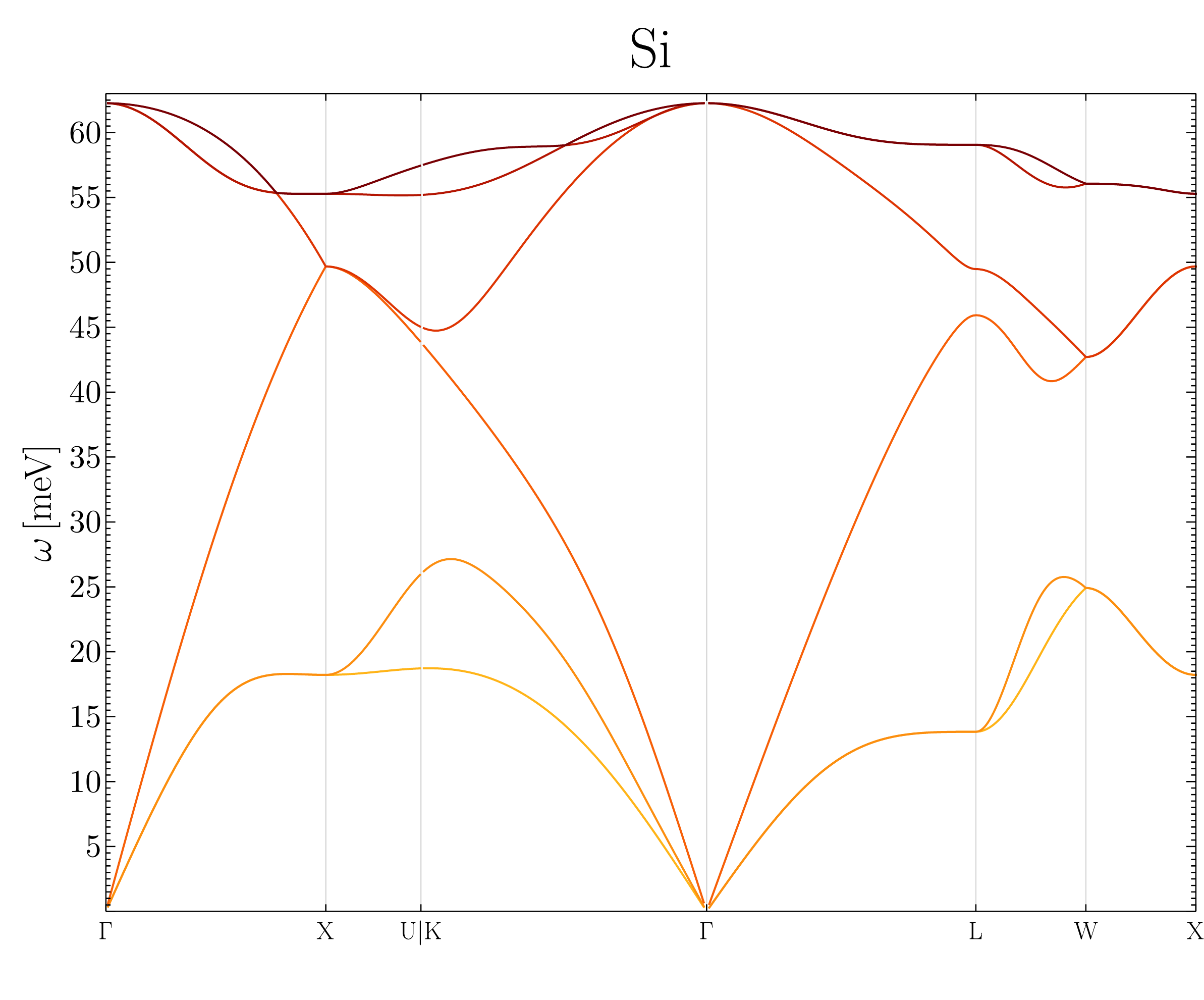} \\
\caption{ \label{fig:bandstructures} (a) Real-space face-centered cubic (FCC) lattice structure for Si. (b) First Brillouin Zone for FCC crystal. (c) Electron band structure for Si (d) Phonon band structure for Si. The reciprocal space primitive vectors are given by ${\bf b}_{1} = 2 \pi/a (-1, 1, 1)$, ${\bf b}_{2}= 2 \pi/a (1, -1, 1)$, and ${\bf b}_{3} =2 \pi/a (1, 1, -1)$. $\Gamma$ labels the origin of the reciprocal space while the capital Roman letters indicate standard high-symmetry points in the BZ. Graphics reproduced from Ref.~\cite{Griffin:2019mvc},  which also gives examples of additional crystals and target materials.
}
\end{center}
\end{figure}

\subsubsection{Electronic band structure}

A first approximation can be made on the basis of the fact that ions are heavy and slow and electrons are light and fast. The \emph{Born-Oppenheimer approximation} postulates that the energy eigenstates of the ion-electron system can be factorized into a piece depending explicitly on the ionic coordinates, and an electronic wavefunction which depends only parametrically on the ionic coordinates. In addition, we will typically consider the core electrons of each atom to be tightly bound to each nucleus, such that the only electrons whose dynamics we are interested in are the outer-shell \emph{valence} electrons. For example, silicon atoms (atomic number 14) in a lattice can be treated as ions of charge $Z = +4$ consisting of the nucleus and its 10 core electrons. Since the ions are much heavier than the electrons, the ion positions can be treated as approximately fixed when determining the electron wavefunctions. If we further ignore the electron-electron repulsion term, the Hamiltonian separates into single-particle Hamiltonians for each electron,
\be
H_1 =  \frac{\vecp^2}{2m_e} + \mathcal{V}_I(\vecr)
\ee
where $\mathcal{V}_I(\vecr)$ is an effective potential experienced by the valence electrons due to the ions, which are located at fixed positions $\vecR_I$.\footnote{Note that in this effective description, both $m_e$ and $\mathcal{V}_I(\vecr)$ must be renormalized, but the important point is that the one-body Hamiltonian will always contain a kinetic energy operator and a periodic potential.}

The attractive Coulomb potential $\mathcal{V}_I(\vecr)$ is periodic, because the ion positions themselves determine the geometric structure of the lattice. Using the fact that the operator for translation by a lattice vector $\vecR$ commutes with $H_1$, we may write the single-particle wavefunctions as a phase factor $e^{i \veck \cdot \vecr}$ times a \emph{cell function} $u_{\veck}$ which has the periodicity of the lattice:
\be
\psi_{\veck}(\vecr) = \frac{1}{\sqrt{V}}e^{i \veck \cdot \vecr}u_{\veck}(\vecx), \qquad u_{\veck}(\vecr + \vecR) = u_{\veck}(\vecr),
\label{eq:Bloch}
\ee
where $V$ is the total volume of the crystal. This result is known as \emph{Bloch's theorem}, and has the important implication that \emph{valence electrons are delocalized}: their wavefunctions have support throughout the entire crystal, thanks to the constant modulus of the phase factor and the periodicity of the cell function. Solving the single-particle Schr\"{o}dinger equation with the ansatz~(\ref{eq:Bloch}) for the wavefunction will yield a discrete set of quantized energy eigenvalues and eigenfunctions for each $\veck$, which may be labeled with an integer $n$, called the \emph{band index}, and restricting to inequivalent solutions restricts $\veck$ to a region of the reciprocal lattice space known as the \emph{first Brillouin zone (1BZ)}. As the number of atoms in the crystal goes to infinity, $\veck$ becomes a continuous parameter in the first BZ, and for fixed $n$, the energies $E^{(n)}_\veck$ trace out curves called \emph{energy bands}. A diagram of such a band structure, along with the associated real-space crystal lattice and BZ diagram is shown in Fig.~\ref{fig:bandstructures}.  The vector $\veck$ is called a \emph{crystal momentum}.\footnote{While it is tempting to interpret $\veck$ as a physical momentum which is conserved, this is often \emph{not} correct: crystal momentum is only conserved up to the addition of an arbitrary reciprocal lattice vector $\vecG$.}

The electron-electron Coulomb repulsion term cannot, of course, be ignored forever. Moreover, electrons are fermions, and thus the Pauli exclusion principle requires antisymmetry of the true wavefunction, which cannot simply be a product of single-electron wavefunctions. For systems with a small number $N$ of electrons (atoms or molecules, for example), an approximate wavefunction may be constructed to obey the exclusion principle as a fully antisymmetric linear combination of $N$-orbital products (the \emph{Hartree-Fock approximation}) where the spin degrees of freedom are treated separately; this construction is known as a Slater determinant.
By using the variational principle, minimization of the total ground-state energy with respect to the basis of single-particle states leads to an effective single-particle Schr\"{o}dinger equation for each state $\psi_i$, $i = 1, \dots, N$:
\begin{align}
\left [-\frac{\nabla^2}{2m_e} + \mathcal{V}_I(\vecr) + \mathcal{V}^H(\vecr) \right] \psi_i(\vecr) + \int \mathcal{V}^X(\vecr, \vecr') \psi_i(\vecr') \, d\vecr' & = \epsilon_i \psi_i(\vecr) \qquad {\rm (Hartree-Fock)}
\end{align}
where the 
\emph{total potential} $\mathcal{V}^H$ and the \emph{exchange potential} $\mathcal{V}^X$ are defined in terms of the single-particle number densities $n_i(\vecr) \equiv |\psi_i(\vecr)|^2$ as
\begin{align}
\mathcal{V}^H & = \alpha \int \frac{n(\vecr')}{|\vecr - \vecr'|}\,d\vecr' \\
\mathcal{V}^X & = -\alpha \frac{\sum_j \psi_j^*(\vecr') \psi_j(\vecr)}{|\vecr - \vecr'|}
\end{align}
Note that these potentials are functions not just of $\psi_{i}$ but also of all the other single-particle wavefunctions $\psi_j$, so these equations must be solved with an iterative trial-and-error process, adjusting the wavefunctions to achieve self-consistency. A key limitation of the Hartree-Fock approximation is that it cannot account for electron correlations.

Rather than dealing with single-particle orbitals directly, one can also dispense with the wavefunction and frame the entire problem in terms of the total electron density $n(\vecr)$, which may be expressed in terms of the exact many-body wavefunction $\Psi$ as
\be
n(\vecr) = N \int \Psi^*(\vecr, \vecr_2, \dots, \vecr_N) \Psi(\vecr, \vecr_2, \dots, \vecr_N) \,d\vecr_2 \cdots d\vecr_N
\ee
The Hohenberg-Kohn-Sham theorem states that an external potential for the electrons (here taken to be the ionic potential $\mathcal{V}_I(\vecr)$) \emph{uniquely} determines the ground state density $n(\vecr)$. Since the potential also determines the many-body wavefunction through the many-body Schr\"{o}dinger equation, the expectation value of the many-body Hamiltonian in the ground state must also be a functional of the density,
\be
\langle \Psi | H | \Psi \rangle \equiv \mathcal{F}[n(\vecr)] + \int \mathcal{V}_I(\vecr) n(\vecr) \, d\vecr
\ee
which by variational arguments is minimized when $n(\vecr)$ is the true density corresponding to the potential $\mathcal{V}_I(\vecr)$. 

This approach to the problem is called \emph{density functional theory}, and is the primary tool used by practitioners to determine the band structure of real solids theoretically, from first principles. To do this, one can work backwards from the density and construct \emph{fictitious} non-interacting single-particle states $\phi_i$ which satisfy $n(\vecr) = \sum_i |\phi_i(\vecr)|^2$. The functional $\mathcal{F}$ then takes the form
\be
\mathcal{F}[n(\vecr)] = \frac{\alpha}{2} \int \frac{n(\vecr) n(\vecr')}{|\vecr - \vecr'|} \, d\vecr \, d\vecr' + \sum_i \left \langle \phi_i   \left | \frac{-\nabla^2}{2m_e} \right | \phi_i \right \rangle + \mathcal{E}^{XC}[n(\vecr)].
\ee
The difficulty is that there is no known exact expression for the last term $\mathcal{E}^{XC}$, known as the \emph{exchange-correlation functional}, so while we have reformulated the problem, we have not evaded the issue of interacting electrons. With suitable approximations for this term, one can solve the single-particle equations for $\phi_i(\vecr)$. Despite their fictitious nature, these wavefunctions can serve as a decent model for the band structure wavefunctions $\psi_\veck$, since the single-particle DFT equations obey the conditions of Bloch's theorem due to the periodic potential $\mathcal{V}_I(\vecr)$.

\subsubsection{Phonons}
Finally, we consider coherent motion of the ions in the crystal, the quantized oscillations of which are known as \emph{phonons}. Unlike electrons, ions are highly localized in the ground state of the crystal at positions $\vecr_{nj} = \vecR_{n} + \vecR^{0}_{j}$, which we write as the sum of a lattice vector labeling the unit cell $n$ and  $\vecR^{0}_{j}$, the equilibrium position of ion $j$ within the unit cell. We consider the amplitudes of small displacements $\vecu_{nj}$ about the equilibrium positions and begin with the classical equations of motion for $\vecu$. Since the potential energy $U(\vecu_{nj})$ is minimized at $\vecu_{nj} = 0$ when the crystal is in its ground state, a Taylor expansion of $U$ begins with the quadratic term, and thus we obtain a coupled system of harmonic oscillator equations
\be
M_j \ddot{\vecu}_{nj} = - \sum_{mi} {\bf F}_{nj, mi} \vecu_{mi}
\label{eq:phononEOM}
\ee
where $M_{j}$ is the mass of ion $j$ and ${\bf F}$ is a matrix of spring constants with indices running over ion labels $nj, mi$ as well in spatial components $x,y,z$. Of course, since the ion labels run over all of the Avogadro's number of ions $N$ in the crystal, this matrix is enormous and we can solve this eigenvalue problem by Fourier transformation, similar to what is done for the electron band structure.

Using the periodicity of the system, we can index the displacements in terms of a crystal momentum $\veck$ restricted to the first BZ, analogous to the application of Bloch's theorem with electron states above. We similarly write the displacement as
\be
	\vecu_{nj}(\veck) = \frac{1}{\sqrt{M_j}} \tilde\vecu_{j}(\veck) e^{i \veck \cdot \vecR_{n} + i \veck \cdot \vecR^{0}_{j}} e^{-i \omega_{\veck} t}
\ee
where the function $ \tilde\vecu_{j}(\veck) $ does not depend on the unit cell; the factor of $\sqrt{M_j}$ is convenient to include here in order to account for the masses in the force equations. The phase factor of $e^{i \veck \cdot \vecR^{0}_{j}}$ is included here to match a convention commonly used in the literature. The periodicity of the system also implies that the force matrix ${\bf F}$ depends only on differences in unit cell position, such that solving Eq.~(\ref{eq:phononEOM}) with $n=0$ is sufficient. Then the equations of motion can be rewritten as
\be
	-\omega_{\veck}^{2}\tilde\vecu_{j}(\veck) = - \sum_{i} \left( \sum_{m} \frac{{\bf F}_{0j, mi}}{\sqrt{M_{j}M_{i}}} e^{i \veck \cdot (\vecR_{m} + \vecR^{0}_{i} - \vecR^{0}_{j})}\right)  \tilde \vecu_{i}(\veck) \equiv - \sum_{i} {\bf D}_{ji}(\veck)  \tilde \vecu_{i}(\veck)
\ee
where in the last line we have defined the dynamical matrix ${\bf D}_{ji}(\veck)$, which is now a $3 n_{c} \times 3 n_{c}$ Hermitian matrix, with $n_{c}$ is the number of ions per unit cell (typically $\mathcal{O}(1-20)$). For each $\veck$, there are therefore $3 n_{c}$ real normal mode frequencies, which we label as $\omega_{\nu, \veck}$ and $\nu$ runs over all $3 n_{c}$ phonon branches. The equilibrium positions and dynamical matrix can again be computed using density functional theory methods, see \cite{RevModPhys.73.515,TOGO20151,osti_1434899} for more details. This allows for a numerical determination of the mode frequencies and polarization vectors ${\mathbf{e}}_{\nu,\veck,j}$ describing a displacement of each ion in the unit cell. The eigenmodes are generally normalized as $\sum_{j} {\mathbf{e}}^{*}_{\nu,\veck,j} {\mathbf{e}}_{\mu,\veck,j} = \delta_{\mu \nu}$ and also satisfy the property that 
\be
{\mathbf{e}}_{\nu,\veck,j} = {\mathbf{e}}^{*}_{\nu,-\veck,j}
\label{eq:phonon_eig}
\ee since the displacement vector ${\bf u}_{nj}$ is real.

The most general classical solution would then be a linear combination of all $\veck,\nu$ modes with arbitrary complex amplitudes $c_{\veck,\nu}$ satisfying Eq.~\ref{eq:phonon_eig}. 
Treating phonons quantum mechanically involves a replacement of the classical amplitudes with creation and annihilation operators, giving
\be
	\vecu_{n, j} (t) = \sum_{\veck,\nu} \frac{1}{\sqrt{2 N_{\rm cell} M_j \omega_{\nu,\veck}}} \left[  \mathbf{e}_{\nu,\veck j} a_{\nu,\veck} e^{i \veck \cdot (\vecR_{n} +  \vecR^{0}_{j}) } e^{-i \omega_{\nu,\veck} t} +  \mathbf{e}^{*}_{\nu,\veck j} a^{\dagger}_{\nu,\veck} e^{-i \veck \cdot (\vecR_{n} +  \vecR^{0}_{j}) } e^{i \omega_{\nu,\veck} t} \right].
\label{eq:udef}
\ee
Here the operators $a_{\nu,\veck}, a_{\nu,\veck}^{\dagger}$ satisfy the commutation relation $[a_{\nu,\veck} , a_{\mu,\vecq}^{\dagger}] = \delta_{\veck,\vecq} \delta_{\nu,\mu}$. $N_{\rm cell}$ is the number of unit cells, and the overall normalization of the operator in Eq.~\ref{eq:udef} was selected to obtain the usual form of the Hamiltonian for harmonic oscillators, $H = \sum_{\veck,\nu} (a_{\nu,\veck}^{\dagger} a_{\nu,\veck}  + 1/2)$.

The eigenvalues  $\omega_{\veck,\nu}$ describes a band structure for phonons in close analogy to those for electrons, and as $N_{\rm cell} \to \infty$, the index $\veck$ takes continuous values in the first Brillouin Zone.
An example band structure is shown in Fig.~\ref{fig:bandstructures}. The difference with electrons is that for phonon modes, the tower of excitations for each $\veck$, $E = n \omega_{\nu, \veck}$, just amounts to larger occupation numbers $n \gg 1$ for a given mode, with classical phonon waves corresponding to coherent states of phonon modes.
Thus the phonon bands in Fig.~\ref{fig:bandstructures} have a maximum energy, while the electron bands in principle continue up to infinite energy. In a 3-dimensional solid, the three lowest-energy phonon branches extends to arbitrarily low energy as $\veck \to 0$, with a linear dispersion relation $\omega = c_s |\veck|$ for small $\veck$. This can be understood since the ground state of an infinite crystal spontaneously breaks continuous translation invariance to a discrete subgroup (in other words, the order parameter is the lattice spacing), and thus there must be massless Goldstone bosons, which in the condensed matter context are known as \emph{acoustic phonons}. The propagation speed $c_s$ is the sound speed with typical values are $\sim 3-10 \ {\rm km/s}$, yielding typical energies
\be
	\omega_{\rm acoustic} = c_{s} k \simeq 8 \  {\rm meV} \left(\frac{c_s}{5 \ {\rm km/s}}\right) \left(\frac{k}{500\ {\rm eV}}\right)
\ee
again for small $|\veck| \ll q_{\rm coh}$.
The physical interpretation is that all $N$ ions in the crystal are oscillating in phase with the same amplitude as $\veck \to 0$, which must have zero energy. In an anisotropic material $c_s$ may differ along different lattice directions, leading to distinct dispersion relations for the three acoustic modes. 

If $n_c > 1$, there are additional sets of normal modes generically corresponding to out-of-phase oscillations within a unit cell. These are known as \emph{optical phonons}, with typical energies $\omega_{\rm optical} \simeq 10-100 \ {\rm meV}$. These modes do not correspond to a broken symmetry and are \emph{gapped}, with approximately constant energy across the entire BZ (or equivalently, approximately constant in $k$). We can understand the energy scale of optical phonons from dimensional analysis: the normal mode frequencies will be proportional to $\sqrt{\kappa/M_I}$ where $\kappa$ is a spring constant and $M_I$ an ion mass. In addition the acoustic branch has a linear dispersion as $k \to 0$, so $\omega_{\rm acoustic} \sim \sqrt{\kappa/M_I} (k a)$ where $a$ is the lattice spacing. Identifying $a \sqrt{\kappa/M_I}$ with $c_s$, we have
\be
\omega_{\rm optical} \simeq \frac{c_s}{a} \simeq 10  \ {\rm meV} \left(\frac{c_s}{5 \ {\rm km/s}}\right)\left(\frac{0.5 \ {\rm nm}}{a}\right).
\ee
We can also estimate optical phonon frequencies based on the electrostatic interactions of ions within the unit cell \cite{AshcroftMermin}, yielding similar values of
\be
\omega_{\rm optical} \simeq \sqrt{\frac{e^2}{M_I a^3}} \simeq 20 \ {\rm meV} \sqrt{\frac{14 \ \GeV}{M_I}} \sqrt{\frac{(0.5 \ {\rm nm})^3}{a^3}}.
\ee
The fact that this energy scale corresponds to the kinetic energy of DM with keV-MeV scale masses, and that they can be excited with a wide range of momentum transfers $\lesssim$ keV, makes the optical phonon branch particularly useful for DM detection.

\subsection{Dynamic structure factor \label{sec:DMstructurefactor}}

Having now determined the spectrum of elementary excitations, we might assume that the states $|i \rangle, | f \rangle$ in Eq.~(\ref{eq:Sqw}) are states with definite numbers of those excitations. This is a good approximation for phonons, which have sufficiently negligible interactions, but it is not so for electrons. Although electron quasiparticles are weakly interacting, the interactions are large enough that they can give rise to collective modes such as the plasmon. Furthermore, it is not entirely obvious how the operator coupling to all electrons in Eq.~(\ref{eq:ODarkPhoton}) acts on the single-particle electron quasiparticle wavefunctions computed with DFT methods. 

To simplify the discussion, here we focus our discussion to the cases when (i) the DM couples only to electrons and (ii) the DM couples only to the nucleus, and on the leading excitations being created. As such, there are various subtleties which we will gloss over, and which will be discussed in detail in the section to follow, where we elaborate on the structure factor for the dark photon model when both electron and nucleus couplings are present. In addition, electrons and ions are not truly decoupled, so that phonons can be created from DM-electron scattering. Electronic excitations may also arise from nuclear scattering (as opposed to electron scattering) via the condensed matter analogue of the \emph{Migdal effect}, which we discuss further in Sec.~\ref{sec:Migdal}.

\subsubsection{Electronic excitations}

Assuming DM only interacts with electrons, and taking the target system to be at zero temperature, the relevant dynamic structure factor is 
\be
S(\vecq, \omega) \equiv \frac{2\pi}{V} \sum_{f} |\langle f | \sum_k e^{i \vecq \cdot \vecr_k} | i \rangle|^2 \delta(E_f - E_i - \omega)
\label{eq:Sqw_electron}
\ee
where $k$ sums over all electrons and the initial and final states are generic many-body states. The first calculations of DM-electron scattering assumed that $|i\rangle,|f \rangle$ are single-excitation Bloch states and that the sum over all electrons could be replaced by the operator $e^{i \vecq \cdot \vecr}$ acting only on a single electron; however, as commented on at the beginning of this section, for low $q < q_{\rm coh}$ it is not clearly justified to assume that only interactions with a single electron dominate. Related to that, it is also not obvious that only Fock states are relevant and electron-electron interactions can be neglected. Indeed, while an important step forward, this single particle approach turns out to miss some important many-body effects. 

A more general approach can be taken, where we do not make any assumptions about the many-body states. The discussion here is largely based off of Refs.~\cite{PinesNozieres_book,girvin_yang_2019,Arovas20}. We instead rewrite the dynamic structure factor in terms of the electron number density operator, 
\be
	n_{\vecq} = \sum_k e^{i \vecq \cdot \vecr_k} = \int d^{3} \vecr \, e^{i \vecq \cdot \vecr} \sum_{k = 1}^N \delta^{(3)}(\vecr - \vecr_k) 
\ee
and focus on understanding the quantity
\be
S(\vecq, \omega) \equiv  \frac{2\pi}{V}  \sum_{f} |\langle f |  n_{\vecq} | i \rangle|^2 \delta(E_f - E_i - \omega).
\ee
The structure factor is related to the rate to produce charge excitations in the medium, which can be rewritten in terms of the imaginary part of a correlation function:
\be
	S(\vecq, \omega) =  - 2 \, {\rm Im} \left( - \frac{i}{V} \int_{0}^{\infty} dt \, e^{i \omega t} \langle \left[ n_{\vecq}(t), n_{-\vecq}(0) \right] \rangle
\right) .
\label{eq:fluctuationdissipation}
\ee
This is a special case of the {\emph{fluctuation-dissipation theorem}} (which is slightly more general, applying to finite-temperature systems as well), but can be understood simply as a consequence of unitarity or an application of the optical theorem. This correlation function determines the linear response of the charge density in a medium to any external perturbation, whether it is dark matter or a SM probe. Indeed, consider subjecting the material to some other external potential $\Phi_{\rm ext}$ which also couples linearly to electron density. A suitable candidate is an electromagnetic potential, and we can think of the potential as being sourced by a weak external charge $\rho_{\rm ext}(\vecq, \omega)$, with $\Phi_{\rm ext} = 4\pi\rho_{\rm ext}(\vecq, \omega)/q^{2}$.  
Gauss's law in a medium tells us the response of the electric fields in Fourier space: 
\be
	i \vecq \cdot {\bf E}(\vecq, \omega) = \frac{\rho_{\rm ext}(\vecq,  \omega) }{\epsilon(\vecq,  \omega) }
\ee
where ${\bf E}(\vecq, \omega)$ is the total field in the medium and the right hand side of the equation is the total charge density. The total charge density is the sum of the external charge density and the induced charge density in the medium, and is related to the external charge density by $1/\epsilon(\vecq,  \omega) $. The fact that $\epsilon(\vecq,\omega) \neq 1$ generically is a manifestation of charge screening, since the external field is scaled down by $\epsilon$ to generate the in-medium field which a test charge in the material would feel. 
Meanwhile, because the induced charge density is due to the response of the medium, it is determined by the correlation function appearing in Eq.~(\ref{eq:fluctuationdissipation}).
This gives a relationship between $S(\vecq, \omega)$ and the dielectric response \cite{PhysRev.113.1254}:
\be
	S(\vecq, \omega) = \frac{q^2}{2 \pi \alpha}  {\rm Im} \left(-\frac{1}{\epsilon(\vecq, \omega)}\right),
\label{eq:Sepsilon}
\ee
which will be investigated in much more detail in Sec.~\ref{sec:dielectric}. Note that we have made the approximation of a homogeneous medium, and ignored some subtleties here regarding the fact that we are in a periodic medium. We also implicitly work with the longitudinal dielectric function everywhere here, since only the longitudinal fields appear in Gauss's law.

The advantage of this point of view is that we have made no assumptions about the nature of the exact eigenstates $|f \rangle$ of the target. At this point, however, we must find some way to calculate or determine $\epsilon(\vecq, \omega)$. We first proceed by close analogy to the calculation of the photon polarization in field theory. The dielectric function $\epsilon(\vecq, \omega)$ is by definition related to the longitudinal photon polarization $\Pi_{L}$, with the exact relationship given by $\epsilon(\vecq, \omega) = 1 - \Pi_{L}(\vecq, \omega)/|\vecq|^{2}$. This suggests an approach to calculating the polarization in terms of single-particle Bloch states as derived above. In Sec.~\ref{sec:dielectric}, we will describe a number of analytic models for the dynamic structure factor, as well as explicit numerical calculations using Bloch wavefunctions. The result from including the full many-body states (or equivalently of including screening in the single-particle picture) leads to matrix elements with qualitatively different behavior. 

For example, in generic solid-state systems (including both semiconductors like silicon and metals like aluminum), there is a resonance for $q \lesssim p_{F}$ called the \emph{plasmon}, which appears in the dynamic structure factor as
\be
S(\vecq, \omega) \propto q^2 \omega \frac{\omega_p^2 \Gamma_p}{(\omega_p^2 - \omega^2)^2 + \omega^2 \Gamma_p^2}.
\label{eq:plasmonsimple}
\ee
with $\Gamma_p$ a finite width which regulates the resonance. The appearance of the \emph{plasma frequency}
\be
\omega_p = \sqrt{\frac{4\pi \alpha n_e}{m_e}}
\ee
suggests an interpretation of this resonance as the collective oscillation of the entire valence electron density $n_e$, which is not visible in a picture of single-particle wavefunctions. The quantized mode corresponding to the collective excitation is also known as a plasmon, where we can interpret this result as this structure factor for producing a single plasmon. Since the ground state is an eigenfunction of the density operator for $\vecq = 0$ (with eigenvalue equal to the total number of particles in the system), the factor of $q^2$ in Eq.~(\ref{eq:plasmonsimple}) can be understood as enforcing that the overlap of the initial ground state and the final state with an excited plasmon should vanish as $q \to 0$.

\begin{figure*}[t!]
\begin{center}
\includegraphics[width=0.45\textwidth]{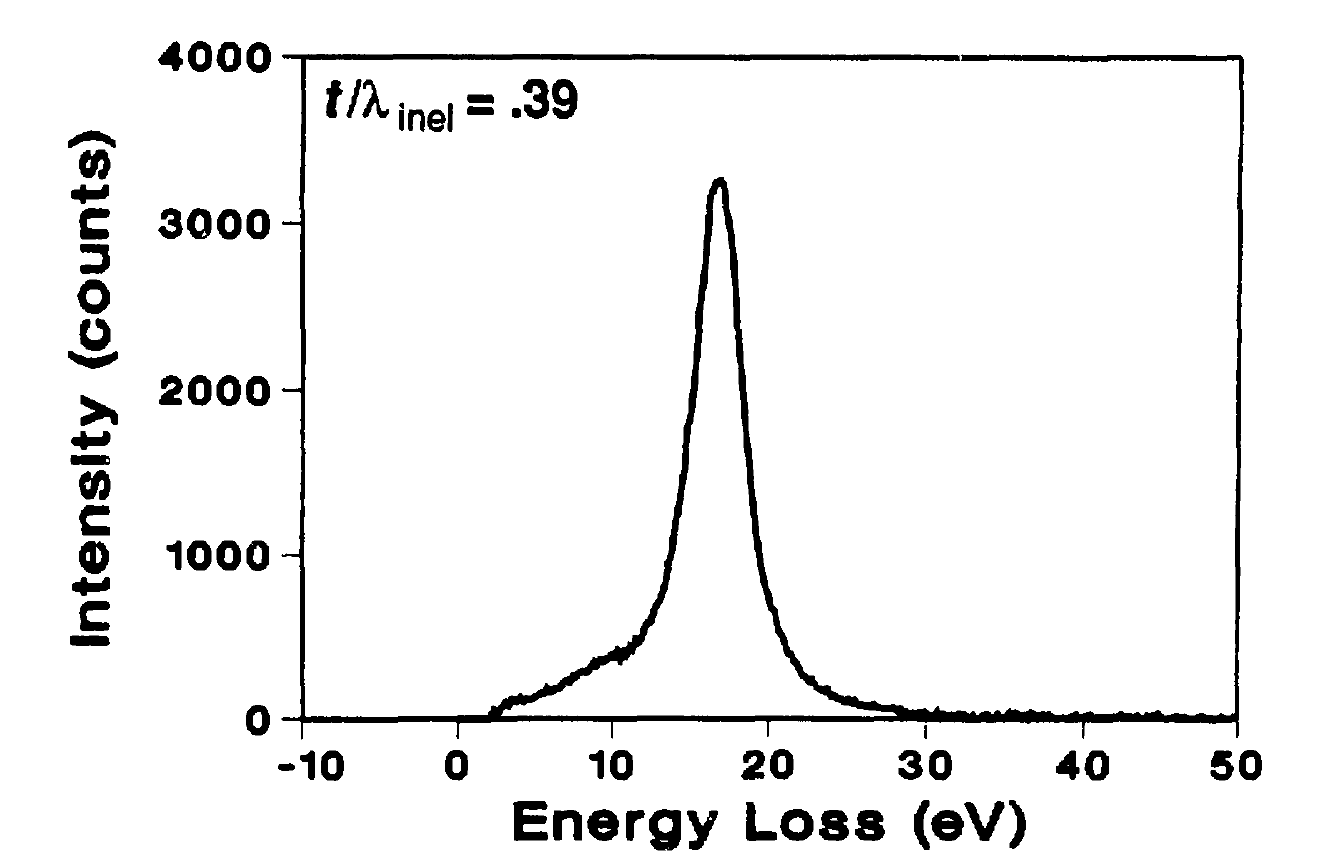}
\caption{ \label{fig:PlasmonLoss}
The measured ELF in silicon at $q \ll p_{F}$, obtained from electron energy-loss spectroscopy (EELS), shows a pronounced plasmon resonance of the form Eq.~(\ref{eq:plasmonsimple}) at $\omega_p \simeq 16 \ \eV$. Reproduced from Ref.~\cite{kundmann1988study}. A contour plot of the dynamic structure factor in Si for $q$ up to 8 keV is provided later in Fig.~\ref{fig:Si_sqw}. }
\end{center}
\end{figure*}

We may also express the DM scattering rate directly in terms of the experimentally-measured \emph{loss function} or \emph{energy loss function} (ELF),
\be
\mathcal{W}(\vecq, \omega) \equiv {\rm Im} \left(-\frac{1}{\epsilon(\vecq, \omega)}\right),
\label{eq:ELFdef}
\ee
without ever using explicit electron wavefunctions. (This also only applies as long as the DM-electron interactions are spin-independent and couple to electron density.) Indeed, the same loss function describes energy loss of electrons in materials and X-ray scattering, providing a way to determine the loss function from experimental data in a way that  automatically accounts for all many-body and screening effects. These measurements are exactly analogous to how deep inelastic scattering experiments can probe the degrees of freedom in the proton, in particular the non-perturbative proton form factors at low momentum transfer, by exploiting the electron coupling to quarks.  Fig.~\ref{fig:PlasmonLoss} shows the measured ELF in silicon with a clear plasmon resonance. One can also use phenomenological models of the dielectric function satisfying various properties with parameters fit to the data; for instance, this is done with the Mermin oscillator model fit to optical data or electron energy loss~\cite{10.1002/sia.6227,PhysRevA.58.357,VOS2019242}, as well we also discuss in Sec.~\ref{sec:dielectric}. The disadvantage of the latter approaches is that the connection to individual quantized excitations, like the number of electron/hole pairs produced in a given scattering event, is less transparent. In addition, it can be difficult to experimentally determine some regimes in energy and momentum transfer that are relevant for DM scattering.

\subsubsection{Phonons}
\label{sec:Sqw_phonon}

The dynamic structure factor for nuclear scattering is given by
\be
S(\vecq, \omega) \equiv \frac{2\pi }{V} \sum_f |\langle f | \sum_{I} f_{I} e^{i \vecq \cdot \vecr_I}| i \rangle|^2 \delta(E_f - E_i - \omega)
\label{eq:Sqw_phonon}
\ee
where $f_{I}$ are again the normalized interaction strengths with the ions. The relative scattering strength will depend on the type of ion, and the factors $f_I$ do not factorize out of the structure factor if the system is composed of different types of ions.  Thus, in contrast to the electron-excitation dynamic structure factor, for multi-atom target materials there is not a single dynamic structure, but a continuous class of structure factors depending on how the external probe couples to the individual atoms. As will be discussed further later, this allows for additional interesting effects in the material-dependence of DM scattering, as a way to distinguish different DM coupling scenarios. Note that in this work we will restrict to spin-independent DM interaction strengths $f_{I}$; if there is a spin-dependent interaction potential, then one must also perform an average over all possible spin states of the ions. An extensive review of the dynamic structure factor for phonons, including such spin-dependent interactions, can be found in Ref.~\cite{Schober2014}. 

For phonon excitations, the final states can simply be written by acting with the phonon creation operators introduced in Eq.~(\ref{eq:udef}) on the vacuum. For a single phonon being created, $| f \rangle = a^\dagger_{\nu,\veck} |0 \rangle$, while multiphonon excitations are also possible. In order to detect the phonon excitations being created, the energy deposited is necessarily well above the operating temperature of the experiment, and it is a good approximation to take $T=0$ and assume $| i \rangle = | 0 \rangle$ for the initial state. In order to compute the matrix elements with these states, we must write the ion positions as $\vecr_I = \vecr_{nj} + \vecu_{nj}$, where $\vecr_{nj} = \vecR_n + \vecR^{0}_{j}$ is the equilibrium position of the ion $j$ in the unit cell labeled by $n$ and the ion displacement $\vecu_{nj}$ contains the phonon creation and annihilation operators. Substituting this into the exponential appearing in the matrix element gives:
\be
	f_{I} e^{i \vecq \cdot \vecr_I}  = f_j e^{i \vecq \cdot \vecr_{nj}} \exp\left(i \vecq \cdot   \sum_{\nu, \veck} \frac{1}{\sqrt{2 N_{\rm cell} M_j \omega_{\nu,\veck}}}  \left[  a^\dagger_{\nu,\veck}  \mathbf{e}^*_{\nu,\veck, j} e^{-i \veck \cdot \vecr_{nj}} + a_{\nu,\veck}  \mathbf{e}_{\nu,\veck, j} e^{i \veck \cdot  \vecr_{nj}}  \right] \right).
	\label{eq:expiu_operator}
\ee
Expanding this operator will contain a 0-phonon contribution, a 1-phonon creation contribution, and so on. (Note that  in Eq.~(\ref{eq:udef}) we gave the time-dependent Heisenberg or interaction picture operator $\vecu_{n,j}(t)$, but the matrix elements given in Eq.~(\ref{eq:Sqw_phonon}) are computed with the Schr\"{o}dinger operators where the time-dependence of the states has already been taken into account in Fermi's Golden rule, leading to the energy-conserving delta function. This is why the time-dependent phase factors have been removed in substituting in Eq.~(\ref{eq:udef}).)

To perform the expansion explicitly in terms of phonon creation and annihilation operators, we can make use of the Baker-Campbell-Hausdorff formula for generic operators $A$ and $B$:
\be
\exp(A) \exp(B) = \exp\left(   A + B + \frac{1}{2}[A,B] + \frac{1}{12}[A,[A,B]] - \frac{1}{12}[B,[A,B]]\dots \right)~,
\label{eq:BCHformula}
\ee
which simplifies in the case of the harmonic oscillator algebra because the first commutator $[a, a^\dagger] = 1$ is a $c$-number and truncates the series at the third term. Applying this gives
\begin{align}
f_{I} e^{i \vecq \cdot \vecr_I}  =  f_j e^{i \vecq \cdot \vecr_{nj}}  e^{ - W_{j}(\vecq)}
	\prod_{\nu, \veck }\exp \left( \frac{i\vecq \cdot \mathbf{e}^*_{\nu,\veck, j} e^{-i \veck \cdot \vecr_{nj}}}{\sqrt{2  N_{\rm cell}  M_j \omega_{\nu,\veck}}} a^\dagger_{\nu,\veck}\right)  \exp \left( \frac{i\vecq \cdot \mathbf{e}_{\nu,\veck, j} e^{i \veck \cdot \vecr_{nj}}}{\sqrt{2  N_{\rm cell}  M_j \omega_{\nu,\veck}}} a_{\nu,\veck} \right),
\end{align}
with 
\be
W_{j}(\vecq) \equiv \frac{1}{2} \sum_{\nu, \veck} \frac{ | \vecq \cdot \mathbf{e}^*_{\nu,\veck, j}|^2}{2 N_{\rm cell}  M_j \omega_{\nu,\veck}}.
\ee 
The factor $e^{-W_{j}(\vecq)}$ is also known as the {\emph{Debye-Waller factor}}, which roughly speaking accounts for the effect of the zero-point motion of the ions in the lattice.
Taking the matrix element with the initial vacuum state, the exponential of the phonon annihilation operators reduces to unity and we can write the structure factor as
\be
	S(\vecq, \omega) = \frac{ 2 \pi}{V} \sum_{f} \left| \langle f | \sum_{n,j } f_{j}e^{i \vecq \cdot \vecr_{nj}}  e^{-W_j(\vecq)}  \prod_{\nu, \veck }\exp \left( \frac{i\vecq \cdot \mathbf{e}^*_{\nu,\veck, j} e^{-i \veck \cdot \vecr_{nj}}}{\sqrt{2  N_{\rm cell}  M_j \omega_{\nu,\veck}}} a^\dagger_{\nu,\veck}\right)  | 0 \rangle \right|^{2}  \delta(E_f - E_i - \omega).
	\label{eq:sqw_phonon_expanded}
\ee
where we have replaced the sum over $I$ with a sum over $n,j$.

Taking only the zeroth-order term in the exponential of phonon creation operators $a^\dagger_{\nu,\veck}$, there are no phonon transitions, and this just corresponds to DM elastically recoiling off the lattice as a whole. The leading nontrivial contribution comes from expanding the exponentials to linear order, which allows for single-phonon creation. Summing over final states $|f \rangle = a^{\dagger}_{\nu,\veck} |0 \rangle$, this leads to the single-phonon structure factor
\begin{align}
	S^{(1-ph)}(\vecq, \omega) &=  \frac{ 2 \pi}{V} \sum_{\nu, \veck} \left| \sum_{n,j } f_{j}e^{i (\vecq - \veck) \cdot \vecr_{nj}}  e^{-W_j(\vecq)}   \frac{i\vecq \cdot \mathbf{e}^*_{\nu,\veck, j} }{\sqrt{2 N_{\rm cell} M_j \omega_{\nu,\veck}}} \right|^{2}  \delta(E_f - E_i - \omega_{\nu,\veck}) 
\end{align}
We next use the fact that for $\vecq$ smaller than any reciprocal lattice vector $\vecG$, the sum over lattice sites simply enforces momentum conservation:\footnote{Since reciprocal lattice vectors are defined by the condition $e^{i \vecG \cdot \vecR} = 1$, momentum is only conserved up to a reciprocal lattice vector, $\veck = \vecq + \vecG$. For nonzero $\vecG$, this is called \emph{Umklapp scattering}.}
\be
\sum_{n} e^{i (\vecq - \veck) \cdot \vecR_n} =  N_{\rm cell} \delta_{\vecq, \veck}
\ee
since phonon modes are only defined for $\veck$ within 1BZ. This implies that we will have excitation of any phonon with the same momentum $\vecq$ and energy $\omega$. The single-phonon structure factor then simplifies to 
\begin{align}
	S^{(1-ph)}(\vecq, \omega) &= \frac{ 2 \pi N_{\rm cell}}{V} \sum_{\nu} \left| \sum_{j } f_{j} e^{-W_j(\vecq)}   \frac{i\vecq \cdot \mathbf{e}^*_{\nu,\vecq, j} }{\sqrt{2 M_j \omega_{\nu,\vecq}}} \right|^{2}  \delta(E_f - E_i - \omega_{\nu,\vecq}) \nonumber \\
	& \equiv \frac{2 \pi}{\Omega} \sum_{\nu} \frac{|F_{\nu}(\vecq)|^{2}}{\omega_{\nu,\vecq}}   \delta(E_f - E_i - \omega_{\nu,\vecq})
	\label{eq:Sqw_1phonon}
\end{align}
where the second line defines a single-phonon form factor $F_{\nu}(\vecq)$ and we defined $\Omega = V/N_{\rm cell}$ as the primitive unit cell volume. This form factor sums over the coupling of the probe with the ions $f_{j}$ in the unit cell, multiplied by the normalized motion of that ion $\propto \mathbf{e}^*_{\nu,\vecq, j}/\sqrt{M_j}$ and is therefore describing an effective coupling of the probe with a particular phonon mode, accounting for interference effects. This structure factor therefore describes coherent scattering off the ions in the lattice. However, we explicitly see the 1-phonon form factor is an intrinsic quantity of the material and does not scale with the size of the system. 

Neglecting the details of this (probe-dependent) form factor, the 1-phonon structure factor has an amplitude that can be estimated as $\sim q^{2}/(2 M \omega_{\vecq,\nu})$, and we see from Eq.~(\ref{eq:sqw_phonon_expanded}) that excitations of more phonons will be expected to give a contribution to the structure factor that roughly scales as
\be
	\left| \langle n | \sum_{I} f_{I} e^{i \vecq \cdot \vecr_{I}} |0 \rangle \right|^{2} \sim \left( \frac{q^{2}}{2 M \omega_0 } \right)^{n} 
	\label{eq:Sqw_roughscaling}
\ee
for a final state with $n$ phonons. Here we have replaced $M_{j}$ with some averaged ion mass $M$, and  $\omega_{\vecq,\nu}$ with a typical phonon energy $\omega_{0}$ that will be some average over acoustic and optical phonons, to give a heuristic scaling.\footnote{Of course, to actually obtain the structure factor $S(\vecq, \omega)$, these matrix elements must be computed with interference effects and integrated over the phase space such that the phonon energies match $\omega$, which can change the scaling for specific final states and/or couplings. Another subtlety is that the single-phonon piece of the operator $e^{i \vecq \cdot \vecr_{I}}$ can also give rise to two (or more phonon) excitations through anharmonic interactions in the phonon Hamiltonian. The next term in the phonon Hamiltonian $\sim(\vecu)^{3}$, which comes from expanding the potential to higher order in the displacements. A more detailed discussion of the various contributions to the two-phonon dynamic structure factor can be found in Ref.~\cite{Campbell-Deem:2019hdx}.} This approximate scaling gives an estimate of the importance of higher-order phonon excitations to the structure factor, depending on the regime of momentum transfer $q$. Note that for a harmonic oscillator mode with energy $ \omega_{0}$, $\sqrt{2 M \omega_{0}}$ corresponds to the momentum spread of the ground-state wavefunction of a particle of mass $M$ in a harmonic oscillator potential with frequency $\omega$. Taking as typical parameters $M \sim 30m_{p}$ and $\omega_{0} \sim 50$ meV, the momentum spread is about 50 keV. We can thus interpret the expansion in $q^{2}/2 M \omega_0 $ as follows: at low $q$ compared to the momentum spread, the dynamic structure factor will be dominated by the 1-phonon contribution since higher-energy excitations have a suppressed overlap with the probe potential. As $q$ becomes comparable to $\sqrt{2 M \omega_0}$, higher order phonon contributions are important, and for $q \gg \sqrt{2 M \omega_0}$, the structure factor will transition to that of free nuclear recoils since the kinetic energy will dominate over potential energy.  In Sec.~\ref{sec:DMN} below, we will illustrate this more explicitly with simple model of a single ion in a harmonic oscillator, and follow the transitions down from the high-$q$ regime of incoherent scattering off a single nucleus, to the regime of low-$q$ coherent scattering which leads to single or few-phonon production.

Finally, we note that the formalism for phonon production through coherent nuclear scattering may apply to systems other than crystal lattices, including superfluid helium which is actively being investigated as a potential experimental target. We will discuss this in Sec.~\ref{sec:helium}.

\section{Dark Matter-Nucleon Scattering}
\label{sec:DMN}

A microscopic theory of DM interactions with quarks and/or gluons yields an effective coupling of DM to nucleons, which may include neutrons as well as protons. This may be parameterized by a fiducial DM-nucleon cross section, $\sigmabar_n$, which by default is assumed to be a spin-independent scattering cross section. Traditional direct detection of WIMP DM has exclusively focused on DM-nuclear scattering by treating the nuclei as free target particles at rest. Here we briefly review the parametrics and the main results to draw a contrast with the phenomenology of sub-GeV DM scattering; a more complete treatment can be found in Ref.~\cite{Lewin:1995rx}.

DM heavier than about 10 GeV carries kinetic energy greater than 10 keV, well in excess of any displacement energy, and likewise carries momentum greater than 10 MeV which exceeds any zero-point lattice momenta. Thus the nuclear target may be treated as a free particle, and in particular as a momentum eigenstate, so that classical 2-body scattering kinematics applies. In addition, the nucleus may be treated as being at rest initially. We can then use the kinematic relationship in Eq.~(\ref{eq:omegaqdef}), setting the energy deposited to $\omega = E_{R} = q^{2}/(2m_N)$ for a nucleus $N$, which gives the relationship
\be
 \vecq \cdot \vecv = \frac{q^2}{2 \mu_{\chi N}} \qquad {\rm(elastic \ nuclear \ scattering)}
\ee
where $\mu_{\chi N}$ is the DM-nucleus reduced mass. 
For incoming DM with speed $v$, the maximum momentum which may be transferred is $q_{\rm max} = 2 \mu_{\chi N} v$, and so the maximum nuclear recoil energy is
\be
E_{R, \rm max} = \frac{q_{\rm max}^2}{2m_N} = \frac{2 \mu_{\chi N}^2 v^2}{m_N}.
\ee
The best kinematic match is obtained when $m_{\chi} \sim m_{N}$, when the free nucleus dispersion relation passes through the region of phase space with $q \sim m_{\chi} v$ and $\omega = q^{2}/(2 m_{\chi})$. For $m_\chi \ll m_N$, as is the case for sub-GeV DM, this energy transfer becomes very inefficient: not only does the incoming DM energy scale as $m_\chi$, but only a fraction $\sim m_\chi/ m_N$ is transferred to the nucleus. Indeed, collective modes like phonons can provide a much better kinematic match to the DM. From Eq.~(\ref{eq:vmingen}), the minimum velocity $\vmin$ required to generate a nuclear recoil of energy $E_R$ is
\be
\vmin = \sqrt{\frac{m_N E_R}{2\mu_{\chi N}^2}},
\label{eq:vminelastic}
\ee
so that a hard upper limit to $E_R$ is obtained by setting $\vmin$ to the maximum speed of DM in the local neighborhood, assumed to be $\vesc + v_\oplus$ in the Standard Halo Model.

Assuming a contact potential between DM and nucleon, the interaction Hamiltonian is given by 
\be
\Delta H_{\chi T} =  \sqrt{ \frac{\pi \bar \sigma_{n}}{\mu^2_{\chi n} }} \int \frac{ d^{3} \vecq}{(2\pi)^{3}} \ e^{i \vecq \cdot (\vecr_{n} - \vecr_{\chi})},
\label{eq:Hnuc_contact}
\ee
where $\sigmabar_n$ is the DM-nucleon scattering cross section and $\mu_{\chi n}$ is the DM-nucleon reduced mass. For simplicity, we assume the same coupling to protons and neutrons. Summing over all available target nuclei $N_{\rm nuc}$, the dynamic structure factor for elastic nuclear recoils is given by
\be
	S(\vecq, \omega) = \frac{ 2 \pi N_{\rm nuc}}{V} A^{2} |F_{N}(q)|^{2} \, \delta \! \left(\omega - \frac{q^2}{2 m_{N}} \right),
	\label{eq:Sqw_elasticNR}
\ee
where $A$ is the mass number and $F_{N}$ is a nuclear form factor,
\be
	F_{N}(q) = \langle N | \frac{1}{A}  \sum_{\alpha = 1}^A e^{i \vecq \cdot \vecr_{\alpha}} | N \rangle.
	\label{eq:Fnuclear}
\ee
Here $| N \rangle$ is the bound state of nucleons in a nucleus and $\alpha$ sums over all nucleons in the nucleus. $|F_{N}(q)|^{2}$ thus captures the strength of response of the system, similar to what we will calculate for a condensed matter target, with the difference that the excited states in a nucleus are usually too high in energy ($\sim$ MeV) to be excited by the light DM which is the focus of this review. In what follows, we will always take the initial and final states are both the ground state of the nucleus. $|F_{N}(q)|^{2}$ can be computed in specific nuclear shell models, but for our purposes we can just model it as the Fourier transform of the nucleon mass distribution. $F_N$ then parameterizes the loss of coherence over all of the nucleons in the nucleus as the momentum transfer becomes comparable to the inverse nuclear radius: this is usually relevant for WIMP DM, but universally irrelevant for sub-GeV DM, so we will always set $F_N = 1$ in what follows. 

We can now use the isotropic rate in Eq.~(\ref{eq:Riso}) to obtain the rate for a target with mass number $A$. Performing the trivial $\omega$ integral, this gives the nuclear recoil rate:
\be
R = N_{T} \frac{\rho_\chi}{m_\chi} \frac{A^2\sigmabar_n}{2\mu_{\chi n}^2} \int q \, dq \, \eta(\vmin(E_{R})) F^2_N(q) 
\label{eq:dRdERElastic}
\ee
where $N_{T} = N_{\rm nuc}/(\rho_{T}V)$ is the number of target nuclei per unit detector mass. Usually the $q$ integral can be replaced with $q\, dq \to dE_{R} m_{N}$ to obtain a differential rate per recoil energy, $dR/dE_{R}$. Apart from $\eta$ and $F_N$, the spectrum is flat in $E_R$: all recoil energies are equally likely. The $E_R$ dependence of the velocity integral is quite important, though: using the Maxwellian ansatz for $f_\chi(\vecv)$, we have
\be
\eta(\vmin) \propto \exp \left(-\frac{m_N E_R}{2\mu_{\chi N}^2 \sigma_v^2}\right),
\ee
so in fact the spectrum is exponentially falling in $E_R$. 

To summarize, the maximum nuclear recoil energy drops as $m_\chi^2$ as the DM mass drops below $m_{N}$, while the spectrum is exponentially falling with $E_R$. Traditional direct detection of WIMP DM uses heavy target nuclei such as Xe and have thresholds for nuclear recoil detection at $\sim$ few keV, severely limiting sensitivity to GeV-scale DM.  Given the exponentially-falling rate, reducing energy thresholds is key to extending sensitivity to lower mass dark matter, and generally results in large increases in rate. For a given detector energy threshold, lighter target nuclei are also preferred for lighter DM, although there are clearly limits to this strategy for sub-GeV DM.

\begin{figure*}[t!]
\begin{center}
\includegraphics[width=0.55\textwidth]{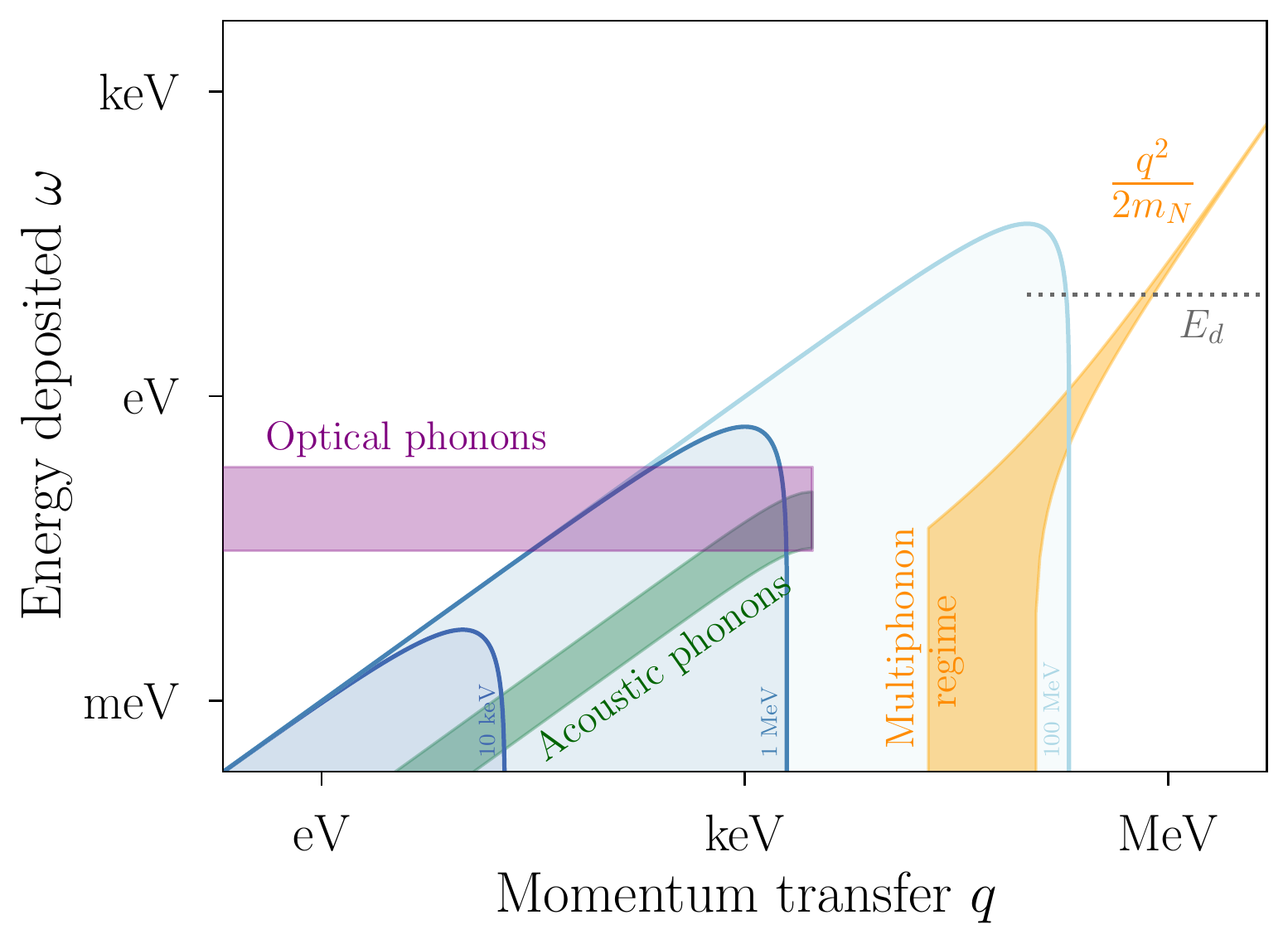}
\caption{ \label{fig:DMN_qw} For DM-nucleus interactions, the response at high $q$ is highly peaked about the free-nucleus dispersion $\omega = q^{2}/(2 m_{N})$. As the energy drops below the displacement energy for a nucleus in a potential, $E_{d}$, the response about the free-nucleus dispersion broadens and we enter the multiphonon regime at $q \sim 10-100$ keV. For $q$ well below $\sim 10$ keV, the dynamic structure is dominated by resonant response on the acoustic and optical phonon dispersions, corresponding to single phonon excitations. Kinematically allowed regions for DM scattering are shown for $m_{\chi} = 10$ keV, 1 MeV, and 100 MeV at $v= 10^{-3}$, as in Fig.~\ref{fig:parabola}.}
\end{center}
\end{figure*} 

Importantly, this description in terms of free nuclear recoils will start to break down for DM below the GeV scale, since we run into the energy and momentum scales relevant for ions in a condensed matter target. There are several assumptions made in Eq.~(\ref{eq:dRdERElastic}) that will no longer hold. One assumption was that any initial motion of the nucleus in the medium could be neglected, and that we could treat the nucleus wavefunction as a plane wave. However, as $q$ approaches  the typical zero point momentum for an ion in a medium, $\sim 30-100$ keV in solid state materials, then the bound nature of the nucleus cannot be neglected and we must account for nontrivial ion wavefunctions. In addition, we assumed that the momentum transfer was high enough that only individual nuclear recoils had to be considered. This led to a total elastic recoil rate which was quadratic in $A^2$, indicative of coherent scattering off of all the nucleons in a given nucleus, and linear in the total number of target nuclei, indicative of incoherent scattering off of all the nuclei in the target.  As $q$ drops below the scale $\lesssim 30$~keV, the DM can probe multiple ions and we must consider final states involving correlated motions of the ions. This will lead to coherent scattering off the medium, or phonon production.   Fig.~\ref{fig:DMN_qw} illustrates these different regimes for the nuclear response in $q, \omega$ as compared with the kinematically allowed region for different sub-GeV DM masses: DM with $m_{\chi} \gtrsim 100$ MeV will dominantly excite free elastic recoils, DM with 10~MeV~$\lesssim m_{\chi} \lesssim 100$ MeV will probe the bound nature of a nucleus (leading to the multiphonon regime), and DM with $m_{\chi} \lesssim$ 10 MeV will mainly excite single phonons.  In the rest of the section, we will go through these latter two momentum regimes in detail. 

\begin{figure*}[t!]
\begin{center}
\includegraphics[width=0.48\textwidth]{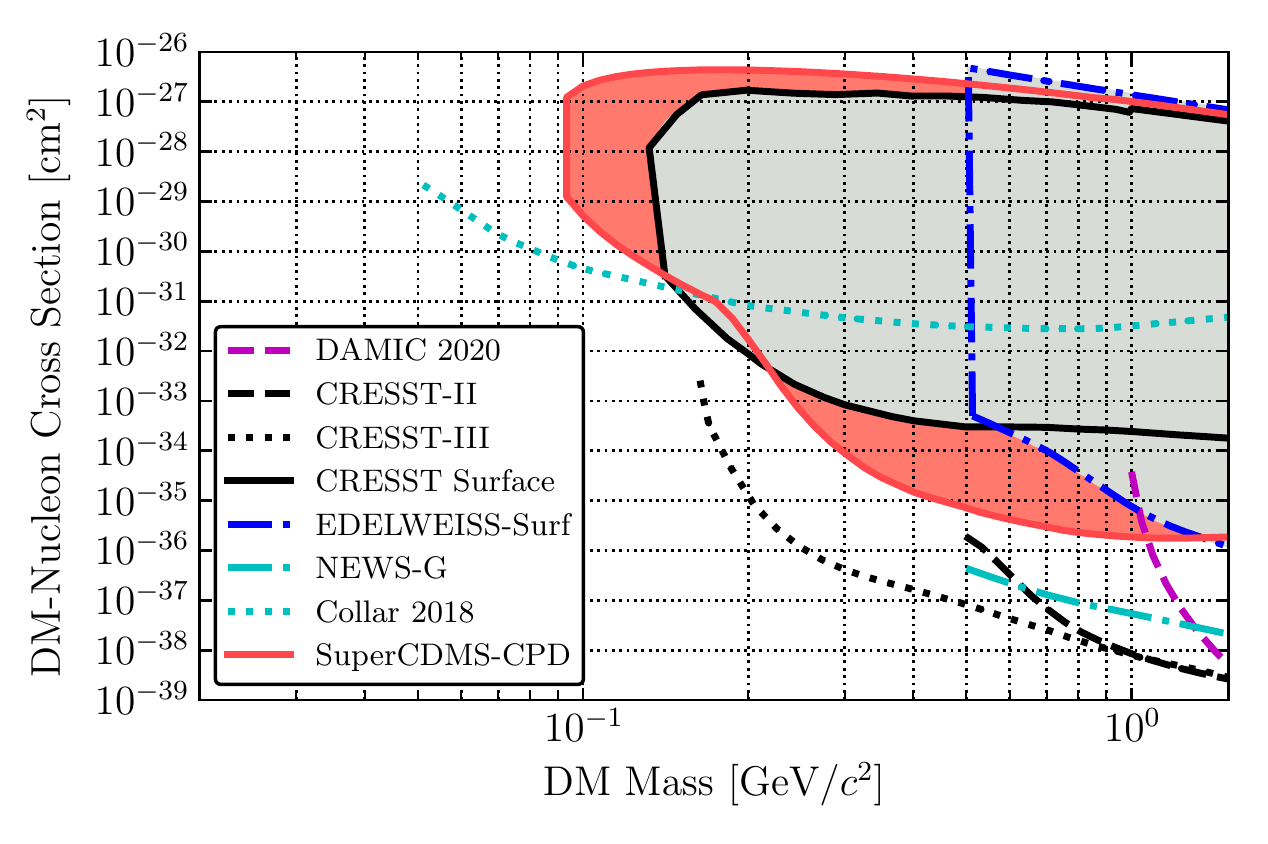}
\includegraphics[width=0.48\textwidth]{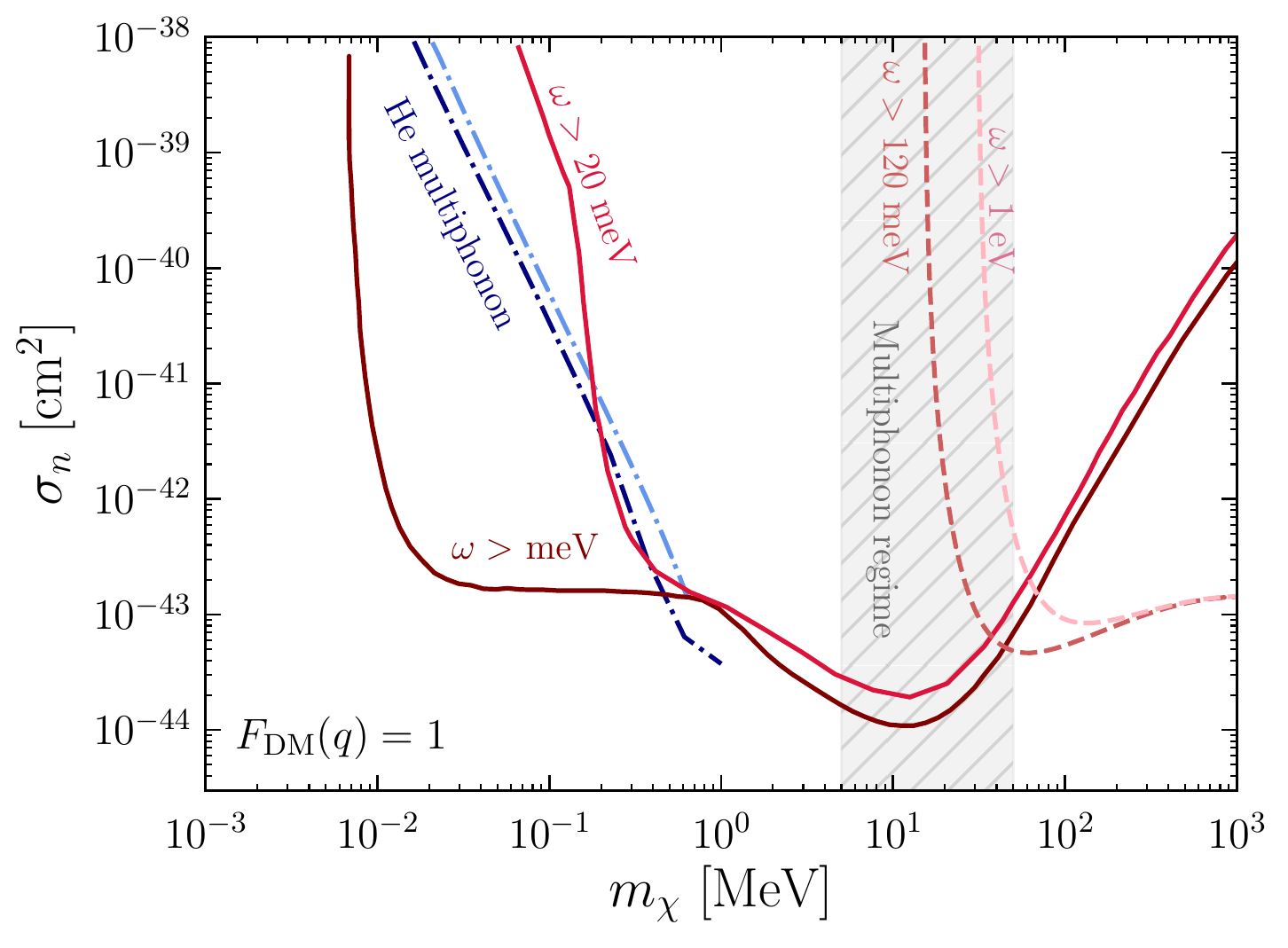}
\caption{ \label{fig:DMNLimits}
(\emph{left}) Leading experimental limits on DM-nucleon cross section $\overline{\sigma}_n$ for scattering through a heavy mediator. Plot reproduced from Ref.~\cite{Alkhatib:2020slm}. Note that no direct detection limits exist below $m_\chi = 50 \ \MeV$.
(\emph{right}) A comparison of the cross sections needed to observe 3 events per kg-year for various energy thresholds. The red lines show the reach in Si from exciting single acoustic ($\omega > $ meV) or single optical and high-energy acoustic  ($\omega > 20$ meV) phonons~\cite{Griffin:2019mvc}, discussed in Sec.~\ref{sec:acousticoptical}. The dashed lines show the result assuming free nuclear recoils of Si for $\omega > $ 120 meV or $\omega > $ eV. Multiphonon contributions are expected to be important in the shaded range, here taken to be the mass range where $q/\sqrt{2 m_{N} \omega_{0}} \approx 0.1-10$ with $\omega_{0}$ the Debye frequency in Si. The actual importance of multiphonons depends also on the energy threshold, see Fig.~\ref{fig:ReachPlotNucleusSHO} for an estimate of the reach with various thresholds in a toy harmonic oscillator model.  The dash-dotted lines are the reach from multiphonon excitations in He, discussed in Sec.~\ref{sec:helium}; the dark blue line is from Ref.~\cite{Knapen:2016cue} with $\omega > 1.2$ meV and the light blue line is from the EFT of Ref.~\cite{Caputo:2019cyg}.
}
\end{center}
\end{figure*} 

The current limits on DM-nuclear scattering are summarized in Fig.~\ref{fig:DMNLimits} (left). The reach at low masses is driven entirely by the detector thresholds, which are currently at the eV scale for solid-state detectors (see Sec.~\ref{sec:Detection}); the curve labeled ``Collar 2018'' refers to a liquid scintillator detector \cite{Collar:2018ydf}, for which the electron scattering reach is discussed further in Sec.~\ref{sec:scintillation}. At higher mass the reach is driven by exposure, which is somewhat limited for the low-threshold prototype detectors which have gram-scale masses. The projections for future kg-year exposure experiments with sub-eV thresholds, sensitive to single phonons produced from a DM-nucleus contact interaction, are summarized in Fig.~\ref{fig:DMNLimits} (right). Searches for DM-nucleus scattering mediated by a dark photon require a more careful choice of target material, and are discussed in Sec.~\ref{sec:DMNdarkphoton}, with reach projections summarized in Fig.~\ref{fig:DMN_darkphoton}.

\subsection{From nucleons to phonons: a harmonic oscillator model}
\label{sec:SHOToy}

As the DM mass drops below $\sim$ GeV, the free-nucleus recoil energy $\sim (m_\chi v)^{2}/m_N$ becomes comparable to the typical energy to break a molecular bond or displace an ion in a crystal, $\mathcal{O}(10)$ eV. Indeed, a possible direct detection signature of sub-GeV might be chemical bond breaking or production of defects in crystals~\cite{Essig:2016crl,Budnik:2017sbu}. However, for MeV-scale DM and below, nuclear scattering might not have enough energy to displace a nucleus or break a bond, and we must then consider the relevant eigenstates and energies of the bound nucleus. For a crystal, this can become complicated since the nuclei (ions) are all coupled, which leads to the phonon excitations at low energies. However, if we restrict to the DM mass range above $\mathcal{O}(10)$ MeV, then the typical momentum transfer $q \sim m_{\chi}v \gtrsim \mathcal{O}(10)$ keV. Then it is still possible to take an incoherent approximation where we can model DM scattering as occurring off of an individual bound nucleus in the crystal, as long as the deposited energy remains below the displacement energy.

Thus, in the DM mass range of tens of MeV up to 1 GeV, a useful toy model for DM-nucleon scattering is provided by considering a single nucleus subject to a simple harmonic oscillator potential sourced by its neighboring ions. The toy model allows us to explicitly see the transition from free nuclear recoils to single or few-phonon transitions in crystals, and gives some additional intuition for the phonon dynamic structure factor discussed in Sec.~\ref{sec:Sqw_phonon} without the complicated details of the phonon band structure and eigenstates. The toy model can also be used to describe a simple system of a diatomic molecule, as an intermediary system between free atoms and condensed matter systems, in the limit of small anharmonic corrections to the potential.  In this section, we will give a description of the potential DM signal that comes from exciting bound states of a harmonic potential, as a prelude to single or few-phonon excitations in condensed matter systems.\footnote{For an interesting and detailed study on the possibility of detecting molecular excitations induced by DM scattering in a gaseous target, see Ref.~\cite{Essig:2019kfe}.}

The harmonic potential we will work with is given by
\be
\mathcal{V}(r) = \frac{1}{2} m_N \omega_0^2 r^2,
\ee
where we can interpret the harmonic oscillator level $n$ as an $n$-phonon state in a crystal (where $r$ is the displacement from the equilibrium position), or a bound state in molecule (where $r$ is the interatomic distance). For a diatomic molecular system, the mass $m_N$ should be replaced by the reduced mass of the system. To solidify the interpretation of $n$-phonon states in a crystal, note that if we take $\omega_0 = 60 \ {\rm meV}$, a typical phonon energy for silicon, and $r_0 = 0.235 \ {\rm nm}$ equal to the nearest-neighbor distance in a Si lattice, we obtain $\mathcal{V}(r_0) = 66 \ \eV$, on the same order as the maximum displacement energy of 47 eV along the [110] direction \cite{jiang2018theoretical}. The motivation for using $\omega_0 = 60 \ {\rm meV}$ is that if a single nucleus is struck with a large momentum transfer, the motion will have a large overlap with the phonon modes at the edges of the 1BZ, where the ions have more random motions and energies $\sim 60$ meV.

We will again assume a contact interaction as in Eq.~(\ref{eq:Hnuc_contact}) for simplicity, with equal proton and neutron couplings. Because we are still in the incoherent regime, the form of the rate will be very similar to that of Eq.~(\ref{eq:dRdERElastic}), except that we must use the bound states of the nucleus for the target-dependent form factors. In this regime, the momentum transfer is too low to probe the nuclear structure, so the nuclear form factor in Eq.~(\ref{eq:Fnuclear}) can be taken to be unity and is not relevant for this discussion. The rate in this model is given by
\be
R = N_{T} \frac{\rho_\chi}{m_\chi} \frac{A^2 \sigmabar_n}{2 \mu_{\chi n}^2} \int q \, dq  \sum_n |f(n, \vecq)|^2 \eta(\vmin(q, \omega_{n})) 
\ee
where
\be
|f(n, q)|^2 \equiv \frac{1}{4\pi} \int d\Omega_{\vecq} \, |f(n,\vecq)|^2  =  \frac{1}{4\pi} \int d\Omega_{\vecq} \sum_{n_x + n_y + n_z = n} | \langle n_x, n_y, n_z | e^{i \vecq \cdot \vecr} | 0 \rangle|^2.
\ee
Here we can take the angular average over all $\vecq$ directions, since we are dealing with an isotropic target system. This quantity is proportional to the dynamic structure factor by
\be
S(\vecq, \omega) = \frac{2\pi}{V} \sum_n |f(n,\vecq)|^2 \delta(n \omega_0 - \omega).
\ee
In this case, since the spectrum of the harmonic oscillator is known exactly, we can enumerate the final states $|f\rangle$ in Cartesian coordinates by eigenvalues $n_x, n_y, n_z$, with total energy $\left(n + \frac{1}{2}\right) \omega_0$ where $n = n_x + n_y + n_z$. Thus $\vmin(q, \omega_{n})$ is evaluated at  $\omega_{n} = n \omega_0$.

To compute $|f(n, \vecq)|^2$, consider first one Cartesian component of the matrix element, $|\langle n_x | e^{i q_x x} | 0 \rangle|^2$. The perturbation operator can be written in terms of creation and annihilation operators:
\be
\exp (i q_x x)  = \exp\left[     \frac{i q_x}{\sqrt{2 m_N  \omega_0}} \left(a_x+ a_x^\dagger \right) \right],
\ee
which has a very similar form to the exponential phonon displacement operator in Sec.~\ref{sec:Sqw_phonon}. Similar to the calculation there, we use the Baker-Campbell-Hausdorff formula, Eq.~\ref{eq:BCHformula}, with $A = \kappa a_x$, $B = \kappa a_x^\dagger$, and $\kappa = \frac{i q_x}{\sqrt{2 m_N \omega_0}}$ to obtain
\be
\exp\left[\frac{i q_x}{\sqrt{2 m_N  \omega_0}} \left(a_x + a_x^\dagger \right) \right]
= 
\exp\left(-\frac{q_x^2}{4 m_N  \omega_0}\right)
\exp\left(\frac{i q_x}{\sqrt{2 m_N  \omega_0}} a_x^\dagger \right) 
\exp \left(           \frac{i q_x}{\sqrt{2 m_N  \omega_0}} a_x   \right)~.
\ee
The first exponential factor is analogous to the Debye-Waller factor introduced in Sec.~\ref{sec:Sqw_phonon}. Taking the matrix element of this operator between the states $\langle n_x |$ and $|0 \rangle$, the only contribution is from the  $n_x$th term of the middle exponential and taking the last exponential to unity. Since these operators satisfy 
\begin{equation}
a^n_x |n_x \rangle = \sqrt{n_x!} | 0 \rangle,
\end{equation}
we obtain
\begin{align}
\left \langle 0 \left | 
\exp\left(-\frac{q_x^2}{4 m_N  \omega_0}\right)
\exp\left(\frac{i q_x}{\sqrt{2 m_N  \omega_0}} a_x^\dagger \right) 
\exp \left(           \frac{i q_x}{\sqrt{2 m_N  \omega_0}} a_x   \right)
\right | n_x \right \rangle = & \nonumber \\
\frac{\sqrt{n_x!}}{n_x!}  \left(\frac{i q_x}{\sqrt{2 m_N  \omega_0}}\right)^{n_x} 
& \exp\left(-\frac{q_x^2}{4 m_N  \omega_0}\right)~.
\end{align}
Taking the modulus squared of this expression gives
\begin{equation}
\label{eq:Pnx}
|\langle n_x | e^{i q_x x} | 0 \rangle |^2 = \frac{1}{n_x !} \left(\frac{q_x^2}{q_0^2}\right)^{n_x}\exp\left(-\frac{q_x^2}{q_0^2}\right),
\end{equation}
where $q_0 = \sqrt{2 m_N \omega_0}$ is introduced for both notational convenience and physical transparency: it is the momentum spread of the ground-state wavefunction, which is given by
\be
\phi_{0}(\vecp)=(\pi m_N \omega_0)^{-3/4} e^{-p^2/q_0^2}.
\ee

Eq.~(\ref{eq:Pnx}) describes a Poisson distribution in $n_x$ with mean $q_x^2/q_0^2$.
Since the harmonic oscillator Hamiltonian is separable in Cartesian coordinates, we may compute the desired matrix element $| \langle n_x, n_y, n_z| e^{i \vecq \cdot \vecr} | 0 \rangle|^2$ by multiplying the three Cartesian matrix elements for $n_x$, $n_y$, and $n_z$. Since the sum of Poisson-distributed variables is also Poisson-distributed, the distribution of $n = n_x + n_y + n_z$ will be Poissonian, as can be verified by explicit computation:
\be
|f(n,q)|^2  = \frac{1}{n!} \left(\frac{q^2}{q_0^2}\right)^{n} \!\! e^{-q^2\!/q_0^2}.
\ee
The mean occupation number for a given momentum transfer $q$ is
\be
\bar{n}_q = \frac{q^2}{q_0^2} = \frac{q^2}{2 m_N \omega_0}  = \frac{E_R}{\omega_0},
\ee
where $E_R = \frac{q^2}{2m_N}$ is the recoil energy which would correspond to free-particle elastic scattering. Equivalently, \emph{the mean energy deposit $\bar{n}_q \omega_0$ is the same as classical elastic scattering $E_R$, even for a bound nucleus.}

\begin{figure}
\hspace{-0.2in}
\includegraphics[width=0.55\textwidth,angle=0]{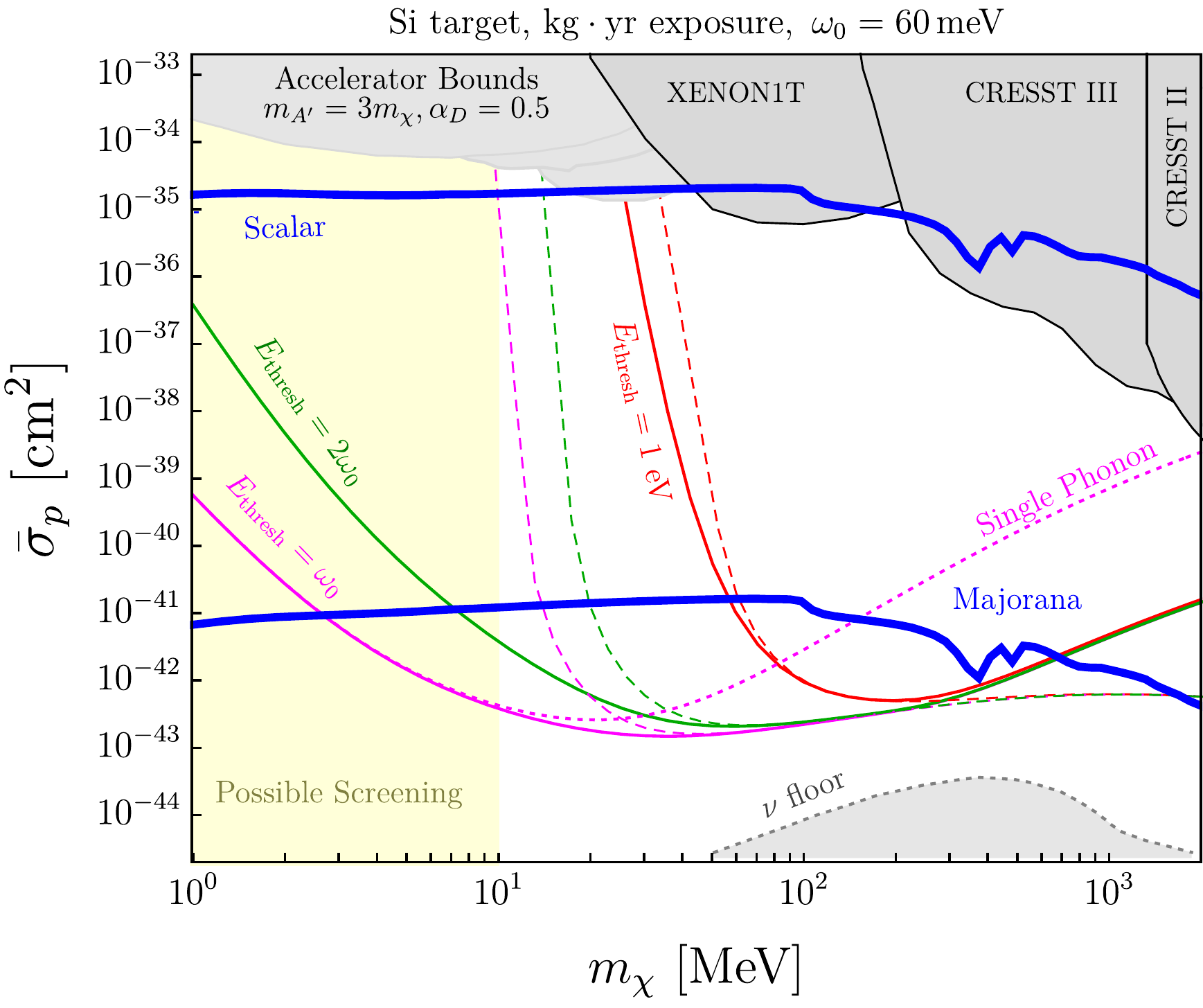}
\caption{ 
Projected exclusion curves (3 events, zero background) for various
 energy thresholds in the simple harmonic oscillator model with a dark photon mediator, along with elastic scattering (long dashed) and single-phonon production (short dashed magenta), reproduced from Ref.~\cite{Kahn:2020fef}. 
Constraints from direct detection experiments \cite{XENON:2019gfn,PhysRevD.100.102002,Abdelhameed:2019hmk} (converted from electron scattering limits to nuclear scattering limits in the dark photon model where necessary) and accelerator experiments are shaded grey, along with the neutrino floor and the thermal target parameter space (blue lines) for various candidate DM spins  \cite{Battaglieri:2017aum}. Since the mediator in this model couples to charge rather than nucleon density, the yellow shaded region $m_\chi < 10$ MeV shows the low-momentum-transfer regime where atomic screening effects may be important, see Ref.~\cite{Kahn:2020fef} for details.}
\label{fig:ReachPlotNucleusSHO}
\end{figure}

In the regime where $q \gg q_0$, the Poisson distribution becomes sharply peaked around $\bar{n}$, approaching a delta function $\delta(n - \bar{n})$. This enforces the kinematic relation for elastic scattering, $q = q_0 = \sqrt{2m_N E_R}$ and permits the change of variables $dq = (m_N/q)dE_R$. Taking the continuum limit with $\sum_n \to \int dn$ and $n \to E_R/\omega_0$, we recover precisely the spectrum for elastic nuclear scattering in Eq.~(\ref{eq:dRdERElastic}) since
\be
\vmin = \frac{\bar{n}_q \omega_0}{q_0} + \frac{q_0}{2m_\chi} = \frac{q_0}{2\mu_{\chi N}} = \sqrt{\frac{m_N E_R}{2\mu_{\chi N}^2}}.
\ee
Thus, at large momentum transfer compared to the characteristic scale of the harmonic oscillator, scattering is nearly elastic and the energy deposited is given by $E_{R}$, up to small fluctuations about $E_{R}$ as well as quantization of the spectrum in units of $\omega_0$. There is an interesting intermediate regime as $q$ approaches $q_0$ from above. Here, $\bar{n}_q \gtrsim 1$ and the Poisson distribution yields order-1 fluctuations in $n$ and thus order-1 fluctuations in the energy deposited from DM scattering. This \emph{multi-phonon regime} smoothly interpolates between large-momentum quasi-elastic scattering and small-momentum single-phonon production, and had not been considered much in the DM direct detection literature until recently \cite{Kahn:2020fef,Knapen:2020aky}. As a practical matter, the fact that $\bar{n}_q \omega_0 = E_R$ means that the intuition for the kinematics of elastic scattering still hold for a bound nucleus in a harmonic potential, but the fact that the energy spectrum is quantized means that the rate for energy deposits of twice the elastic energy, $2 \omega_0$, is of the same order as the rate for $E_R$. This implies that for a given detector threshold, part of the multiphonon signal may be above threshold even when $E_{R}$ is not. This allows for new parameter space to be probed by detectors which cannot yet achieve single-phonon sensitivity, as illustrated in Fig.~\ref{fig:ReachPlotNucleusSHO}. Another approach to the multi-phonon regime that does not use the simple harmonic oscillator model can be found in Ref.~\cite{Knapen:2020aky},  which instead uses a phonon density of states, yielding similar results with the nuclear response peaked about the free recoil $E_{R}$ with a width of $\sim \sqrt{\omega_{0} E_{R}}$

Finally, let us consider the regime $q \ll q_0$. Here the factor $e^{-q^2/q_0^2} \approx 1$, and the probability for the production of $n$ phonons is suppressed by $(q^2/q_0^2)^n$. This is the same scaling as Eq.~(\ref{eq:Sqw_roughscaling}),  following our calculation there with detailed phonon modes. However, we can see by explicit comparison with Eq.~(\ref{eq:Sqw_1phonon})  that the $n=1$ term in this simple model will not capture the target-dependent details of a solid state system. As $q$ drops down to keV scales and below, the particular phonon spectrum and modes of the material will play an important role in the form factors, and the assumption of a single phonon energy $\omega_{0}$ breaks down. Furthermore, the DM couplings also enter into Eq.~(\ref{eq:Sqw_1phonon}). The simple $q^{2}/q_0^2$ scaling will not necessarily be present, depending on the DM model. We will turn to this regime next.

\subsection{Acoustic and optical phonons}
\label{sec:acousticoptical}

For DM mass below $\sim$ 1 MeV, the maximum momentum transfer is $q < 1 \ \keV < q_{\rm coh}$, such that the incoherent approximation is no longer valid. 
Furthermore, because $q \ll q_{0}$, the leading rate will generally be from single-phonon excitations, as argued above due to the $(q^2/q_0^2)^n$ scaling of $n$-phonon excitations. In this section, we will give some intuition for the rate for DM to produce single phonons in solid state systems, and summarize the results of detailed numerical calculations from the literature.

First, it is worth commenting on how the kinematics of single-phonon excitations compares to elastic nuclear recoils.  As discussed in Sec.~\ref{sec:CMexcitations}, there are two basic branches of phonons we could consider, acoustic and optical phonons. The comparison of DM scattering kinematics with the dispersion relations of these phonons is illustrated schematically in Fig.~\ref{fig:DMN_qw}. For the acoustic phonon branch, the energy deposited must be $\omega \sim c_{s} q$ with $c_{s} \sim10^{-5}$. Because the sound speed is so low compared to the DM speed, energy conservation (Eq.~(\ref{eq:omegaqdef})) leads to the solution $q \sim 2 m_{\chi} v$, which is the same as for sub-GeV elastic nuclear recoils. However, for acoustic phonons the energy deposited will be
\be
	\omega \sim 2 c_{s} m_{\chi} v \sim 2 \,  {\rm meV} \times \frac{m_{\chi}}{100 \, {\rm keV}}.
	\label{eq:acoustic_kinematics}
\ee
which is well above the elastic recoil energy for the same DM mass, $E_{R} \sim 10^{-7}-10^{-6}$ eV depending on target mass. The energy deposited into a single acoustic phonon could be made larger by using a relatively hard target material, such as SiC \cite{Griffin:2020lgd} or diamond \cite{Kurinsky:2019pgb} where $c_{s} \approx 4-5 \times 10^{-5}$. However, this energy deposition is still quite small: a DM particle of mass 100 keV has a typical kinetic energy of 100 meV, suggesting that acoustic phonons are not ideal in terms of matching DM kinematics. 

The optical phonon branch offers a potential solution to the problem of kinematic matching. Since the dispersions are fairly flat in momentum across the BZ, Eq.~(\ref{eq:omegaqdef}) gives the approximate solution
\be
	q \sim m_{\chi} v \pm \sqrt{ (m_{\chi} v)^{2} - 2 m_{\chi} \omega_{\rm optical}}.
\ee
The typical energies are around $\omega_{\rm optical}\sim 30-150$ meV, which matches well with the total kinetic energy of DM with mass $\sim 10$ keV -- 1 MeV. The higher energies are also favorable for experimental implementation. From this discussion, we might expect that an ideal target material would likely have a broad spectrum of optical phonon energies in the range of $10$ meV up to 150 meV, to allow kinematic matching with a broad range of DM masses. In fact, one can go further and consider systems with some amount of disorder, which further smears out the phonon spectrum and leads to broad spectrum of available modes;  this idea was introduced in Ref.~\cite{Bunting:2017net}, which looked at single molecular magnet crystals as a possible direct detection target. Treating the DM coupling to specific modes is more challenging in this type of system, however. In this section, we will study only ordered crystalline lattices, where we can next specify how DM couples to individual phonon branches.

Assuming the same contact interaction as above, Eq.~(\ref{eq:Hnuc_contact}), and again applying our main rate formula Eq.~\ref{eq:Rfactorized}, the rate is given by
\be
	R_{\chi}^{(1-ph)} = \frac{1}{\rho_T}\frac{\rho_\chi}{m_\chi} \int  \, d^3 \vecv  f_\chi(\vecv) \int \frac{d^3 \vecq}{(2\pi)^3} \,  d\omega \, \frac{\pi \bar \sigma_{n}}{\mu_{\chi n}^2}  \delta( \omega + E'_{\chi} - E_\chi) \, S^{(1-ph)}(\vecq, \omega).
\ee
For convenience, we reproduce the single phonon dynamic structure factor:
\begin{align}
	S^{(1-ph)}(\vecq, \omega) &= \frac{ 2 \pi}{\Omega} \sum_{\nu} \left| \sum_{j } f_{j} e^{-W_j(\vecq)}   \frac{i\vecq \cdot \mathbf{e}^*_{\nu,\vecq, j} }{\sqrt{2 M_j \omega_{\nu,\vecq}}} \right|^{2}  \delta(\omega - \omega_{\nu,\vecq})
\end{align}
with $f_{j} \to A_{j}$ for equal couplings to protons and neutrons. 

While the eigenmodes and dispersions of the phonon branches must be solved by DFT methods for arbitrary $\vecq$, it is possible to obtain approximate results in the long-wavelength limit $q \ll \pi/a$ where $\pi/a$ is the typical size of the first BZ. In this limit, we know that the acoustic phonon modes are Goldstone bosons of broken translation invariance, and that as $q \to 0$ all the ions are displaced by the same amount for a zero-energy mode. Comparing with Eq.~(\ref{eq:udef}) for the displacement ${\bf u}$, we see that in order for the $\sqrt{M_{j}}$ dependence to drop out, the eigenmodes for the acoustic phonons must be given by $|\mathbf{e}_{\nu,\vecq, j}| = \sqrt{M_{j}}/\sqrt{\sum_{d} M_{d}} $ as $q \to 0$; here the factor of $\sqrt{\sum_{d} M_{d}}$ is just to give a normalized eigenvector, where $d$ sums over all ions in the unit cell. In addition, we can restrict only to the longitudinal acoustic (LA) phonon branch where $\mathbf{e}_{LA,\vecq,j} = \hat \vecq  \sqrt{M_{j}}/\sqrt{\sum_{d} M_{d}} $ because of the  $\vecq \cdot \mathbf{e}^*_{\nu,\vecq, j}$ dot product. This gives the LA single-phonon form factor in the long-wavelength limit:
\begin{align}
	\lim_{q\to 0} S^{(1-ph, LA)}(\vecq, \omega) \approx \frac{ 2 \pi}{\Omega} \frac{q^{2} | \sum_{j } A_{j} |^2}{2 \, (\sum_d M_d ) \, c_{s} q} \delta(c_{s} q - \omega) 
	\label{eq:Sqw_LA}
\end{align}
where we have also taken $e^{-W_j(\vecq)}  \approx 1$ and an isotropic speed of sound. The form of this structure factor is similar to that computed for the harmonic oscillator model, $\propto q^{2} A^{2}/(2 m_{N} \omega_{0}) \delta (\omega-\omega_{0})$ for single-phonon excitations. The difference here is that we are taking coherent sum of the couplings over the ions in the unit cell, $| \sum_{j } A_{j} |^2$, as well as dividing by unit cell mass. In addition, replacing $\omega_{0}$ with the linear dispersion of the acoustic phonons leads to a $\sim q$ scaling for the single phonon structure factor, rather than the $\sim q^{2}$ scaling that was found in the toy harmonic oscillator model. Eq.~(\ref{eq:Sqw_LA}) shows there is a coherent coupling enhancement over ions in a unit cell for acoustic phonons, and in total scales as the number of ions in the unit cell. As noted above, the kinematic matching is not ideal. If the energy threshold can be lowered to $\omega \ge 1$ meV, however, there is still potentially a very strong cross section reach from single acoustic phonons, as can be seen for the $\omega \ge 1$ meV line in Fig.~\ref{fig:DMNLimits}.

We next turn to the optical phonon branch.  For DM models where $f_{j} = A_{j}$, there is instead a destructive interference for the optical phonon coupling. To see why, let us assume $M_{j} = A_{j} m_{n}$. Then we can rewrite the structure factor as
\begin{align}
	S^{(1-ph)}(\vecq, \omega) &= \frac{ 2 \pi}{\Omega} \sum_{\nu} \left| \sum_{j }  e^{-W_j(\vecq)}   \frac{i \sqrt{M_{j}} \vecq \cdot \mathbf{e}^*_{\nu,\vecq, j} }{\sqrt{2 m_n^2\omega_{\nu,\vecq}}} \right|^{2}  \delta(\omega - \omega_{\nu,\vecq}).
\end{align}
In the long-wavelength limit, we can exploit the scaling of the LA phonon mode and rewrite the dot product $\sqrt{M_{j}} \vecq \cdot \mathbf{e}^*_{\nu,\vecq, j} $ as a dot product with the LA mode, $\mathbf{e}_{LA,\vecq, j}  \cdot \mathbf{e}^*_{\nu,\vecq, j} $. However, by definition for normal modes, the optical phonon eigenmodes are orthogonal to the acoustic phonon eigenmodes, so this dot product vanishes! Thus, the rate to produce optical phonons in this model (where the DM coupling is proportional to mass) is highly suppressed compared the acoustic phonon rate, despite the kinematic advantages. This effect was observed in the first calculations of scattering into optical modes for specific materials~\cite{Knapen:2017ekk,Griffin:2018bjn}, and shown to be true in general using the orthogonality argument in Ref.~\cite{Cox:2019cod}.  

Since optical modes are so kinematically well-matched, it is worth obtaining the leading nonzero contribution to the rate. First, $M_{j} = A_{j} m_{n}$ is not exactly true, since a nucleus is a bound state of nucleons and there is a small binding energy. Protons and neutrons also have slightly different masses. This implies a small nonzero dot product, but the effect is quite small since the deviation from $A_{j} m_{n}$ is at the $10^{-3}$ level~\cite{Cox:2019cod}, and we will continue to assume $M_{j} = A_{j} m_{n}$. A larger effect results from finite $q$ corrections to the eigenmodes. We can see this by taking a unit cell with only two ions of mass $M_{1}, M_{2}$, and approximating the optical phonon eigenmode at low but finite $q$:
\be
	\mathbf{e}_{LO,\vecq,1} \approx \hat \vecq\frac{ \sqrt{M_2}}{\sqrt{M_{1} + M_{2}} }, \quad 
	\mathbf{e}_{LO,\vecq,2} \approx - \hat \vecq \frac{ \sqrt{M_1}}{\sqrt{M_{1} + M_{2}} } e^{-i \vecq \cdot \vecR_{2}^{0}}
	\label{eq:eigenmodeLO}
\ee
where we took only the longitudinal optical (LO) mode again due to the appearance of $\vecq \cdot \mathbf{e}^*_{\nu,\vecq, j}$ in the structure factor.  As $q \to 0$, this eigenmode is orthogonal to the LA mode, and the two ions oscillate exactly out of phase. At finite $\vecq$, however, the different equilibrium positions of the ions start to be resolved compared to the wavelength of the oscillation and there is an additional relative phase factor (see also the toy model with a diatomic molecule in Ref.~\cite{Cox:2019cod}). In this unit cell, one ion is at the origin and the other ion is located at position $ \vecR_{2}^{0}$. For a cubic lattice, for instance, $ \vecR_{2}^{0} = (a/4, a/4, a/4)$ with $a$ the lattice constant. For $q \ll \pi /a$, we can expand the phase factor in powers of $qa$, which gives rise to the dynamic structure factor
\be
	S^{(1-ph, LO)}(\vecq, \omega)\approx \frac{ 2 \pi}{\Omega} \frac{q^{2}A_1 A_2}{2(M_1 + M_2)\omega_{LO}}  \frac{q^{2} a^{2}}{16}  \delta(\omega - \omega_{LO}).
	\label{eq:Sqw_LO}
\ee
where we have taken the angular average $|\vecq \cdot \vecR_{2}^{0}|^{2} \approx q^{2}a^{2}/16$ and assumed a $q$-independent LO phonon energy $\omega_{LO}$. This has a similar form to the previous single-phonon excitation factors derived, but there is an additional $(qa)^{2}$ suppression when $q \ll \pi/a$ due to the destructive interference in the coupling. The structure factor thus scales as $\sim q^{4}$ for sub-MeV DM scattering. This behavior has been confirmed in numerical calculations~\cite{Cox:2019cod,Griffin:2018bjn}, and leads to the reduced cross-section sensitivity for producing a single optical phonon. This can be seen in Fig.~\ref{fig:DMNLimits} for $m_{\chi} \lesssim 0.2$ MeV and $\omega > 20$ meV, where only scattering into a single optical phonon is possible. For masses $m_{\chi} \gtrsim 0.2$ MeV, larger momentum transfers are accessible and single acoustic phonon excitations also contribute to the $\omega > 20$ meV line, leading to a similar sensitivity as the $\omega > 1$ meV line. Given that the single-optical-phonon rate scales as $q^{4}$, it is interesting to consider whether the 2-phonon contribution to the rate is comparable, since it is expected to have the same scaling. This question was studied in Ref.~\cite{Campbell-Deem:2019hdx}, where it was found that the 2-phonon contribution does indeed scale as $q^{4}$, but is still smaller than the single-phonon rate, at least for sub-MeV DM.

\subsubsection{Dark photon couplings}
\label{sec:DMNdarkphoton}

\begin{figure*}[t!]
\begin{center}
\includegraphics[width=0.55\textwidth]{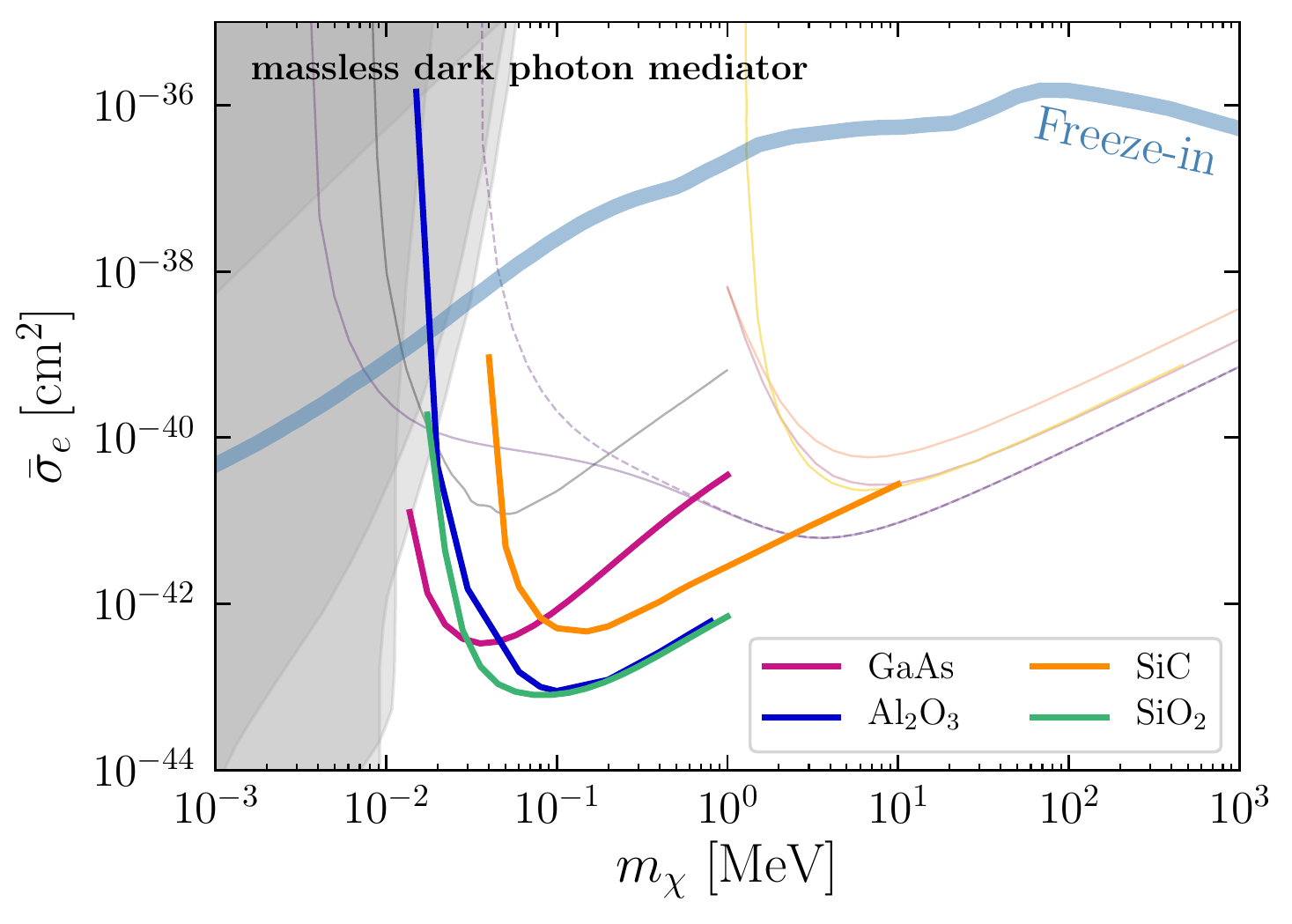}
\caption{ \label{fig:DMN_darkphoton} Cross section for 3 events/kg-year for single optical phonon excitations in various polar materials, and assuming a massless dark photon mediator. For this mediator, the convention in the literature is to show projections in terms of the DM-electron cross section $\bar \sigma_{e}$ even when the scattering is into phonons. This is for easier comparison with experiments searching for DM-electron scattering, which can probe the same model; in particular, the different faint lines in this plot are the various projections for DM-electron scattering proposals shown in Fig.~\ref{fig:DMelectron_all}.  The thick blue line is the predicted cross section if all of the relic DM is produced by freeze-in interactions~\cite{Essig:2011nj,Dvorkin:2019zdi} and the shaded regions are constraints from stellar emission~\cite{Vogel:2013raa,Chang:2018rso}.}
\end{center}
\end{figure*} 

Up to this point, we have dealt with equal proton and neutron couplings (and zero electron coupling), but it is instructive to also consider a dark photon mediator for single phonon excitations. The scaling for the acoustic and optical structure factors above is not universal and depends on the DM model couplings. The situation is quite different for dark photons, with enhanced couplings to optical phonons and a destructive interference with acoustic phonons. With this example, we will see the possibility of selecting target materials to optimize for a certain DM model.

For dark photon mediators, the DM will couple equally and oppositely to electrons and protons, similar to the ordinary photon. The electron coupling introduces some additional complication, since as the ions undergo displacements, the electrons will respond on a rapid time scale. With the same Born-Oppenheimer approximation allowing decoupling of ion and electron motion, the electron response to ion motion can be calculated with first-principles approaches, such that one can determine an effective {\emph {dynamical}} ion charge. This leads to the definition of the {\emph{Born effective charge}}, which is the dynamical ion charge in the long-wavelength limit. Formally, it is a charge tensor for each ion $j$ in the unit cell, defined as the change in polarization $\mathbf{P}$ resulting from a displacement to ion $j$:
\be
	\mathbf{Z}^{*}_{j} \equiv \frac{\Omega}{e} \frac{\partial \mathbf{P}}{\partial \mathbf{u}_{j}}.
\ee
The Born effective charges are nonzero for polar materials, while they vanish for standard non-polar semiconductors such as Si and Ge.  Let us take as a simple example of a polar material GaAs, which has a unit cell of just two ions. The Born effective charges can be approximated to be diagonal and isotropic, so that $\mathbf{Z}^{*}_{\rm Ga} \approx {\rm diag}(2.27, 2.27, 2.27)$ and $\mathbf{Z}^{*}_{\rm As} \approx {\rm diag}(-2.27, -2.27, -2.27)$~\cite{Griffin:2018bjn}, describing an effective charge sharing/splitting between the two ions. If one had modeled the Ga as donating all 3 outer shell electrons to the As, the electric charges would be $+3$ and $-3$ of the two ions, but the actual Born effective charges of $+2.27$ and $-2.27$ account for the deformation of the electron wavefunctions as the ion is displaced. Since we are dealing with a net neutral target, the sum of Born effective charge tensors for the ions must  also be equal to zero.  Note that aside from determining the structure factor, the polarization induced $\mathbf{P}$ implies that the phonon energies must be re-calculated including the electrostatic energy of this polarization. This leads to an additional contribution to the dynamical force matrix, and an increase of the LO phonon energy. For further discussion of the Born effective charges and their effect on the LO energies, see discussion in Refs.~\cite{Griffin:2018bjn,Trickle:2019nya}.

Recalling the out of phase oscillations for LO modes, Eq.~(\ref{eq:eigenmodeLO}), and the fact that Ga and As have opposite Born effective charges, we see that the LO mode in a polar material can be thought of as a coherently oscillating dipole in the $q\to0$ limit. Thus, we can expect that the dark photon mediator primarily couples to the LO mode. To relate the dynamical ion charge to the DM couplings $f_{j}$, we must further use the fact that the ion charge will be screened in a medium, where the relevant screening factor is given by $\epsilon_{\infty}$.
Here $\epsilon_{\infty}$ is the long-wavelength dielectric screening at frequencies below the electron band gap but well above the optical phonon frequencies, such that $\epsilon_{\infty}$ only receives contributions from valence electrons. (At frequencies below the optical phonon frequencies,  the optical phonons also contribute to dielectric screening, giving rise to a low-frequency dielectric constant $\epsilon_{0}$, with $\epsilon_{0} > \epsilon_{\infty}$ in a polar material.)
Then, taking $f_{j} \to Z^{*}_{j}/\epsilon_{\infty}$ and using Eq.~(\ref{eq:eigenmodeLO}), we accordingly find
\be
	S^{(1-ph, LO)}(\vecq, \omega)\approx 
	\frac{ 2 \pi}{\Omega} \frac{(Z^{*})^{2}}{\epsilon_\infty^{2}} \frac{ q^{2} }{2 \mu_{12} \omega_{LO}}  \delta(\omega - \omega_{LO}).
\ee
where $\mu_{12}$ is the reduced mass of $M_{1}$ and $M_{2}$. Because of the opposite signs of both the phonon eigenmodes and the Born effective charges, there is a coherent sum over the ions in the unit cell, in contrast to the case in Eq.~(\ref{eq:Sqw_LO}). We find the same form and $q^2$ scaling as the harmonic oscillator toy model, if we make the identification of the nucleus mass with $\mu_{12}$ and the replacement of the nucleus coupling with $Z^{*}/\epsilon_\infty$. The structure factor in the case is also sometimes written in terms of a Fr\"{o}hlich interaction which characterizes electron-phonon interactions, since the interaction of the electron is very similar to that of DM through a dark photon mediator~\cite{Knapen:2017ekk,Griffin:2018bjn}. Finally, for the acoustic phonons, the opposite Born effective charges implies a destructive interference when we sum coherently over ions in the unit cell, with the structure factor going to 0 in the limit $q \to 0$. 

From these examples, we see that polar materials with large effective charges and a range of optical phonon energies are nearly-ideal target systems for DM interacting through a dark photon, since they enjoy both the kinematic matching and a coherent sum over the ions in the unit cell. For crystals with multiple optical phonon energies, it is also often the case that the highest-energy mode gives the strongest coupling~\cite{Griffin:2018bjn}. This is because large effective charges also implies larger electrostatic energies associated with the phonon.  For DM which couples equally to protons and neutrons, instead the rate is determined primarily through a combination of the sound speed $c_{s}$, target nucleus masses, and optical phonon energies of the target system, depending on the energy threshold. Fig.~\ref{fig:DMN_darkphoton} shows cross section sensitivities for example polar materials experiments, including GaAs and Al$_{2}$O$_{3}$ which are planned to be used in experimental collaborations. Studies of additional target materials can be found in Refs.~\cite{Griffin:2018bjn,Griffin:2019mvc,Griffin:2020lgd,Coskuner:2021qxo,Knapen:2021bwg}. Due to the resonant response, it can also be seen that single phonon excitations can give a much larger rate than DM-electron scattering (faint lines) for sub-MeV dark matter, at least in the materials studied so far; we will explore the DM-electron response more in the following section. Note that in Fig.~\ref{fig:DMN_darkphoton}, a massless dark photon mediator has been assumed where $F_{\rm DM}(q) = (\alpha m_{e}/q)^{2}$. For sub-MeV DM with $ q < $ keV, this form factor can be quite large. For massive dark photon mediators with $F_{\rm DM}(q) = 1$, the rate is much smaller and there is very limited sensitivity to cosmologically interesting parameter space from optical phonon excitations. 

While we have mainly taken an isotropic approximation for the dynamic structure factors, another advantage of condensed matter systems is the potential directional dependence in $S(\vecq, \omega)$. If the DM-phonon couplings or the phonon dispersions are highly anisotropic, this will lead to a modulation of the DM scattering rate as the Earth (and thus crystal) rotates relative to the typical direction of the incoming DM. The modulation is also sensitive to the DM model details. Combined with the fact that the scattering form factor depends on the DM model, it might be possible to obtain some signal-to-background discrimination in phonon-based detection schemes, or in the case of multiple targets and a positive signal, to deduce information about the DM candidate. The directionality of single-phonon excitation rates is explored further in Refs.~\cite{Griffin:2018bjn,Griffin:2020lgd,Coskuner:2021qxo}.

\subsection{Superfluid helium}
\label{sec:helium}

Historically, the first efforts to exploit the particular properties of phonon modes for sub-GeV DM scattering began with an investigation of superfluid helium, a system which is different enough from solid-state lattices that it merits its own discussion. Superfluid helium-4 had been identified as an excellent candidate detector material as early as 1988 \cite{lanou1988superfluid}, just three years after the first concept for direct detection via nuclear recoil \cite{Goodman:1984dc,Drukier:1986tm}, though the theoretical priors on WIMP DM at the time were strong enough that the prospects for sub-GeV DM were not considered. From the kinematics of nuclear scattering discussed above, it is clear that helium, as the second-lightest nucleus, is ideal for maximizing the energy transfer from GeV-mass DM \cite{Guo:2013dt}. To pack enough atoms into a compact volume to maximize the detection rate, one wants to use the liquid phase: \He\ remains a liquid even down to absolute zero, and in fact becomes a superfluid at 2.2 K. In contrast to a solid-state system with a fixed lattice spacing $a$, the disordered liquid has only an average interparticle spacing $\lambda$. For DM with mass above 5 MeV, the typical momentum transfer exceeds the inverse interparticle spacing, $1/\lambda \gtrsim 5 \ \keV$, and the scattering process can be treated as quasi-elastic, similar to the regime in Sec.~\ref{sec:SHOToy}. However, for lighter DM, the small momentum transfers instead imply that we are in the regime of coherent scattering. The DM will then couple to collective density oscillations in the superfluid, which are quantized into acoustic phonons. Superfluid \He\ is a strongly-correlated liquid, and so the formalism in Sec.~\ref{sec:Sqw_phonon} and Sec.~\ref{sec:acousticoptical} does not apply, however. We will review here the excitations in superfluid \He\, along with a few different strategies for treating this system.

\begin{figure}[bt]\centering
\includegraphics[width=0.48\textwidth]{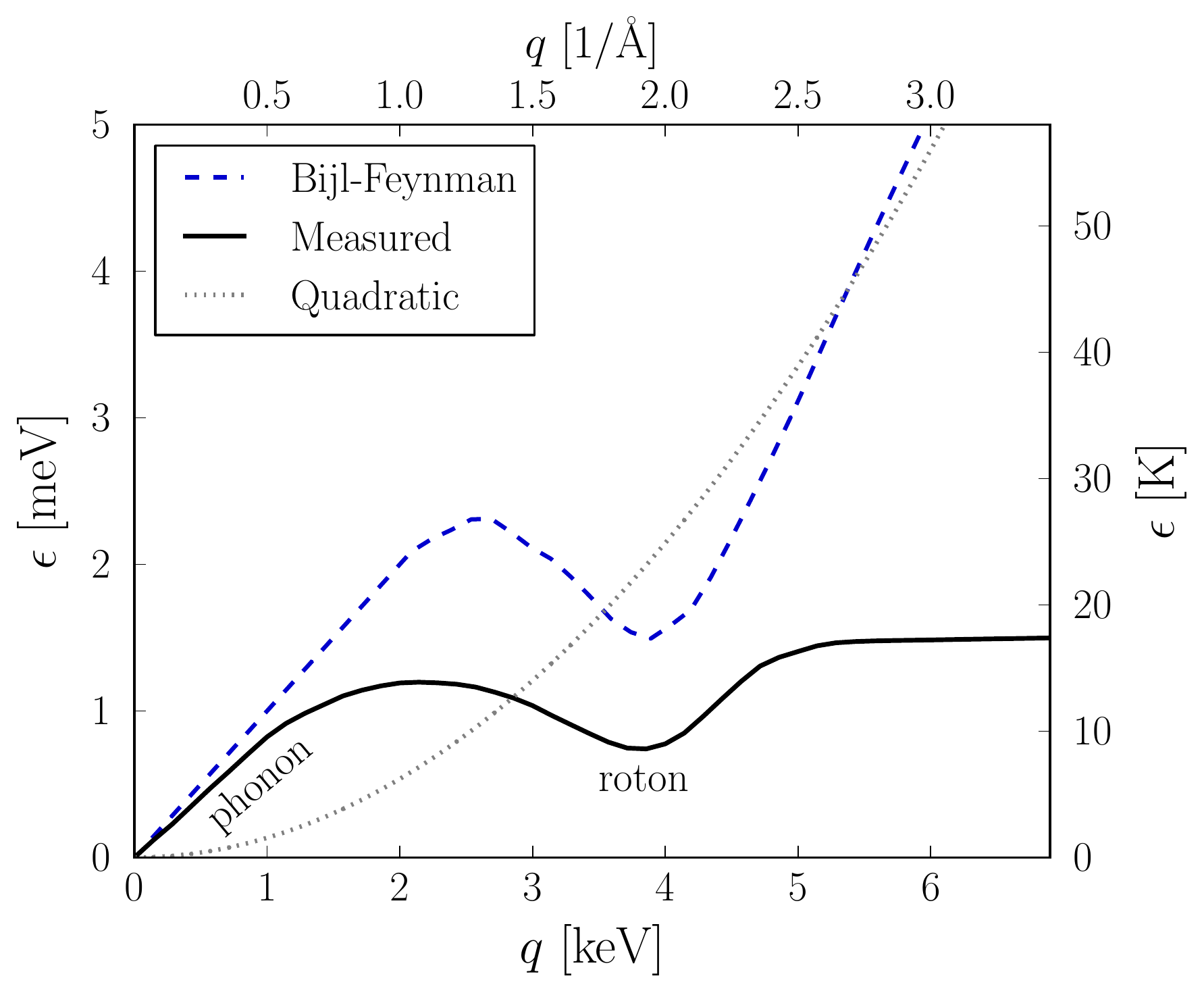}
\includegraphics[width=0.48\textwidth]{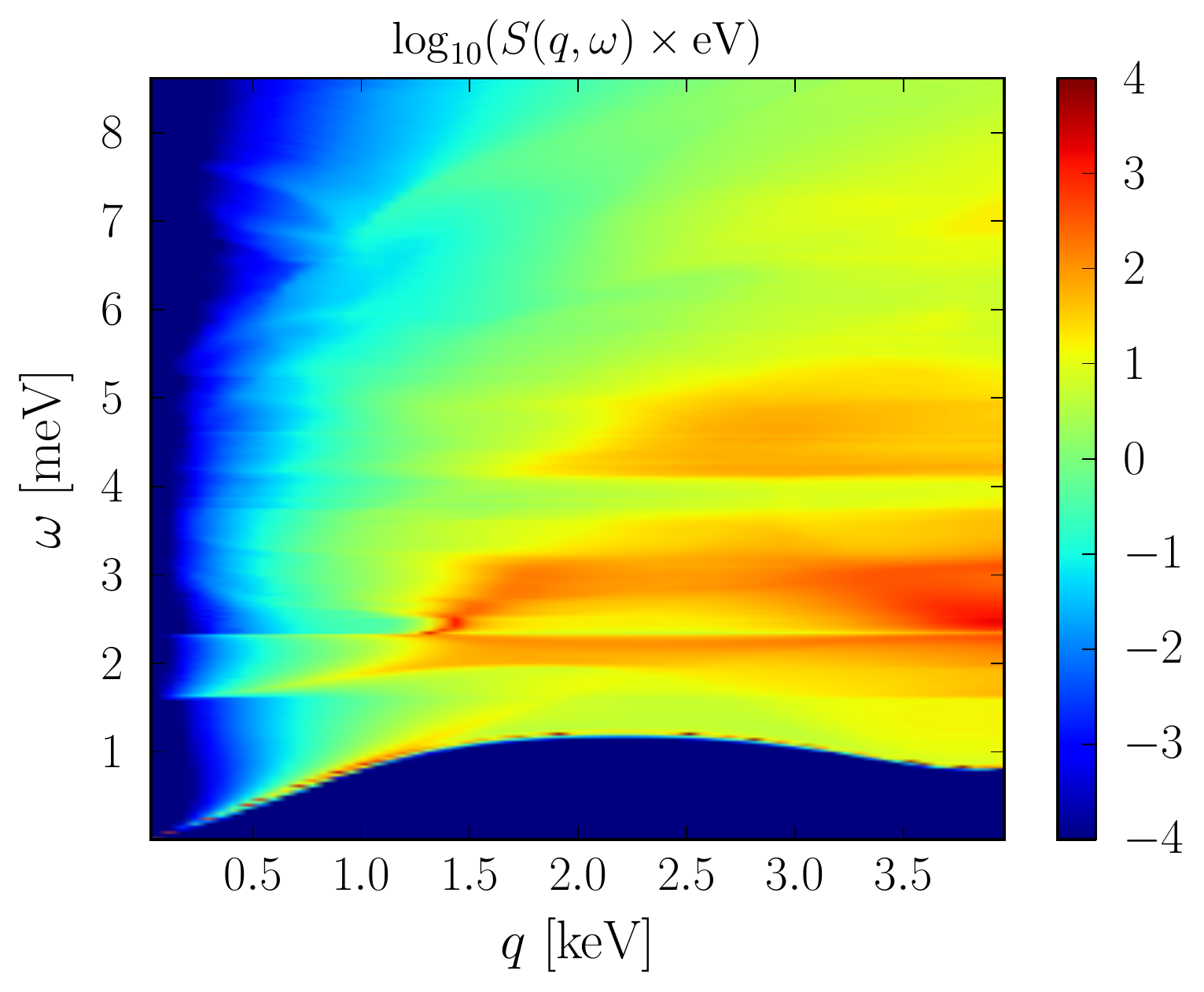}
\caption{Plots reproduced from \cite{Knapen:2016cue}.
   (\emph{left}) The measured dispersion curve for single excitations in superfluid helium (solid black line) has phonon modes at low $\vecq$ and the maxon and roton at high $\vecq$. At high $q$, there is also a broad multi-excitation response centered around the free-particle dispersion (dotted line). The dashed blue is the Bijl-Feynman relation for excitations, Eq.~(\ref{eq:BijlFeynman}), which approximates the phonon-roton dispersion at low $q$ and converges to the free-particle excitations at high $q$. 
    (\emph{right}) Self-consistent calculation of the dynamic structure function $S(q,\omega)$, obtained from Ref.~\cite{Krotscheck2015}. There is a resonant response on the single excitation dispersion, corresponding to the response at the minimum $\omega$ for a given $q$. The response at larger $\omega$ is the multi-excitation component, where the structures at 2 meV and above arise from multi-excitations of rotons/maxons. In the experimental data these  structures are less prominent, which is expected once additional interactions are included (see figures 21-22 and discussion in Ref.~\cite{Krotscheck2015}.)
 \label{fig:Helium}  }
\end{figure}

The measured single-excitation spectrum $\epsilon(\vecq)$ of \He\ (Fig.~\ref{fig:Helium}) contains a piece with linear dispersion at small $\vecq$, $\epsilon(\vecq) \approx c_s |\vecq|$ where $c_s \approx 2.4 \times 10^2$ m/s is the speed of sound, extending out to about $q \simeq 1 \ \keV$. This mode is interpreted as an acoustic phonon. DM can create a single acoustic phonon if   $\omega = c_s |\vecq|$, and the kinematics is similar to that of acoustic phonons in crystals. Analogous to the discussion around Eq.~\ref{eq:acoustic_kinematics}, the typical recoil energy is
\be
	\omega \sim 2 c_{s} m_{\chi} v \sim 0.2\,  {\rm meV} \times \frac{m_\chi}{100 \, {\rm keV} }
\ee
which, unfortunately, is even more poorly kinematically matched to the DM since $c_s$ is an order of magnitude smaller for \He\ than in solid-state targets. The rest of the dispersion curve contains a local maximum at $q \simeq 2 \ \keV$ and $\epsilon(\vecq) \simeq 1.1 \ {\rm meV}$ (the \emph{maxon}) and a local minimum at $q \simeq 4 \ \keV$ (the \emph{roton}), asymptoting to $\epsilon(\vecq) \simeq 1.5 \ {\rm meV}$ at large $q$. We see that single-excitation production can only ever yield an energy deposit of at most $\sim 1$ meV, regardless of the DM mass. Eventually, at large $q \gg 1/\lambda$, the dispersion is expected to asymptote to $\epsilon(\vecq) \sim q^2/(2m_{\rm He})$, with the DM effectively scattering off single helium atoms. However, as discussed throughout this section, the dynamic structure factor also includes multi-excitation rates which grow in importance as $q$ grows, until they eventually converge to the quasielastic regime in Sec.~\ref{sec:SHOToy}. Thus, it is also possible to consider the multi-excitation rate in the regime $\omega > 1 \ {\rm meV}$ and $q \ll 1 \ \keV$, which is better kinematically matched to the DM and which does not require as low of a threshold as detecting single acoustic phonons. This was indeed the idea behind the recent renewal of interest in \He\ as a DM detector~\cite{Schutz:2016tid,Knapen:2016cue}.

The dynamic structure factor for \He\ (using conventions from \cite{Knapen:2016cue} which differ slightly from our previous definitions) is defined as
\be
S(\vecq, \omega) = \frac{1}{n_0} \sum_\beta |\langle \Psi_\beta | n_{\vecq} | \Psi_0 \rangle |^2 \delta(\omega - \omega_\beta)
\ee
where $n_0 = N/V$ is the average number density, and the density operator in Fourier space is
\be
n_{\vecq} = \frac{1}{\sqrt{V}}\sum_{i = 1}^{N} e^{i \vecq \cdot \vecr_i}
\ee
where $\vecr_i$ is the position operator for helium atom $i$. As discussed in Sec.~\ref{sec:CMReview}, DM which couples to nucleons through a scalar or vector mediator will couple to the helium density operator. However, for a dark photon mediator,  in the small-$q$ limit the DM will couple to induced dipole moments since the overall helium atom is neutral (see Eq.~(\ref{eq:ODarkPhoton})), exactly as in the case of polar crystals.  Then the correct response function will also include the polarizability of \He, which is small. This generally results in a much less competitive reach for dark photon mediators~\cite{Knapen:2016cue}.

For superfluid \He\, there are various strategies for determining the dynamic structure factor and DM scattering rate:
\begin{enumerate}
\item Directly tackle the microscopic physics to the extent possible, up to some assumptions needed to improve agreement with experimental data. There has been some recent progress in determining the dynamic structure factor including multiphonon contributions in this way, based on a many-body theory for strongly-correlated systems (e.g.~\cite{Krotscheck2015} and references therein). This calculation roughly agrees with the measured dynamic structure factor where data is available, but extends it to the regimes needed for DM scattering. This was applied to calculate the excitation rate from DM in Ref.~\cite{Knapen:2016cue}.
\item Use measurements of the structure factor to directly determine the rate to produce single excitations. The single-excitation rate shows up as a resonance in the experimentally measured dynamic structure factor, which can then be applied directly for the DM rate~\cite{Baym:2020uos}. Using sum-rule arguments, it is possible to estimate the maximum size of the multi-excitation rate and obtain measurements of the multiphonon excitation in some regimes, but it is more challenging to accurately obtain the multi-excitation rate in this approach.
\item Write down an effective field theory (EFT) based on the spontaneous breaking of particle number symmetry, interpreting the acoustic phonon as a Goldstone mode and using measurements to fix the unknown coefficients in the effective action \cite{Acanfora:2019con,Caputo:2019cyg,Caputo:2019xum,Caputo:2020sys}. This approach can be used for single and multi-excitation rates, but is limited to treating only the acoustic phonons in the excitation spectrum.
\end{enumerate}
In particular, the data-driven approach of the second strategy has a close analogy in the use of the ELF to determine the response to DM-electron scattering, as we will discuss in Sec.~\ref{sec:dielectric} below.

While the details of these approaches are beyond the scope of this review, we introduce some of the main ideas to illustrate the behavior of the dynamic structure factor. One perturbation theory approach starts from the original Bijl-Feyman theory for excitations in superfluid He \cite{bijl1940lowest,feynman1954atomic}. The fact that the system has a continuous translation invariance allows the possibility of a general analysis based on symmetry considerations alone, without requiring knowledge of the microscopic Hamiltonian. The Bijl-Feynman theory postulates that the approximate wavefunction for single excitations of momentum $\vecq$ may be constructed from the exact many-body ground state as
\be
|\vecq \rangle = \frac{1}{n_0 S(\vecq)} n_{\vecq} | \Psi_0 \rangle.
\label{eq:BFq}
\ee
One can show that $|\vecq \rangle$ is orthogonal to the ground state, and in fact approaches an exact eigenstate of the many-body Hamiltonian as $\vecq \to 0$, so this theory accurately describes the acoustic phonon part of the dispersion curve but begins to fail for $q \gtrsim 1 \ \keV$. The normalization of the state is enforced by the \emph{static structure factor}
\be
S(\vecq) = \int d\omega \, S(\vecq, \omega) = \frac{1}{n_0} \langle \Psi_0 |n_{-\vecq} n_{\vecq} | \Psi_0 \rangle.
\ee
Experimentally, $S(\vecq)$ has been measured to be linear as $q \to 0$, with $S(\vecq) = q/(2 m_{\rm He} c_{s})$, and approaches  $S(\vecq) \to 1$ in the high $q$ limit. The leading order energy of the state $|\vecq \rangle$ is given by
\be
\epsilon_0(\vecq) = \langle \vecq | \delta H | \vecq \rangle = \frac{\vecq^2}{2m_{\rm He} S(\vecq)},
\label{eq:BijlFeynman}
\ee
where $\delta H = H - E_0$ is the exact many-body Hamiltonian with the ground state energy $E_0$ subtracted; see Ref.~\cite{Knapen:2016cue} for more details. As can be seen from Fig.~\ref{fig:Helium}, this Bijl-Feynman energy approaches the acoustic phonon dispersion as $q \to 0$, while at high $q$ it approaches the free particle dispersion $q^{2}/(2 m_{\rm He})$. However, it clearly does not reproduce the correct phonon and roton dispersion, and it is necessary to compute the full energy $\epsilon(\vecq)$ of the state accounting for quasiparticle interactions, such as the three-excitation vertex. These interactions also lead to multi-excitation states in the dynamic structure factor.

In general, the dynamic structure factor contains a single-quasiparticle peak and a continuum of multi-excitation states,
\be
S(\vecq, \omega) = Z(\vecq) \delta(\omega - \epsilon(\vecq)) + S_m(\vecq, \omega)
\ee
where $Z(\vecq) \approx S(\vecq)$ in the low $q$ region, where Eq.~(\ref{eq:BFq}) is a good approximation to the true single-particle eigenstates. In this region, $S_m(\vecq, \omega)$ will be subdominant and the single-phonon structure factor is given by \be
	S(\vecq, \omega) \approx \frac{q^{2}}{2m_{\rm He} c_{s}q}  \delta(\omega - \epsilon(\vecq)) \approx  \frac{q^{2}}{2m_{\rm He} \epsilon(q)}  \delta(\omega - \epsilon(\vecq))
\ee
which is precisely the same form as the single acoustic phonon structure factor computed in Eq.~\ref{eq:Sqw_LA}, up to the factor of $2\pi/\Omega$ corresponding to different normalization conventions. 

To obtain $S_m(\vecq, \omega)$ by perturbation theory, one must next construct a set of basis states from repeated applications of the density operator, $|\veck, \vecq \rangle \sim n_{\veck} n_{\vecq} |\Psi_0 \rangle$; the difference from the previous treatment of phonons is that the states are not automatically orthogonal, and one must orthogonalize the states by subtracting off overlaps with states of different phonon number. Unfortunately, to compute the overlap term relies on knowing the strongly correlated ground state, but a simple approach is to use the ``convolution approximation'' to postulate its form. With these ingredients, it is then possible to compute the  three-excitation vertex $\langle \vecq - \veck, \veck | \delta H | \vecq \rangle $. One can compute the renormalized energies $\epsilon(\vecq)$, as well as write the leading contribution to $S_m(\vecq, \omega)$ in terms of two-excitation states:
\be
S^{(2)}_m(\vecq, \omega) = \frac{S(\vecq)}{2} V \int \frac{d^3 \veck}{(2\pi)^3} \frac{|\langle \vecq - \veck, \veck | \delta H | \vecq \rangle |^2}{(\epsilon_0(\vecq) - \omega)^2} \delta(\omega - \epsilon_0(\veck) - \epsilon_0(\vecq - \veck)).
\label{eq:S2He}
\ee
The approach which led to Eq.~(\ref{eq:S2He}) may be refined and improved, but even from the leading order result it is possible to reproduce the rough behavior of $S_m(\vecq, \omega)$. In the small $q$ limit with fixed $\omega$, $S_m(\vecq, \omega)$ will be dominated  by nearly back-to-back excitations with momenta $\veck$ and $\vecq - \veck \approx -\veck$, with $|\veck| \gg |\vecq|$ \cite{Schutz:2016tid}. In this limit the 2-excitation structure factor scales as
\be
S_m^{(2)}(\vecq, \omega) \propto q^4
\label{eq:S2KZ}
\ee
at fixed $\omega$, which is physically reasonable and in agreement with the arguments of Sec.~\ref{sec:Sqw_phonon} as well as in this section.  The expected sensitivity of a \He\ experiment sensitive to energy deposits $\omega \in [1.2, 8.6] \ {\rm meV}$ is shown in Fig.~\ref{fig:DMNLimits}, based on theoretical calculations. 

Another option to determine DM rates might be to simply use experimental data on $S(\vecq, \omega)$ in the general rate formula Eq.~(\ref{eq:Rfactorized}). The dynamic structure factor may be directly measured with neutron scattering because neutrons also couple to nucleon density: by arguments very similar to the derivation of Eq.~(\ref{eq:Rfactorized}), the double differential neutron scattering cross section is
\be
\frac{d^2 \sigma}{d\Omega d\omega} = b_n^2 \frac{p_f}{p_i} S(\vecq, \omega)
\ee
where $p_i$ ($p_f$) is the initial (final) neutron momentum and $b_n$ is the neutron scattering length from an individual helium nucleus. Unfortunately, experimental data on $S_{m}(\vecq, \omega)$, for $q$ below $\sim$ keV is quite limited because of the small rates, which is precisely the region needed for sub-MeV DM. The peculiar kinematics of DM make it particularly difficult to probe certain kinematic regions with standard techniques, and one must therefore use theoretical approaches validated at higher $q$ in these cases. Indeed, as we emphasize throughout this review, the kinematic regime relevant for sub-GeV DM has not previously been of interest to the condensed matter community, and even theoretical data may not exist at the required $q$.

\begin{figure}
\includegraphics[width=0.5\textwidth]{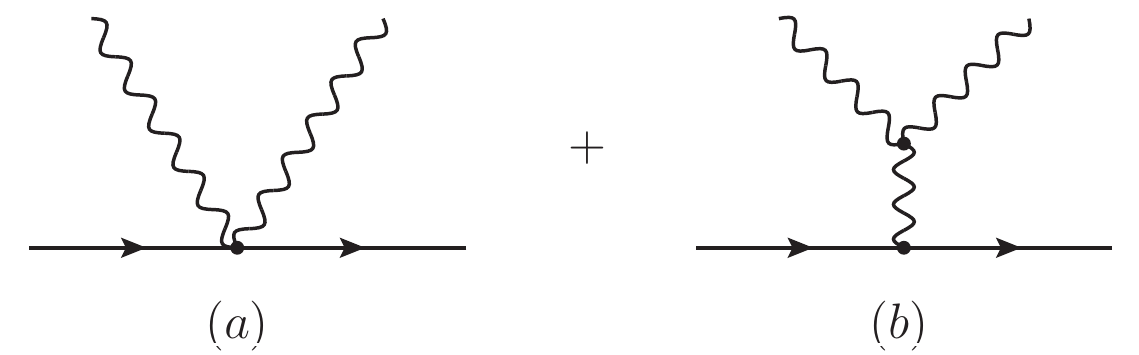}
\caption{Feynman diagrams contributing to two-phonon production in the \He\ EFT. Reproduced from Ref.~\cite{Caputo:2019cyg}.}
\label{fig:2phononHe}
\end{figure}

One can obtain qualitatively similar results for the dynamic structure factor from an EFT perspective. The presence of the acoustic phonon mode with approximately linear dispersion can be understood as a Goldstone mode for the spontaneous breaking of both particle number and time translations, via a real field $\psi$ whose vacuum expectation value (vev) represents a time-dependent chemical potential, $\langle \psi \rangle = \mu t$, where $\mu$ is the relativistic chemical potential which includes the rest mass $m_{\rm He}$ in addition to the non-relativistic piece. Parameterizing the fluctuations about the expectation value in terms of a real field $\pi$ as
\be
\psi(\vecr, t) = \mu t + c_s \sqrt{\frac{\mu}{n_0}}\pi(\vecr, t),
\ee
one can write down an effective action for $\pi$ whose coefficients are fixed entirely in terms of the superfluid equation of state, $P = P(\mu)$ or $c_s = c_s(P)$. This effective action has a cutoff at the momentum scale $\Lambda \simeq 1 \ \keV$ where the dispersion $\epsilon(\vecq)$ deviates from linear, and thus it does not describe the maxon or roton modes, and moreover it neglects the fact that the dispersion of the phonon mode is not exactly linear but has a small positive curvature (\emph{anomalous dispersion}, to be discussed further below). Since in the large-volume limit the system is translationally invariant, one can write down a relativistic EFT invariant under Poincar\'{e} transformations and the internal U(1) particle number symmetry. If we model the DM field as a complex scalar $\chi$ with a dark U(1) symmetry to ensure its stability (a discrete $Z_2$ symmetry would also accomplish the same thing), the most general effective action which generates a spin-independent interaction between DM and \He, at lowest order in momenta, is
\be
S_{\rm eff} = \int d^4 x \, \left \{ P(X) + Z(X) |\partial_\mu \chi|^2 - m^2(X) |\chi|^2 \right \}, \qquad X = \sqrt{\partial_\mu \psi \partial^\mu \psi}.
\ee

By computing the stress tensor for $\psi$, one can verify that $P(X)$ does indeed represent the pressure of the superfluid. Expanding the Lagrangian to cubic order in $\pi$, canonically normalizing $\pi$ and $\chi$, and taking the non-relativistic limit fixes all of the coefficients of the Lagrangian in terms of derivatives of $P$, and can thus be matched to data. The result contains terms of the form $\dot{\pi} (\nabla \pi)^2$ and $\dot{\pi}^3$, which represent 3-phonon vertices, as well as $|\chi|^2 \dot{\pi}$ and $|\chi|^2 (\partial_\mu \pi)^2$ terms, which represent DM-phonon 3- and 4-point interactions, respectively. In the EFT language, the 2-phonon creation is described by two Feynman diagrams shown in Fig.~\ref{fig:2phononHe}. In the limit $\vecq \to 0$, the intermediate phonon in the second diagram is highly off-shell, and integrating it out leads to a seemingly fine-tuned cancellation between the two amplitudes that results in the 2-phonon matrix element scaling as
\be
|\mathcal{M}_{\rm 2-phonon}|^2 \propto \frac{q^4}{\omega^4},
\ee
which is consistent with the result (\ref{eq:S2KZ}) from the dynamic structure factor extrapolation.\footnote{Of course, fine-tuned cancellations in low-energy Lagrangians are often manifestations of symmetries, and in this case the cancellation may be understood as a consequence of the Ward identity \cite{Caputo:2019ywq,Baym:2020uos}. The cancellation breaks if DM does not couple linearly to density, but the non-relativistic limit of all spin-independent scalar or vector exchange yields a coupling to density. While there may be unusual non-perturbative models for DM-SM interactions which do not couple linearly to density, all of the perturbative UV completions of spin-independent scattering yield this cancellation, which is independent of the details of the \He\ EFT.}
Integrating the 2-phonon rate yields a sensitivity curve in Ref.~\cite{Caputo:2019cyg} which is within a factor of 2 to the one from Ref.~\cite{Knapen:2016cue} that includes rotons and maxons, see Fig.~\ref{fig:DMNLimits}.

While meV energy thresholds are already well beyond current capabilities (though they are a subject of active investigation, see Sec.~\ref{sec:Heat}), one can also explore the idea of lower energy thresholds which opens up the possibility of single-excitation production \cite{Baym:2020uos}. Indeed, multi-phonon production is inefficient at low $q$ and for sub-MeV dark matter, in the sense that it does not fully exploit the fact that the normalization of the dynamic structure factor is fixed by conservation of mass, and thus the DM scattering rate can be maximized by ensuring that the the integral in Eq.~(\ref{eq:Rgeneral}) contains the region where $S(\vecq, \omega)$ has largest support.\footnote{We will see an exactly analogous situation with DM-electron scattering in semiconductors in Sec.~\ref{sec:dielectric} below.}. More precisely, the ``$f$-sum rule'' is given by
\be
\int_0^\infty \, d\omega \, \omega \, S(\vecq, \omega) = \frac{q^2}{2m_{\rm He}},
\ee
which is valid for any momentum $\vecq$ and derives from the analytic properties of $S$ as a causal density-density correlation function. While directly measuring $S_{m}(\vecq, \omega)$ at low $q$ is difficult with neutron probes, the single-excitation rate can be directly computed using experimentally measured quantities, with the sum rule used to constrain the remainder of the spectral weight. At low $q$, the wavefunction renormalization can be expanded as
\be
Z(\vecq) = \frac{q^2}{2m_{\rm He} \epsilon(\vecq)}\left(1 - 1.63 \frac{q^2}{(m_{\rm He}c_s)^2} + \dots\right),
\ee
where the coefficient of the $q^4$ term is fit to neutron scattering data. The sum rule implies that 
\be
\frac{\int_0^\infty d\omega \, \omega \, S_m(\vecq,\omega)}{\int_0^\infty d\omega \, \omega S(\vecq,\omega)} = 1.63 \frac{q^2}{(m_{\rm He}c_s)^2} \simeq 0.1 \left(\frac{q}{0.7 \ \keV}\right)^2.
\ee
In other words, over 90\% of the spectral weight in scattering of sub-MeV DM lies in the single-phonon part of the dynamic structure factor at $q = 0.7$ keV.

An interesting property of superfluid \He\ is that the low-energy dispersion is not exactly linear, but is slightly convex,
\be
\omega(q) \simeq c_s q(1 + \zeta_A q^2 + \cdots)
\ee
with $\zeta_A > 0$.\footnote{This effect may be captured in the EFT through the inclusion of higher-derivative operators.}
 At standard volume and pressure, the dispersion curve has an inflection point at $q \simeq 0.4 \ \keV$, such that phonons with $q < q_c \simeq 0.8 \ \keV$ ($\omega(q_c) = 0.68 \ {\rm meV}$) are unstable against decay into two lower-energy phonons. This regime is less important for the multi-excitation phonons, which have total energy well above 1 meV. For sub-meV single phonons, this will lead to a ``phonon cascade'' phonons reminiscent of jets or electromagnetic showers. This makes detecting individual sub-meV excitations effectively impossible, as the mean free path for splitting is on the order of cm or smaller, but if the total energy in the cascade cone could be detected, anomalous dispersion offers the intriguing possibility of localizing the event within the sample volume based on the ellipticity of the cone projected onto the surface of the detector.

\section{Dark Matter-Electron Scattering}
\label{sec:DMe}

The phenomenology of DM-electron scattering is dominated by the fact that electrons are bound in atomic, molecular, and solid-state systems, with wavefunctions that are very far from momentum-eigenstate plane waves. The typical length or momentum scale for electronic wavefunctions is set by the Bohr radius:
\be
a_0 = \frac{1}{\alpha m_e} = 5.29 \times 10^{-11} \ {\rm m}, \qquad p_0 \equiv \frac{1}{a_0} = 3.73 \ \keV.
\ee
In the ground state of the hydrogen atom, $\langle r \rangle = \frac{3}{2}a_0$; in an larger atom, the larger value of the principal quantum number $n$ is partially compensated by the increased screened nuclear charge, giving a parametrically similar answer. In a molecule, the interatomic distance is set by minimizing the total energy of covalently-bonded atoms, and since the atomic wavefunctions must overlap to bond, the bond length is also of order $a_0$; for example, the carbon-carbon bond length in organic molecules is $0.14 \ {\rm nm} \simeq 2 a_0$. The same logic holds for solid-state lattices (silicon has a minimum interatomic distance of $\sim 4.4 a_0$ and a lattice constant of $\sim 10 a_0$), and even the Fermi momentum $k_F$ for delocalized electrons in a metal is of order $p_0$, since it depends on the number density of electrons and hence is set in part by the lattice spacing. The ground state position-space orbitals are localized as $\exp(-r/a_0)$, yielding momentum-space orbitals which fall off at large $p$ as a power law. The energies of electronic states are parametrically set by the Rydberg energy: 13.6 eV for the ionization energy of hydrogen, $\mathcal{O}(10) \ \eV$ for outer-shell binding energies in noble atoms, $\mathcal{O}(5)$ eV for excitation gaps in organic molecules, and $\mathcal{O}(1-5) \ \eV$ for semiconductor gaps. As mentioned in Sec.~\ref{sec:Intro}, these binding energies decrease from isolated atoms to molecules to solid-state systems.

However, electron interactions can give qualitatively different behavior in specific condensed-matter systems. A key example is a superconductor, where phonons mediate a weak attractive force between electrons, binding them into Cooper pairs with a gap of $\mathcal{O}({\rm meV})$. There are also numerous examples of narrow-gap materials with sub-eV excitation energies: ZrTe$_5$ (an example of a Dirac material) with a 30 meV gap, SmB$_6$ (a Kondo insulator) with a 10 meV gap, and InSb (an otherwise ordinary semiconductor) with an accidentally small gap of 200 meV. In addition, the plasmon mode, a collective oscillation of all the valence electrons in a solid, has typical energies of 15 eV but exists solely at momenta \emph{below} the scale $p_0$. 

\begin{figure*}[t!]
\begin{center}
\includegraphics[width=0.55\textwidth]{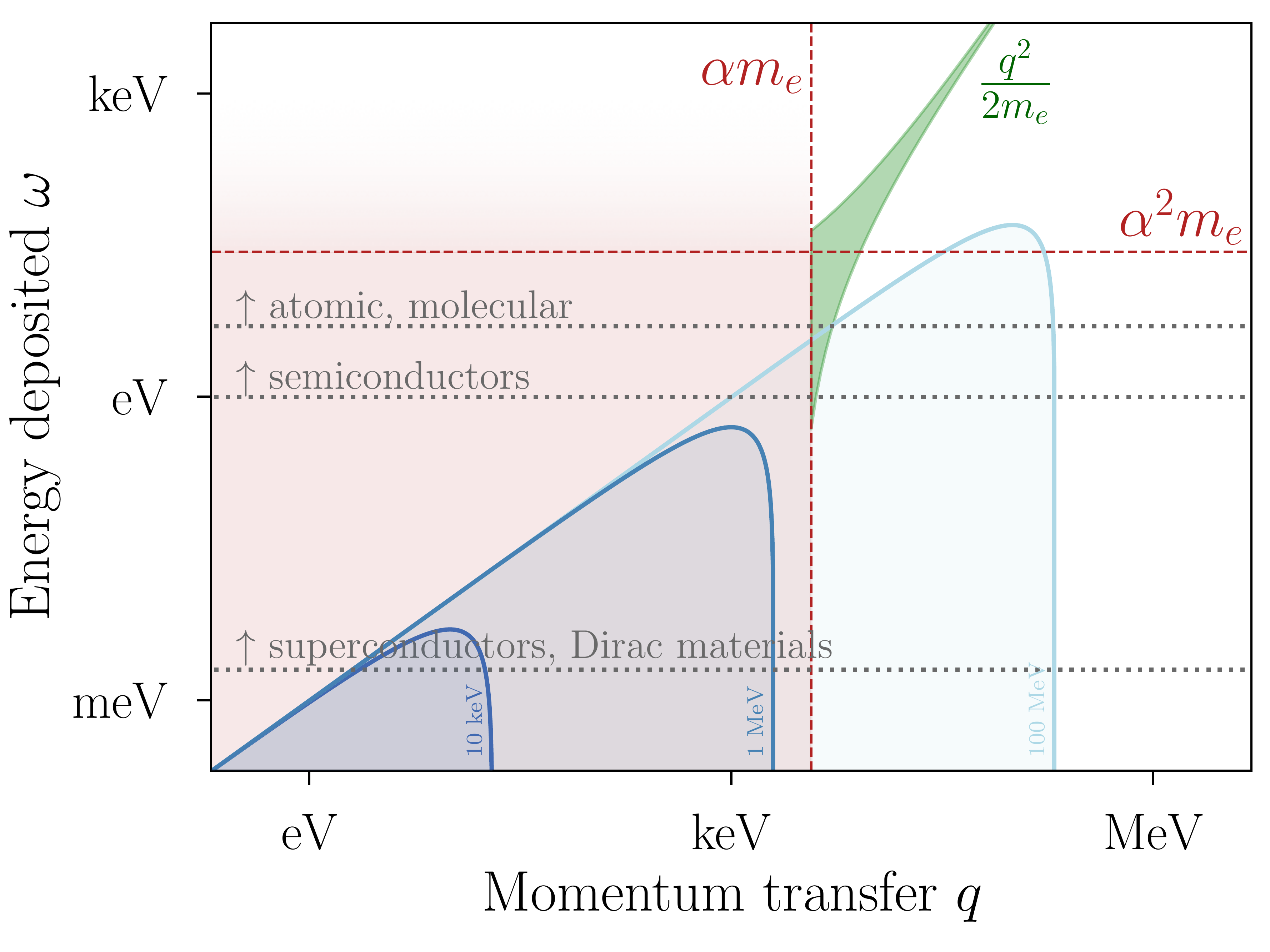}
\caption{ \label{fig:DMe_qw}
For DM-electron interactions, the response at high $q$ is peaked about the free-electron dispersion $\omega = q^{2}/(2 m_{e})$. The dashed lines at $q \sim p_0 = \alpha m_{e}$ and $\omega \sim \alpha p_0 = \alpha^2 m_e$ indicate typical scales for the wavefunction spread and energies of bound electrons. Many-body effects are expected to be particularly important at lower $q, \omega$, indicated by the shaded region. Depending on the target material and detection method, the relevant response function will be cut off at low $\omega$;  we show typical gaps for ionization in atomic systems, scintillation in molecules, electron-hole excitations in semiconductors, and gapped excitations in superconductors and Dirac materials. Kinematically allowed regions for DM scattering are shown for $m_{\chi} = 10$ keV, 1 MeV, and 100 MeV at $v= 10^{-3}$, as in Fig.~\ref{fig:parabola}.
}
\end{center}
\end{figure*}

Comparing these scales to the typical energy and momentum scales of DM, we can refer to Fig.~\ref{fig:parabola} and Eq.~(\ref{eq:omegaqdef}) to see that the momentum transfer required to create an excitation $\omega$ must exceed
\be
q_{\rm min} = \frac{\omega}{v_{\rm max}},
\ee
where $v_{\rm max}$ is the maximum DM velocity, and we have taken the large $m_\chi$ limit (for smaller $m_\chi$, $q_{\rm min}$ is strictly larger). Taking $v_{\max} = v_{\oplus} + v_{\rm esc} \simeq 800 \ {\rm km/s} = 2.67 \times 10^{-3}$, we find
\be
q_{\rm min} \simeq p_0 \left(\frac{\omega}{10 \ \eV}\right)
\label{eq:qminDMe}
\ee
while $q_{\rm max} = 2 m_{\chi} v_{\rm max} \simeq p_{0}  (m_{\chi} / 0.7 \, {\rm MeV} )$.
From this superficially trivial observation, we learn a number of things:
\begin{itemize}
\item By pure (cosmic!) coincidence, MeV--GeV DM has the correct kinematics to access the electronic response for atomic and molecular systems near where they have strong support, as shown schematically in Fig.~\ref{fig:DMe_qw}. However, it can also be seen from Fig.~\ref{fig:DMe_qw} that the DM scattering kinematics is not necessarily ideally matched to the response. For example, for atomic ionization, the rate will be strongly peaked at low ionization energies, since larger $\omega$ requires accessing the high-momentum tail of the electron wavefunctions which is power-law suppressed. Thus, while it is kinematically permitted for (say) 100 MeV DM to deposit \emph{all} of its $\sim 50$ eV of kinetic energy on an atomic electron, it is extremely unlikely to do so. Similarly, this will favor scattering on the high-velocity tail of the DM velocity distribution, which implies a large increase in rate as the gap is lowered \cite{Graham:2012su}.
\item For conventional semiconductors with $\mathcal{O}(\eV)$ gaps, DM \emph{necessarily} probes distance scales smaller than the lattice constant. Taking silicon as an example, with  $\omega = 1 \ \eV$, we find from Eq.~(\ref{eq:qminDMe}) $q_{\rm min} \simeq p_0/10 \simeq (10 a_0)^{-1}$, which is the inverse lattice spacing. Thus, while it is true that the valence electrons are delocalized across the lattice, the particular kinematics of DM scattering weights the the quasi-localized portion of the electronic wavefunctions. Collective modes or effects are therefore less important for conventional semiconductors than for lower-gap materials.
\item The peak of the plasmon mode, with $q \ll p_0$ and $\omega \simeq 15 \ \eV$, is kinematically inaccessible to halo DM. As we will see in Sec.~\ref{sec:dielectric} below, the plasmon is responsible for the bulk of the support of the loss function at low momentum, while the loss function at high momentum is spread out over a larger $\omega$ range. In this sense, conventional materials are ``inefficient'' with respect to DM scattering. This is analogous to the case of multi-phonon excitations in superfluid helium discussed in Sec.~\ref{sec:helium} above, rather than the resonant single-phonon excitations discussed in Sec.~\ref{sec:acousticoptical}. 
\item Accessing the true long-range behavior of delocalized electrons \emph{requires} a narrow-gap material, most of which have rather exotic electronic properties. Since these narrow gaps are mandatory to probe sub-MeV DM which carries sub-eV kinetic energies, the search for novel materials with the required electronic properties, involving close collaboration with condensed matter physicists, is a key component of the active research in light DM detection.
\end{itemize}

The peculiar kinematics of DM also mean that the regime of the electronic response function accessible to DM is quite under-explored experimentally. For example, the plasmon has been extremely well-studied at energies near the peak, but much less attention has been given to the low-energy tail which is the only part accessible to DM in conventional materials. For heavier DM, the high momentum regime dominates, for which there is limited experimental data. Similarly, atomic wavefunctions are often studied using photoabsorption, where $\omega = q$, but DM probes the kinematically-distinct regime $\omega \simeq 10^{-3} q$. At present, almost all of the predictions for DM-electron scattering rates rely on primarily theoretical determinations of the response, using the tools described in Sec.~\ref{sec:CMexcitations}, with limited direct experimental input for calibration. That said, the arguments of Sec.~\ref{sec:DMstructurefactor} show that, at least for the dark photon model, the relevant dynamic structure factor (namely, the ELF) may be \emph{directly} measured with electromagnetic probes such as electron energy-loss spectroscopy (EELS) or X-ray scattering, which can be configured to access the correct kinematic regime. Dedicated measurements in the coming years, on novel and conventional materials alike, will greatly reduce the systematic uncertainty associated with DM-electron scattering.

\begin{figure*}[t!]
\begin{center}
\includegraphics[width=0.48\textwidth]{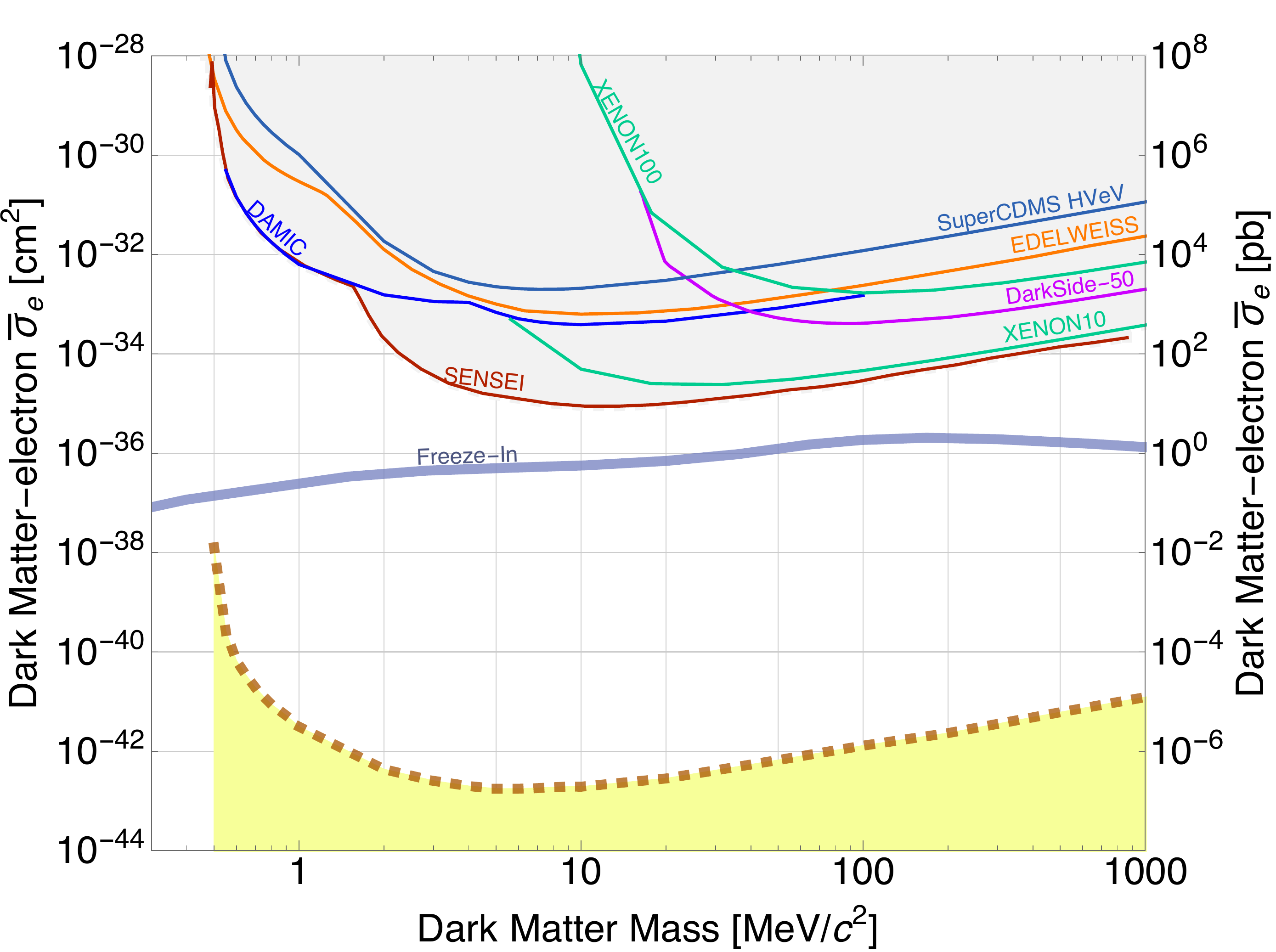}
\includegraphics[width=0.48\textwidth]{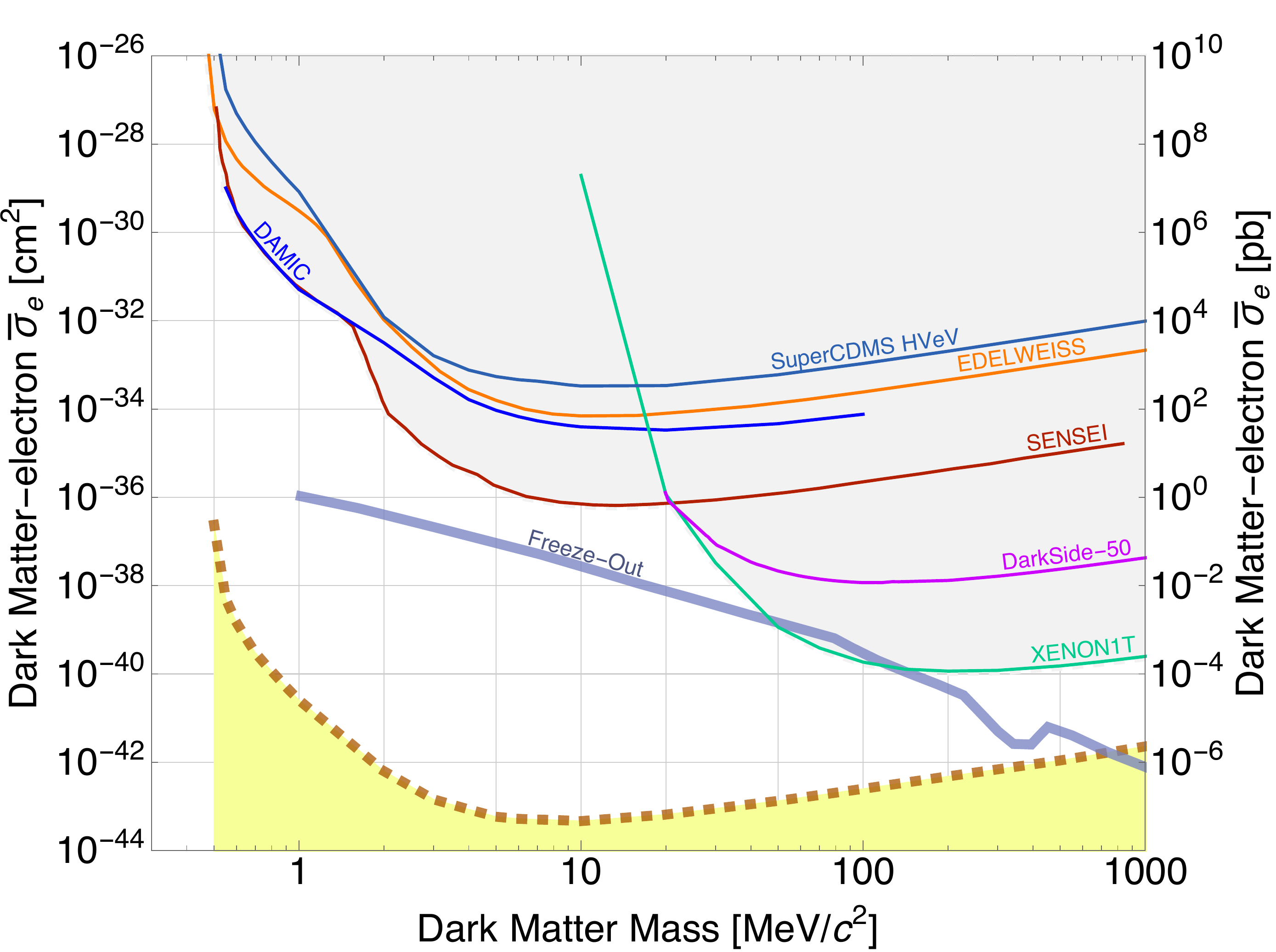}
\caption{ \label{fig:DMeLimits}
Leading experimental limits on DM-electron cross section $\overline{\sigma}_e$ for scattering through a light mediator (left) and heavy mediator (right). Plots reproduced from Dark Matter Limit Plotter, \texttt{https://supercdms.slac.stanford.edu/dark-matter-limit-plotter}. Note that no direct detection limits exist below $m_\chi = 500 \ \keV$.}
\end{center}
\end{figure*}

\begin{figure}
\includegraphics[width=0.95\textwidth]{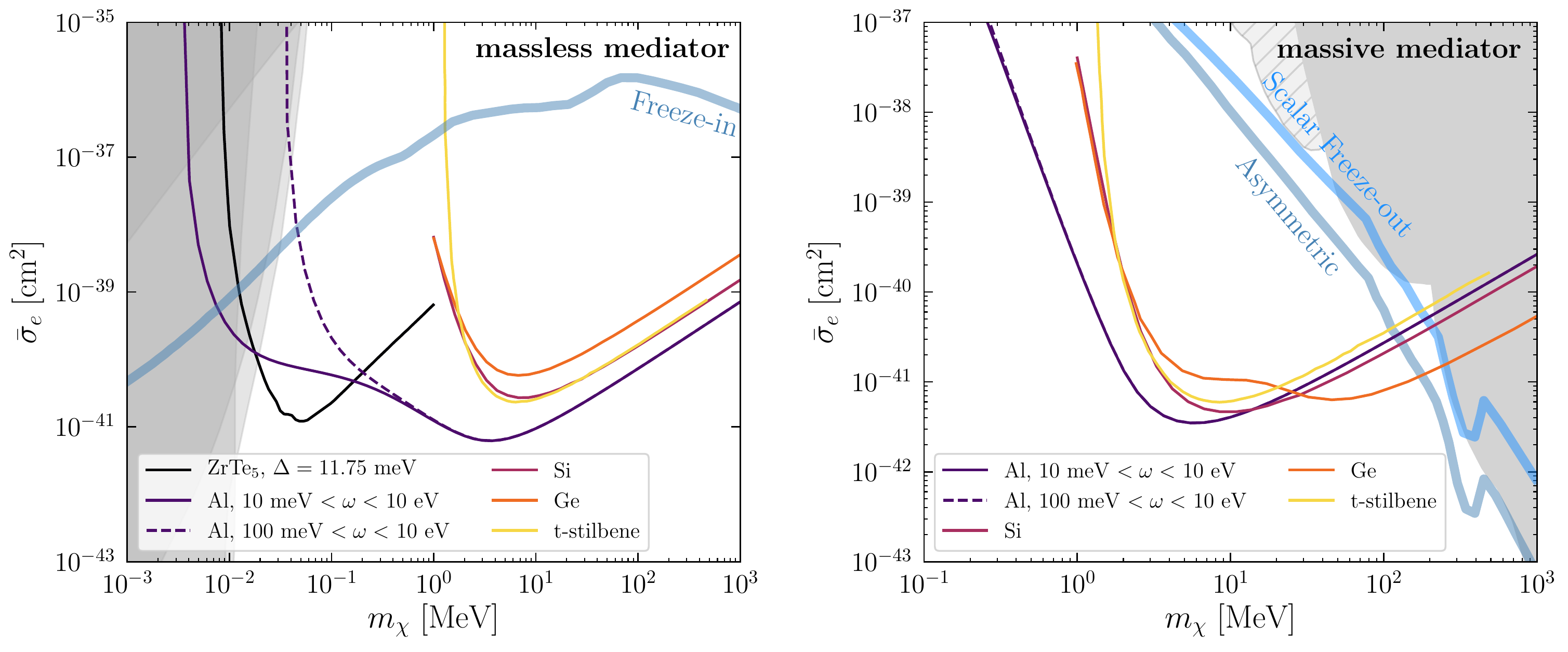}
\caption{Reach to DM-electron cross section in different solid state targets for scattering through a heavy mediator (left) and light mediator (right), assuming kg-year exposure and zero background. t-stilbene gives an example of a molecular crystal target; the reach shown is from Ref.~\cite{Blanco:2021hlm}.  Si and Ge have O(eV) gaps and have been studied extensively as target materials. The reach for Ge with a massless mediator and Si in both cases is from Ref.~\cite{Knapen:2021run}. For Ge and a massive mediator, the reach shown is from Ref.~\cite{Griffin:2021znd}; for DM masses above $\sim20$ MeV, the reach is dominated by the excitation of semi-core electrons, see Fig.~\ref{fig:all_electron}.  For sub-MeV DM, lower gap materials are needed and we show projections for an example Dirac material from Ref.~\cite{Coskuner:2019odd} and for Al from Refs.~\cite{Hochberg:2021pkt,Knapen:2021run}. In the left plot, the thick blue line is the predicted cross section if all of the relic DM is produced by freeze-in interactions~\cite{Essig:2011nj,Dvorkin:2019zdi} and the shaded regions are constraints from stellar emission~\cite{Vogel:2013raa,Chang:2018rso}. In the right plot, the thick blue lines are cross sections for freeze-out of scalar DM or asymmetric DM, and the shaded region shows combined direct detection bounds (solid grey) and model-dependent accelerator bounds  when the dark photon mass is $m_{A'} = 3 m_{\chi}$ (hatched grey)~\cite{Kahn:2020fef}. All bounds and relic density lines assume a dark photon mediator.}
\label{fig:DMelectron_all}
\end{figure}

To give a sense of the available parameter space for DM-electron scattering, Fig.~\ref{fig:DMeLimits} shows the current leading limits from several experiments, each of which will be discussed more below and in Sec.~\ref{sec:Detection}.\footnote{These limits are derived from published results, but many of them are based on theoretical rate predictions which neglect important many-body effects, as pointed out in \cite{Pandey:2018esq,Hochberg:2021pkt,Knapen:2021run,Knapen:2021bwg} and to be discussed further below.}  The two models correspond to spin-independent scattering through a heavy mediator (left panel) or light mediator (right panel). Constraints are shown alongside the relic density targets from the freeze-out and freeze-in mechanisms and the ``neutrino floor'' (see Sec.~\ref{sec:neutrino}) for a silicon target with a 100 kg-yr exposure and a 1-electron threshold. Fig.~\ref{fig:DMelectron_all} shows the background-free projected limits from a number of theory proposals, many of which would allow access to the remaining cosmologically-allowed sub-GeV parameter space. 

\subsection{Atomic ionization}
\label{sec:atomic}

\subsubsection{Hydrogen toy model}
The first application of dark matter-electron scattering involved electron ionization in liquid xenon \cite{Essig:2011nj,Essig:2012yx}, but to illustrate the essential features we will start first with a simpler toy example, dark matter scattering off a single hydrogen atom \cite{Chen:2015pha,Lin:2019uvt}. Throughout this section, we will take the benchmark dark photon model, where DM couples to both protons and electrons with a spin-independent potential. Summing over all target atoms (or nuclei), the dynamic structure factor is
\be
S(\vecq, \omega) =  \frac{2\pi N_{\rm nuc}}{V} \sum_f | \langle f | e^{i \vecq \cdot \vecr_N} - e^{i \vecq \cdot \vecr_e} | 0 \rangle |^2 \delta (E_f - E_0 - \omega)
\ee
where $\vecr_N$ and $\vecr_e$ are the nuclear and electronic coordinates, respectively, and $|f \rangle$ represents an excited electronic state of the atom. In the approximation where the nucleus is infinitely heavy and thus stationary, we may set $\vecr_N = 0$. In that limit, nuclear scattering does not contribute to electronic excitation when $|f \rangle \neq |0 \rangle$; we will see in Sec.~\ref{sec:Migdal} below how this picture is modified when the finite mass of the nucleus is taken into account.

The excited electron state $|f \rangle$ may be either a bound state or a continuum state, the latter of which corresponds to an ionized electron. Since the formalism of atomic electron scattering is typically applied to liquid noble element detectors which are sensitive to ionized electrons, we will focus on the continuum states and return to the discrete spectrum of bound states in Sec.~\ref{sec:scintillation} below. The initial state is simply the ground state of the hydrogen atom, $\psi_{100}(\vecr) = 2 a_0^{-3/2} e^{-r/a_0}$. Excited states are labeled by a wavevector $\veck$, so the sum over $|f \rangle$ turns into an integral, and the structure factor becomes 
\be
S(\vecq, \omega) =  \frac{2\pi N_{\rm nuc}}{V}  \int \frac{d^3 \veck}{(2\pi)^3} \delta(E_\veck - E_0 - \omega) |f_{0 \to \veck}(\vecq)|^2
\ee
where 
\be
f_{0 \to \veck}(\vecq) =  \int d^3  \vecr \ \psi^*_{\veck}(\vecr) \psi_{100}(\vecr) e^{i \vecq \cdot \vecr}
\label{eq:TFFHydrogen}
\ee
is the \emph{atomic form factor} for transitions between the ground state and the continuum state $\veck$. Asymptotically far away from the nucleus, the final-state electron behaves as a free particle, so we have absorbed an extra factor of $\sqrt{V}$ inside  $\psi^{*}_{\veck}(\vecr)$, such that it behaves as $\psi^{*}_{\veck}(\vecr) \propto e^{i \veck \cdot \vecr}$ as $r \to \infty$. Similarly, we may define $\veck$ through the energy of the continuum state as $E_\veck = \frac{k^2}{2m_e}$. Typically we are interested in the energy spectrum of ionized electrons $dR/dE_{\rm er}$, so using $dE_{\rm er} = dE_\veck = k dk/m_e$, we may trade the integral over $k$ for an integral over $E_{\rm er}$. Collecting the various normalization factors, and decomposing the outgoing wavefunction into spherical waves with angular quantum numbers $l'$ and $m'$, it is convenient to define an \emph{ionization form factor},
\be
|f_{\rm ion} (k, q)|^2 = \sum_{l', m'} \frac{2 k^3}{(2\pi)^3} |f_{0 \to k, l', m'}(\vecq)|^2
\ee
where the factor of 2 accounts for spin degeneracy. The radial part $\widetilde{R}_{kl}$ of the ionized wavefunctions is normalized as
\be
\int dr \, r^2 \, \widetilde{R}^*_{kl}(r)\, \widetilde{R}_{k'l'}(r) = (2\pi)^3 \frac{1}{k^2}\delta_{ll'} \delta(k-k')\,,
\ee
so that $\widetilde{R}_{kl}(r)$ itself is dimensionless, and therefore so is $f_{\rm ion}$. Using Eq.~\ref{eq:Riso} since the target system is spherically symmetric, we obtain the experimental quantity of interest, the total differential rate per unit detector mass, 
\be
\frac{dR}{d \ln E_{\rm er}} = N_{T} \frac{\rho_\chi}{m_\chi} \frac{\bar{\sigma}_e}{8 \mu_{\chi e}^2} \int dq \, q \, |F_{\rm DM}(q)|^2 |f_{\rm ion}(k, q)|^2 \eta(\vmin),
\ee
where for electron scattering
\be
\vmin = \frac{E_{\rm er} + |E_0|}{q} + \frac{q}{2m_\chi}
\ee
and the form factor for a dark photon mediator $F_{\rm DM}(q)$ was defined in Eq.~(\ref{eq:FDMDef}). $N_{T}$ is again the number of target nuclei per unit detector mass, and $E_0 = -13.6 \ \eV$ is the binding energy of hydrogen.

If we approximate the outgoing state $\psi_{\veck}$ as a pure plane wave $e^{-i \veck \cdot \vecr}$, we may use the decomposition of plane waves into spherical wavefunctions to compute the form factor analytically for $l' = m' = 0$ \cite{Lin:2019uvt}: 
\be
 \frac{2 k^3}{(2\pi)^3} |f_{0 \to k, 0,0}(\vecq)|^2  = \frac{64 (k a_0)^3}{\pi [a_0^4(k^2 -q^2)^2 + 2 a_0^2 (k^2 + q^2)+1]^2} \quad {\rm(hydrogen, \ plane-wave \ final \ state)}
\label{eq:fionplanewave}
\ee
This function has the most support at $k \simeq q \simeq 1/a_0$, reflecting the fact that the characteristic momentum scale for electrons in the hydrogen atom is $1/a_0$. Of course, since DM must have at least $|E_0|$ of kinetic energy to ionize hydrogen, ionization is only kinematically allowed for $m_\chi \gtrsim 10 \ \MeV$, for which $q = 1/a_0$ is kinematically allowed as well. This intuition typically holds for more complicated atoms -- the peak of the structure factor is kinematically accessible -- but not necessarily for molecular or solid-state excitation, as we will see below. 

Note also that $f_{\rm ion}(k, q)$ in (\ref{eq:fionplanewave}) does not vanish as $q \to 0$, as required by orthogonality of initial and final states. This unphysical behavior is simply because the plane-wave approximation is sufficient far from the origin, but due to the localization of the initial state, the integrand of the transition form factor (\ref{eq:TFFHydrogen}) is dominated by the region close to the origin. Indeed, the Coulomb potential of the nucleus distorts the continuum wavefunction, which we can account for by choosing the final-state wavefunction to be the exact positive-energy solution of the Coulomb potential. Writing $\psi_{\veck}(\vecr)$ in spherical coordinates as $Y^m_l(\theta, \phi) \widetilde{R}_{kl}(r)$, the radial function is \cite{bethe2013quantum}

\be
\widetilde{R}_{kl}(r) =(2\pi)^{3/2}\frac{\sqrt{\frac{2}{\pi}}\left|\Gamma\left(l+1 + \frac{iZ}{ka_0}\right)\right| e^{\frac{\pi Z}{2ka_0}}}{(2l+1)!}e^{ikr}\,{}_1F_1\left(l+1 + \frac{iZ}{ka_0},2l+2, 2ikr\right),
\label{eq:continuumexact}
 \ee
 where $\Gamma$ is the gamma function, $ {}_1F_1$ is the confluent hypergeometric function,  and we leave the nuclear charge $Z$ arbitrary to facilitate later comparison with larger hydrogenic atoms. Since this is an eigenstate of the same potential which determines the ground state, it is automatically orthogonal to $\psi_{100}$ when $Z = 1$. 
 The ratio of the wavefunction at the origin to the wavefunction at infinity is 
\be
\left | \frac{\widetilde{R}_{kl}(r=0)}{\widetilde{R}_{kl}(r=\infty)}\right|^2 = \frac{2\pi Z}{k a_0}\frac{1}{1-e^{-2\pi Z/(k a_0)}},
\ee
where the right-hand side is also known as the \emph{Fermi factor}. It diverges as $k \to 0$, reflecting the infinite range of the pure Coulomb potential. This increases the form factor at small $k$ compared to the plane-wave approximation \cite{Essig:2011nj,Vergados:2016niz}, an effect known as \emph{Sommerfeld enhancement} in the context of beta decay. 

\begin{figure*}[t!]
\begin{center}
\includegraphics[width=0.55\textwidth]{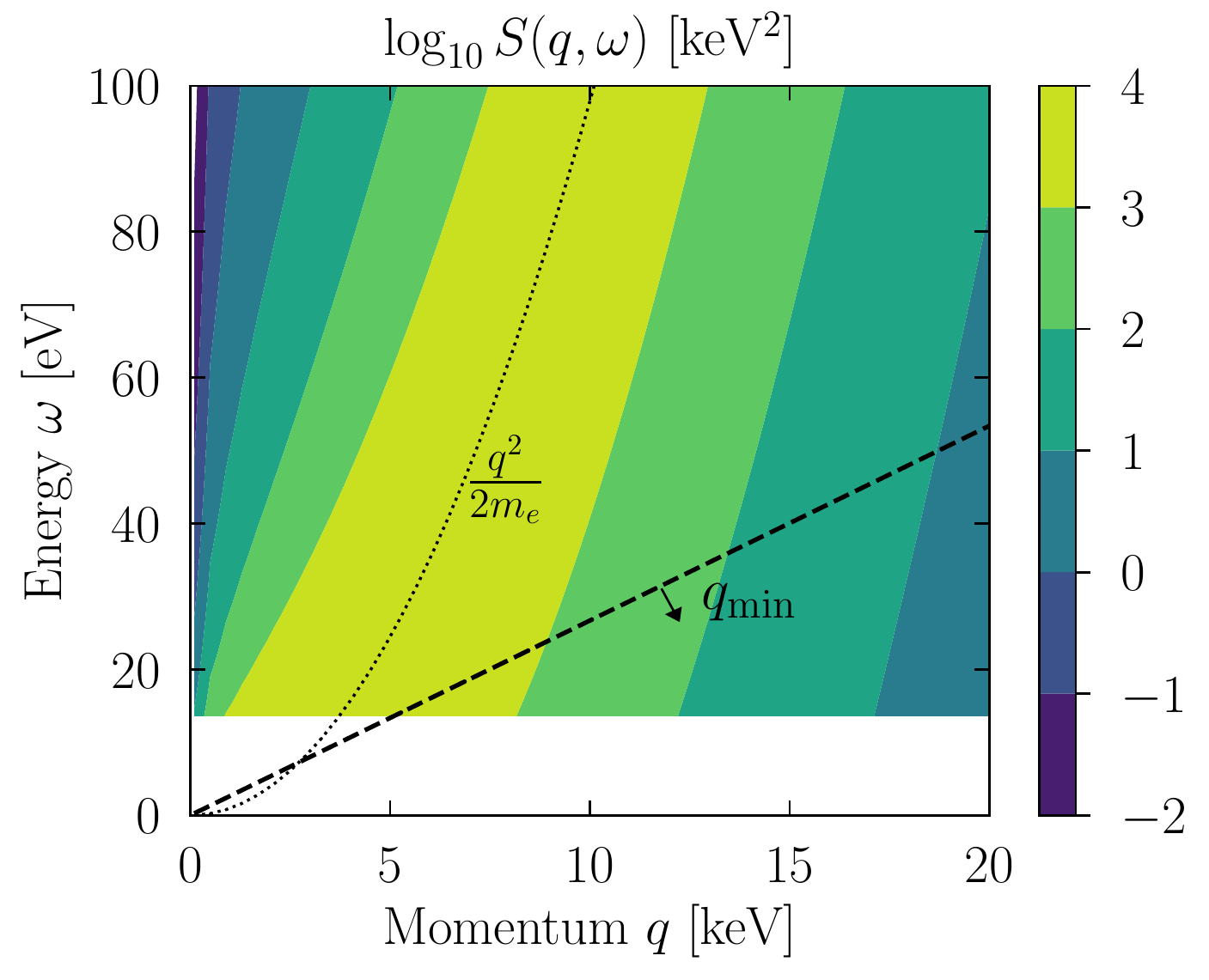}
\caption{ \label{fig:hydrogen_sqw} The dynamic structure factor for ionization of hydrogen, obtained using Eq.~(\ref{eq:fionHydron_exact}) with an arbitrary reference $n_{\rm atom} = 1/(a_{0})^{3}$. For $q \gg p_{0}$, the bound nature of the electron becomes less important and the peak of the structure factor converges to the free-particle dispersion $\omega = q^{2}/(2 m_{e})$, indicated by the dotted line. The dashed line is the minimum $q$ for DM scattering, Eq.~(\ref{eq:qminDMe}), such that the kinematically-accessible region is below and to the right of the dashed line. The region of largest support for the structure factor is inaccessible to DM scattering.}
\end{center}
\end{figure*}

However, using the exact radial wavefunction in Eq.~(\ref{eq:continuumexact}), the ionization form factor can be obtained as \cite{Chen:2015pha}
\be
|f_{\rm ion} (k, q)|^2 = \frac{512 Z^6 k^{2} q^2 a_0^4 ((3 q^2 + k^2)a_0^2 + Z^2) \exp \left [-\frac{2 Z}{k a_0} \tan^{-1}\left(\frac{2 Z k a_0}{(q^2 - k^2)a_0^2 + Z^2}\right)\right]}{3  ((q + k)^2 a_0^2 + Z^2))^3((q-k)^2 a_0^2+ Z^2)^3(1 - e^{-\frac{2 \pi Z}{k a_0}})} \quad {\rm (hydrogen, \ exact)}
\label{eq:fionHydron_exact}
\ee
which vanishes as $q \to 0$ as expected. The dynamic structure factor for ionization of atomic hydrogen is then given by 
\be
	S(\vecq,\omega) = \pi n_{\rm atom} \frac{ m_{e} }{k^{2}} \times |f_{\rm ion}(k, q)|^2, \quad k = \sqrt{2 m_{e} (\omega - |E_{0}|)}.
\ee 
We plot this in Fig.~\ref{fig:hydrogen_sqw}, as compared with the kinematic restriction for DM scattering, which requires us to be to the right of the dashed line. As discussed in the introduction to this section, this implies that we obtain the largest rates for DM on the tail of the velocity distribution and favoring low $\omega$. Furthermore, it is clear that atomic hydrogen is not an ideal direct detection target given that the region where the dynamic structure factor is largest is not entirely accessible to DM.

\subsubsection{Noble atoms}
For DM scattering in many-electron atoms, we must contend with the fact that there are no longer exact solutions for the wavefunctions, and electron interaction and exchange effects can generically no longer be ignored. One approach is to use the Hartree-Fock approximation to construct the approximate many-electron bound-state wavefunctions as a Slater determinant of single-particle orbitals $\psi_i$, $i = 1, \dots, N_e$:
\be
\Psi(\vecr_1, \vecr_2, \dots, \vecr_{N_e}) = |\psi_1 \cdots \psi_{N_e}| \equiv \frac{1}{\sqrt{N_e!}}\sum_{\sigma \in S_{N_e}} (-1)^\sigma \psi_{\sigma(1)}(\vecr_1) \cdots \psi_{\sigma(N_e)}(\vecr_{N_e})
\ee
where $S_{N_e}$ is the permutation group on $N_e$ elements and $\sigma$ is a permutation with sign $(-1)^\sigma$. The matrix element for a single ionized electron will involve a new Slater determinant state $\Psi' = |\psi_1 \dots \psi_{i-1} \psi' \psi_{i+1} \dots \psi_{N_e}|$ with one of the $\psi_i$ replaced by some continuum wavefunction $\psi'$. As long as all of the single-particle orbitals are orthonormal, the only nonvanishing terms will contain transitions between $\psi_i$ and $\psi'$:
\be
\langle \Psi' | \sum_{i = 1}^{N_e} e^{i \vecq \cdot \vecr_i} | \Psi \rangle = \langle \psi' | e^{i \vecq \cdot \vecr} | \psi \rangle.
\ee
Thus, in the Hartree-Fock approximation one can compute the scattering rate using single-particle orbitals, with many-body effects included in the chosen form of the orbitals. 

To generalize the formalism for hydrogen to noble atoms with filled shells (the case most relevant for experiments probing atomic electron scattering, all of which are currently using noble liquid targets), we may define the atomic form factor for the $(n,l)$ orbital as
\be
f_{nl \to \veck}(\vecq) = \sum_m \left | \int d^3 \vecr \, \psi^*_{\veck} (\vecr)\,\psi_{nlm}(\vecr)\,e^{i\vecq \cdot \vecr} \right|^2,
\label{eq:fsqdefm}
\ee
where we have assumed approximate degeneracy in $m$ and summed incoherently over transitions of individual orbitals. Using spherical symmetry and decomposing the outgoing wavefunction into spherical waves, we can then perform the sum over $m$ and write the ionization form factor as an integral over the radial functions,
\be
|f^{nl}_{\rm ion} (k, q)|^2 = \frac{4k^3}{(2\pi)^3}\sum_{l'= 0}^{\infty} \sum_{L = |l'-l|}^{l'+l} (2l+1)(2l'+1)(2L+1)  \left[  \begin{matrix} l & l' & L \\
      0 & 0 & 0 \\
   \end{matrix} \right]^2 \left |\int dr \, r^2 \widetilde{R}_{kl'}(r) R_{nl}(r)j_L(qr)\right|^2\,,
 \label{eq:fsqFull}
\ee
where $\widetilde{R}_{kl'}(r)$ and $R_{nl}$ are the final-state and initial-state radial wavefunctions, respectively, $j_L$ is the spherical Bessel function of order $L$, and the term in brackets is the Wigner-$3j$ symbol evaluated at $m_1 = m_2 = m_3 = 0$.\footnote{Here we have included the factor of 2 for spin degeneracy in Eq.~(\ref{eq:fsqFull}) rather than Eq.~(\ref{eq:fsqdefm}), but this is simply a matter of convention.} In practice, the sum over $l'$ may be cut off at some moderate value $l' \simeq 10$ once the sum converges to the desired accuracy. The total scattering rate is an incoherent sum over all $|f^{nl}_{\rm ion} (k, q)|^2$ for different $(n,l)$,
\be
\frac{dR}{d \ln E_{\rm er}} =  N_{T} \frac{\rho_\chi}{m_\chi} \frac{\bar{\sigma}_e}{8 \mu_{\chi e}^2} \sum_{n,l} \int dq \, q \,  |F_{\rm DM}(q)|^2 |f^{nl}_{\rm ion} (k, q)|^2 \eta(\vmin),
\label{eq:dRdEAtomic}
\ee
with
\be
\vmin = \frac{E_{\rm er} + |E_{nl}|}{q} + \frac{q}{2m_\chi}
\ee
where $E_{nl}$ is the binding energy of the $(n,l)$ shell.

\begin{figure*}[t!]
\begin{center}
\includegraphics[width=0.45\textwidth]{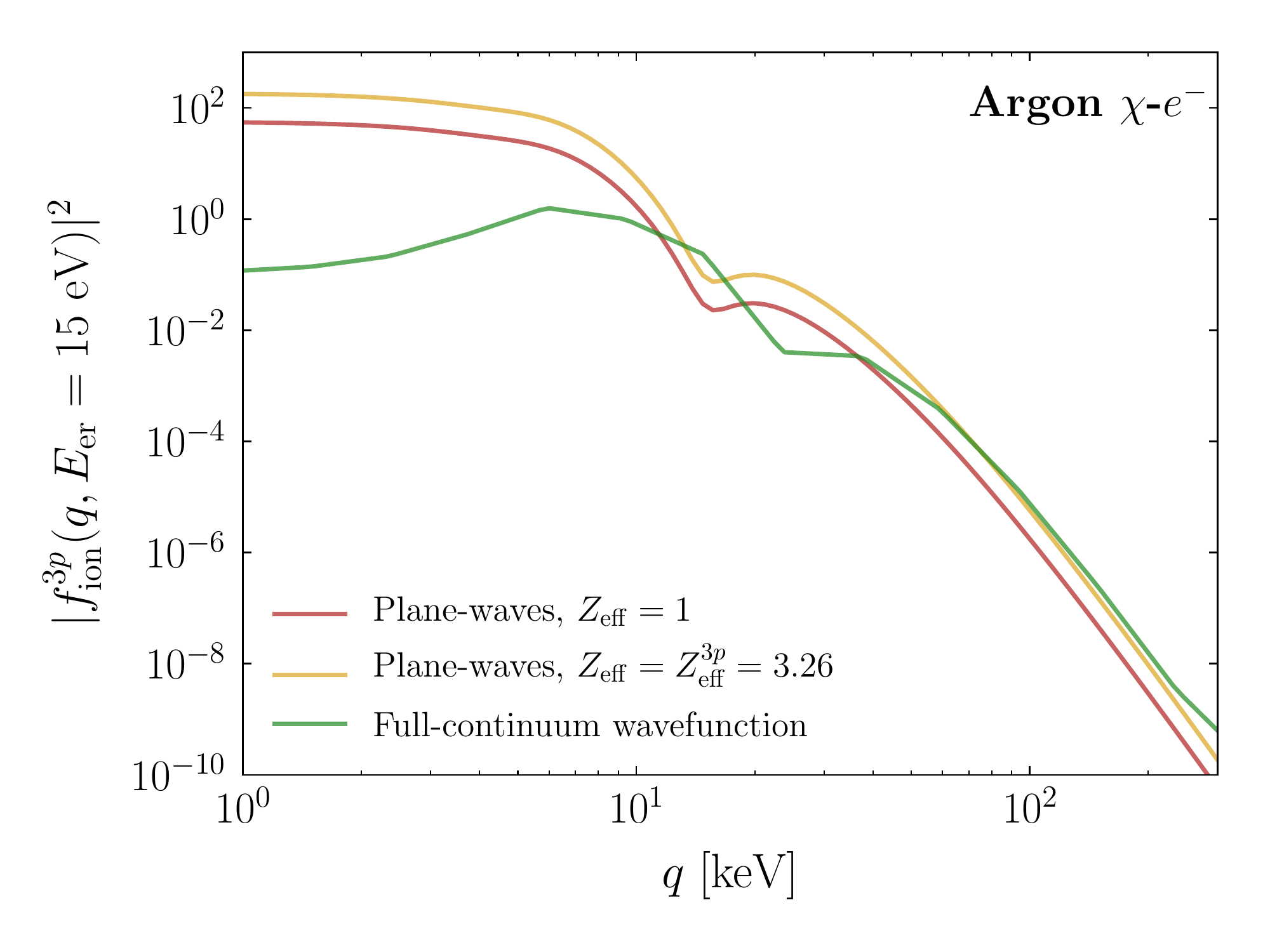}
\includegraphics[width=0.45\textwidth]{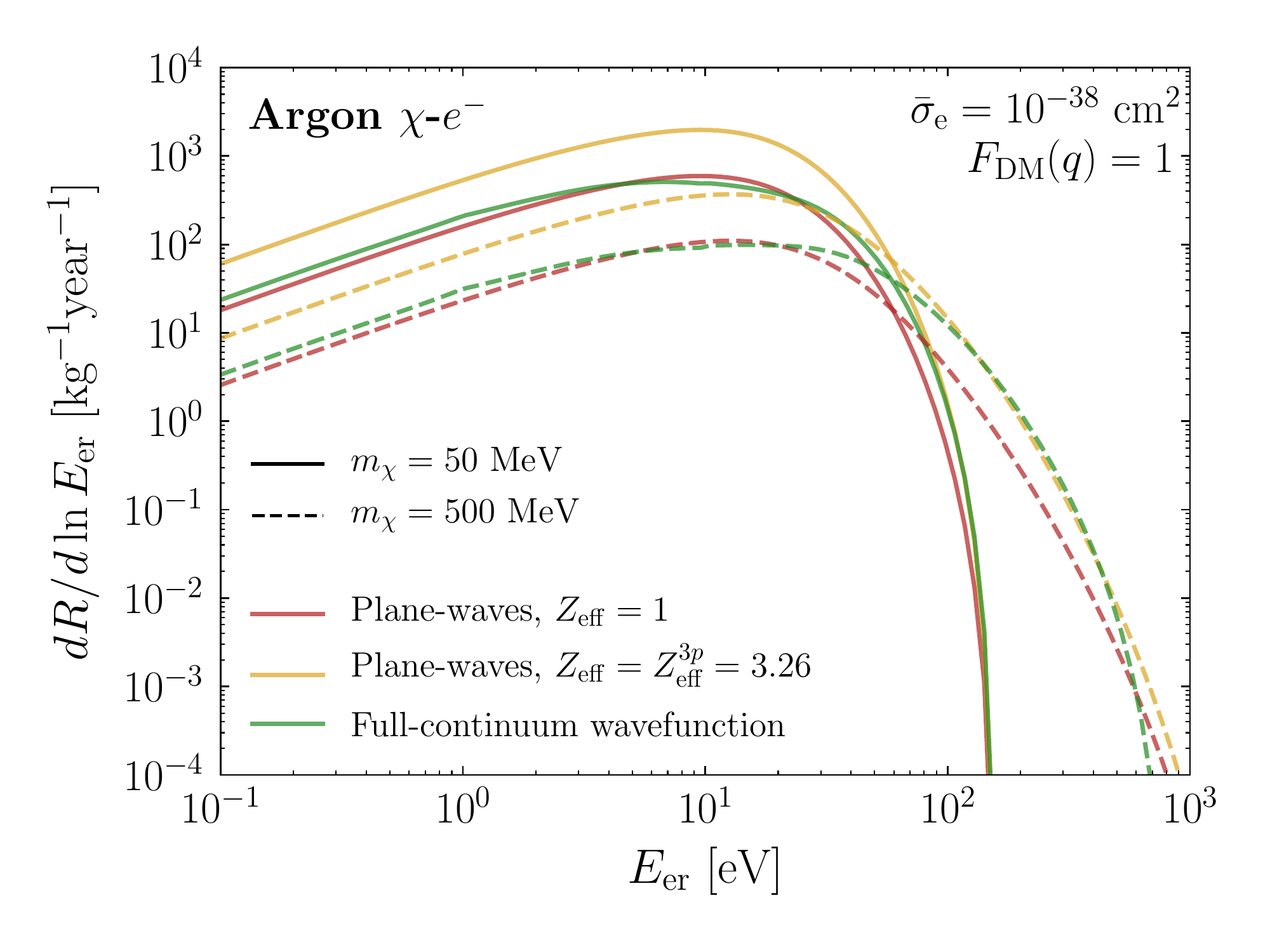}
\caption{ \label{fig:FFCompare}
Ionization form factor (left) and ionization spectrum (right) for the outermost 3p shell in argon, shown using plane-wave outgoing states  with $Z_\text{eff}=1$ (red) and $Z_\text{eff}^{3p} =3.26$ (yellow), as well as the full-continuum wavefunction (green). The form factor is shown at fixed recoil energy $E_\mathrm{er} = 15$ eV; note that Eq.~(\ref{eq:qminDMe}) requires $q \gtrsim 5 \ \keV$, so the large differences at small $q$ between the different models do not enter into the DM scattering rate.}
\end{center}
\end{figure*}

In the Hartree-Fock approximation, the radial wavefunctions for each orbital can be expressed in a basis of Slater-type orbitals with effective charges $Z_{jl}$ and coefficients $C_{jln}$ as \cite{bunge1993roothaan}
\be
R_{nl}(r) = a_0^{-3/2} \sum_{j} C_{jln} \frac{(2Z_{jl})^{n'_{jl}+ 1/2}}{\sqrt{(2n'_{jl})!}}\left(\frac{r}{a_0}\right)^{n'_{jl}-1}e^{-Z_{jl} r/a_0}\,.
\label{eq:RHFdef}
\ee
These wavefunctions are known as Roothaan-Hartree-Fock (RHF) wavefunctions after a standard technique in quantum chemistry for solving the Hartree-Fock equations. Since Ref.~\cite{bunge1993roothaan} does not provide parameterizations of the continuum wavefunctions, the earlier applications of this formalism \cite{Essig:2012yx,Essig:2017kqs} had to supply the final-state wavefunctions externally, which were not guaranteed to be orthogonal to the bound states and thus neglect many-body effects in a possibly important way.\footnote{In the earliest literature \cite{Essig:2011nj}, a plane-wave approximation was used for computational simplicity, along with choosing a rather ad-hoc $\Zeff$ and adding in a Fermi factor by hand.}  In the most recent analyses used by the XENON \cite{XENON:2019gfn} and Panda-X \cite{PandaX-II:2021nsg} (liquid Xe), and DarkSide \cite{Agnes:2018oej} (liquid Ar) collaborations, the following prescription for the outgoing wavefunctions was used: pretend the bound-state orbital $R_{nl}$ is a bound state of a pure Coulomb potential $-\Zeff^{nl}/r$ (rather than the self-consistent potential giving rise to the RHF wavefunctions), and determine $\Zeff^{nl}$ by matching the Coulomb energy eigenvalue to the RHF eigenvalue. For all transitions from the $nl$ state, one then uses this $\Zeff^{nl}$ to construct the outgoing radial functions $\widetilde{R}_{k'l'}(r)$; note this does imply different outgoing radial functions used for different transitions. Fig.~\ref{fig:FFCompare} shows the ionization form factors~(\ref{eq:fsqFull}) and ionization spectrum~(\ref{eq:dRdEAtomic}) for the outermost shell in argon, for different choices of the outgoing wavefunctions and/or Fermi factors. The rate peaks at $E_{\rm er} \simeq p_0^2/(2m_e) = 9 \ \eV$, largely independent of the DM mass, as anticipated from the arguments at the beginning of this section.

As the atomic number of an atom increases, relativistic effects become more important. These can be incorporated \cite{Ibe:2017yqa,Pandey:2018esq} with the Dirac-Hartree-Fock approximation, for which a public code, \texttt{FAC}, exists to calculate both the bound and continuum wavefunctions from a self-consistent potential \cite{gu2008flexible}, ensuring orthogonality.\footnote{That said, \texttt{FAC} is unable to reproduce measured binding energies of outer-shell electrons in xenon at the 20\% level, which strongly affects the sensitivity near threshold; the private code of Ref.~\cite{Pandey:2018esq} utilized the relativistic random-phase approximation and benchmarked the wavefunctions by matching the measured ionization energies of each shell.} The combination of relativistic and many-body effects (which are expected to be important for excited states) in xenon produces a spectrum which differs by almost an order of magnitude at both small and large ionization energies, suppressing the rate at small energies but drastically increasing the tail at large energies, which can potentially have a large impact on experimental searches \cite{Pandey:2018esq}. As of yet there is no direct measurement of the ionization form factors of xenon and argon in the relevant kinematic regime for sub-GeV DM-electron scattering, and thus each of the above approximations for the wavefunctions must be considered to carry some unquantified systematic uncertainty.

\subsubsection{General atomic response}

All of the above discussion has assumed that the DM-electron interaction Hamiltonian $\Delta H_{\chi e}$ is independent of both the DM and electron spins. As shown in Eq.~(\ref{eq:DarkPhotonHamiltonian}), this is true in the $\vecv \to 0$ limit for the dark photon model, but in a bottom-up approach there are 14 non-relativistic operators at linear order in $\vecv$ and quadratic order in $\vecq$ consistent with Galilean invariance and momentum conservation which can depend on various products of momentum and spin, as shown schematically in Eq.~(\ref{eq:BottomUpHamiltonian}). Among the incoming and outgoing DM and electron states, there are two independent momenta, which can be taken to be $\vecq$ and
\be
\vecv_{\rm el}^\perp = \vecv - \frac{\vecq}{2\mu_{\chi e}} - \frac{\veck}{m_e},
\ee
where $\veck$ is the initial-state electron momentum (or the relevant Fourier component of the wavefunction, if the initial state is not a momentum eigenstate). The notation and definition are inherited from the case of elastic nuclear scattering, where $\veck = 0$ for a nucleus at rest, and $\vecv_{\rm el}^\perp \cdot \vecq = 0$ for elastic scattering. 

Specializing to the case of DM-electron scattering in atoms \cite{Catena:2019gfa}, the possible dependence of $\Delta H_{\chi e}$ on $\vecv_{\rm el}^\perp$ leads to a second \emph{vectorial atomic form factor}
\be
\mathbf{f}_{1 \to 2}(\vecq) = \int \frac{d^3 k}{(2\pi)^3} \psi_2^*(\veck + \vecq) \left (\frac{\veck}{m_e}\right) \psi_i(\veck)
\ee
which appears in the total scattering rate through three additional atomic response functions. Schematically, these are
\be
W_2 \sim \frac{\vecq}{m_e} f_{nl \to \veck}(\vecq) \mathbf{f}_{nl \to \veck}(\vecq), \ \ W_3 \sim |\mathbf{f}_{nl \to \veck}(\vecq)|^2, \ \ W_4 \sim \left | \frac{\vecq}{m_e} \cdot \mathbf{f}_{nl \to \veck}(\vecq) \right|^2,
\ee
where $W_1$ is proportional to the ionization form factor $|f_{nl \to \veck}(\vecq)|^2$ and is the sole response function which governs spin-independent scattering. Since $|\vecq| \lesssim m_e$ for sub-GeV DM, and atomic wavefunctions have support peaked at $|\veck| \simeq 1/a_0 =  \alpha m_e \ll m_e$, these additional form factors are typically strongly suppressed compared to the standard atomic form factor. In addition, these response functions are multiplied in the rate by additional powers of $\vecv$, further suppressing the rate. The only exception is the standard ``spin-dependent'' operator $\mathbf{S}_e \cdot \mathbf{S}_\chi$, which does not depend on any powers of momentum and hence scales parametrically like spin-independent scattering. Recently, Ref.~\cite{Liu:2021avx} provided a many-body treatment of these spin-dependent operators.

Depending on the particle nature of the DM, spin-independent scattering may not be the leading contribution to DM-electron interactions, and thus these general atomic responses may offer the possibility of distinguishing between DM models (though at the cost of a suppressed rate compared to the spin-independent expectation) \cite{Catena:2020tbv}. One example is if DM couples to electromagnetism through electric dipole, magnetic dipole, or ``anapole'' moments; if DM is a Majorana fermion, a spin-1/2 particle that is its own antiparticle, the anapole moment is the only nonzero coupling because the electric and magnetic dipoles vanish identically. In this scenario, the mediator between DM and electrons is just the photon itself, but DM couples directly to the Maxwell field strength tensor $F_{\mu \nu}$, giving additional spin and velocity dependence to $\Delta H_{\chi e}$. By exploiting the differences in the ionization spectra in xenon inherited from the various spin-dependent atomic response functions, Ref.~\cite{Catena:2020tbv} showed that observation of $\mathcal{O}(100-1000)$ events can reject the Majorana DM hypothesis. 

\subsection{Molecular excitation and scintillation}
\label{sec:scintillation}

Molecular systems offer several unique advantages for DM-electron scattering. First, consistent with the general principle that closer interatomic spacing leads to smaller binding energies, molecules can have electronic states with smaller separation than the ionization energies of the isolated outer-shell atomic electrons. Second, unlike noble atoms, molecules are naturally anisotropic, and the breaking of spherical symmetry can lead to pronounced daily modulation of the event rate which is a smoking-gun signal of DM.\footnote{Similar effects are present when electrons are ionized from two-dimensional materials like graphene \cite{Hochberg:2016ntt} or carbon nanotubes \cite{Cavoto:2017otc}.} Third, organic compounds in particular have a long history of use as \emph{scintillation} detectors for high-energy particles \cite{birks2013theory}: when a molecule de-excites from an excited electronic state, it emits a photon, which is wavelength-shifted with respect to the excitation energy through internal relaxation (for example, via vibrational levels) such that the probability of the photon exiting the sample (rather than being reabsorbed) is large. This property has previously been proposed to detect nuclear scattering events \cite{Shimizu:2002ik,Collar:2018ydf}, where the electronic excitation occurs due to collisions between molecules, but it was pointed out in \cite{Blanco:2019lrf,Blanco:2021hlm} that the same scintillation effect may be used to search for direct electron scattering.

\begin{figure*}[t!]
\begin{center}
\includegraphics[width=0.45\textwidth]{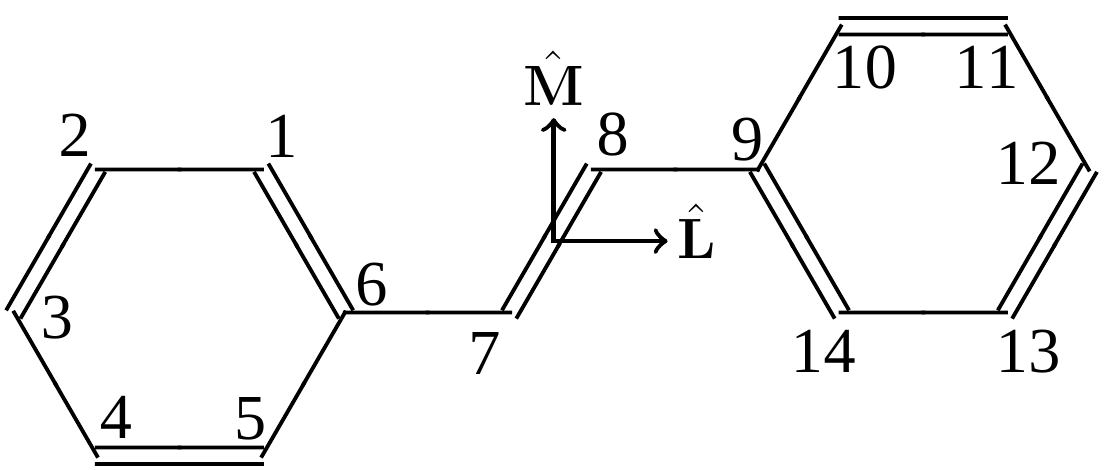}
\caption{ \label{fig:tstilbenediagram}
Chemical structure of \emph{trans}-stilbene, a planar organic molecule which exhibits scintillation. Each vertex hosts a carbon atom, and single and double lines represent single and double bonds, respectively. The 2p$_z$ orbitals responsible for the scintillation transition have electron density concentrated in the direction normal to the plane of the molecule (out of the page). Reproduced from Ref.~\cite{Blanco:2021hlm}.}
\end{center}
\end{figure*}

The key property which differentiates a condensed-matter molecular system (either a liquid, or a molecular crystal with very weak intermolecular correlations) from a generic semiconductor or insulator is that the electrons are not delocalized across the entire sample but are confined to individual molecules. Focusing on \emph{organic aromatic compounds} in particular, the general structure for molecules with delocalized electron systems is a planar network of carbon-carbon double and single bonds in which hexagonal rings are present. Fig.~\ref{fig:tstilbenediagram} shows as an example the structure of \emph{trans}-stilbene (t-stilbene). The partially-filled 2p shell in carbon can accommodate a handful of arrangements of 4 bonds and, in particular, the sp$_2$ hybridized configuration (formed from linear combinations of 2s and 2p orbitals) allows for two single bonds and one double bond. The three sp$_2$ hybrid orbitals responsible for the so-called $\sigma$-bonding, which runs along the interatomic axis, gives these molecules their planar nature. By contrast, the unhybridized 2p$_z$ orbitals extend out of the plane and form $\pi$-bonds through lateral overlap. The alternating $\pi$-bonds form a delocalized electronic system whose dynamics can be considered independently from the tightly bound $\sigma$-bonds that make up the rigid skeleton of the molecule; the electrons in the $\pi$ system are the analog of the valence electrons in a solid-state system. 

Ignoring electron correlation and repulsion effects for the moment, and considering only nearest-neighbor bonding interactions, the ground state and lowest-lying electronic states of the many-body system are well-described by linear combinations of 2p$_z$ orbitals, which represent $\pi$ electrons delocalized across the molecule:
\be
\Psi_i = \sum_j c_i^j \phi_{2p_z}(\vecr - \vecR_j), \qquad \phi_{2p_z}(\vecr) = \sqrt{\frac{Z_{\rm eff}^5}{2^5 \pi a_0^3}} \frac{r \cos \theta}{a_0} \exp \left(\frac{-Z_{\rm eff} r}{2a_0}\right),
\ee
where $j$ runs over all the carbon atoms in the molecule with positions $\vecR_j$ and $Z_{\rm eff} \approx 3.15$ is the effective nuclear charge of carbon seen by the 2p$_z$ orbital. These states are known as H\"{u}ckel molecular orbitals (HMOs). Typical excitation energies from the ground state to the lowest excited state are $\Delta E \simeq 4.5 \ \eV$ in benzene (6 carbon atoms) and $\Delta E \simeq 4.2 \ \eV$ in t-stilbene (14 carbon atoms), about a factor of 2 lower than the outer-shell ionization energy in xenon, which allows organic scintillators to probe lower-mass DM compared to noble liquids.

Due to the relatively small number of electrons involved in the low-lying states (one per carbon atom), a fairly complete treatment of the correlation and repulsion effects is feasible for molecules with $\mathcal{O}(10)$ carbon atoms, using the Hartree-Fock approximation in the HMO basis. Indeed much of the work in understanding organic compounds was done in the quantum chemistry literature in the 1960's with results that can be borrowed almost verbatim, as detailed in \cite{Blanco:2019lrf,Blanco:2021hlm}. Another useful result from this literature is a semi-empirical method which may be used to avoid the computational complexity of Hartree-Fock for larger molecules \cite{Pariser1953,Pariser1953a,pople1955electronic,pople1970molecular}. First, the coefficients $c_j^i$ of the HMOs are fixed by assuming that nearest-neighbor interactions dominate and empirically fitting the coefficients such that the eigenspectrum matches the observed spectrum of the desired molecule; off-diagonal elements are related to bond lengths and diagonal elements can be interpreted as an effective nuclear charge. Degeneracies are resolved by requiring each HMO to transform in an irreducible representation of the symmetry group of the molecule in question, known as the \emph{point group} which keeps the origin fixed. To obtain the many-body wavefunction, antisymmetry is enforced with a variant of the Hartree-Fock approximation, where the ground state is constructed as a Slater determinant of the lowest-energy HMOs in a spin-singlet configuration, and the excited states are constructed by replacing one of the HMOs with a higher-energy orbital in the antisymmetrized wavefunction. Finally, electron-electron interactions are treated as perturbations which mix states with the same symmetries under the point group. This procedure, which focuses on just the bound-electron spin singlet configurations, suffices for practical applications because the singlet states are overwhelmingly responsible for the experimentally-observable scintillation transitions. 

As with the case of atomic scattering, the transition form factor can be reduced to calculating matrix elements with single-particle orbitals at different sites, because each excited electron state can be written as linear combinations of states where one electron has been excited from an occupied HMO $\Psi$ to an unoccupied HMO $\Psi'$. Now, though, the form factor involves taking linear combinations of the single-particle matrix elements and then squaring, reflecting the fact that the system can be treated with the full many-body states. Because all orbitals are simply spatial translations of the same 2p$_z$ orbital, the linearity of the Fourier transform means that there is only a single primary integral to compute, involving the momentum-space orbital
\be
\widetilde{\phi}_{2p_z}(\veck) = \frac{Z_{\rm eff}^{7/2} a_0^{3/2}}{\pi}  \frac{k_z a_0}{(k^2 a_0^2 + (Z_{\rm eff}/2)^2)^3}
\label{eq:phi2pzk}
\ee
weighted by phase factors and coefficients which can be taken from the literature and/or determined from symmetry considerations for a large number of organic scintillators. From there on, the formalism for computing the scattering rate is identical to atomic ionization, with the small modification that the continuous spectrum of ionized states is replaced by a sum over the discrete spectrum of bound states. The first limits on DM-electron scattering from 1.3 kg of organic scintillator EJ-301 (doped p-xylene) using a photomultiplier tube readout were set in 2019 \cite{Blanco:2019lrf}, and are competitive with current constraints from low-mass semiconductor detectors in the few-MeV mass range for $m_\chi$, though many improvements on this setup are possible.

\begin{figure*}
    \centering
    \includegraphics[width=0.98\textwidth]{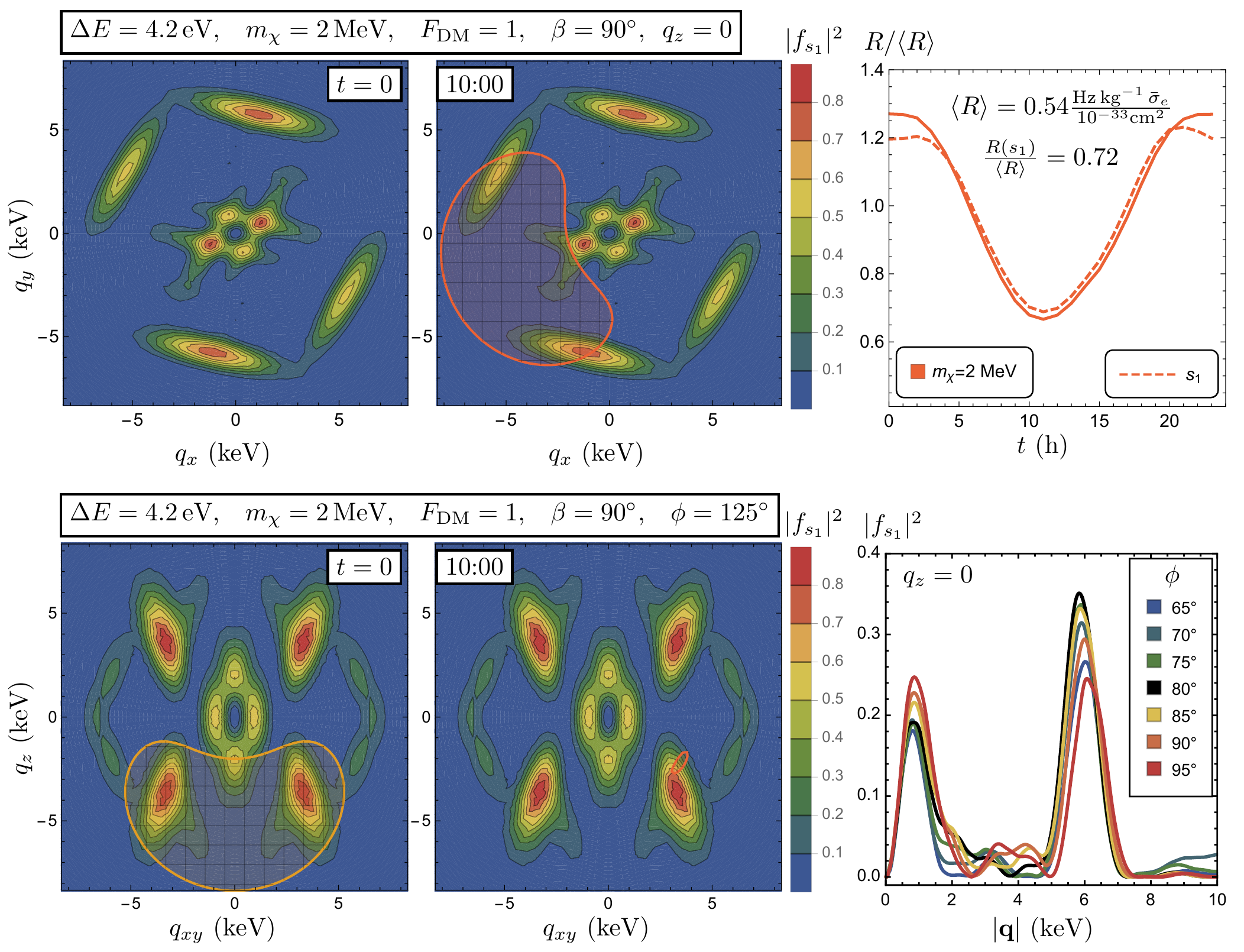}
    \caption{Slices though the molecular form factors for the lowest-energy ($s_1$) transition in t-stilbene and modulating rates for DM masses near threshold, $m_\chi = 2 \ {\rm MeV}$, adapted from Ref.~\cite{Blanco:2021hlm}. The 4 different molecular orientations in the unit cell are apparent in the 4-fold symmetry of the form factor. The gridded bean-shaped region outlined in red and orange indicates the kinematically-accessible momentum transfers $\vecq$ at two different times of day ($t=0$ and $t=10\,\text{h}$); for the chosen orientation of the detector with respect to the DM wind, the kinematically-allowed region at $t = 0$ does not extend into the plane $q_z = 0$ (upper left), and is only barely visible at $t=10\,\text{h}$ for an azimuthal angle of $\phi =125^\circ$ (bottom center), leading to the large modulation amplitude shown in the upper-right plot.}
    \label{fig:beanplots}
\end{figure*}

As can be seen from Eq.~(\ref{eq:phi2pzk}), the electrons in the 2p$_z$ orbitals break spherical symmetry, leading to an anisotropic form factor for transitions where these orbitals dominate. Define the scintillation transition form factors from the (many-body) ground state $|\psi_G \rangle$ to scintillation state $|\psi_{s_i} \rangle$, with energy $\Delta E_i$ with respect to the ground state, as
\be
f_i(\vecq) = \langle \psi_{s_i} | \sum_{k = 1}^{N_C} e^{i \vecq \cdot \vecr_k} | \psi_G \rangle
\ee
where the sum runs over the $k$ delocalized electrons, equal to the number $N_C$ of carbon atoms in the molecule. This notation differs somewhat from the atomic case since we are only indexing the transition by its energy, rather than labeling the initial and final states by individual orbitals, due to the fact that the electron-electron interactions in the molecule have caused these orbitals to mix. Using Eqs.~(\ref{eq:gdef}) and~(\ref{eq:Rdaily}) for the time-dependent rate, the rate for observing a scintillation photon is
\be
R(t) = \xi  N_{\rm mol} \frac{\rho_\chi}{m_\chi} \frac{\bar\sigma_e}{\mu_{\chi e}^2} \sum_{i} \int\! \frac{d^3 \vecq}{4\pi} g(\vecq, \Delta E_i; t) F_\text{DM}^2(q) |f_{i}(\vecq)|^2,
 \ee
where $N_{\rm mol}$ is the number of molecules per unit detector mass, $\xi$ is a quantum efficiency factor representing the probability that the de-excitation of a scintillation state yields a photon which escapes the sample, and the sum runs over all transitions to singlet states. Some examples of form factors for the lowest $s_1$ transition in t-stilbene (which can be grown as high-quality single crystals, preserving the directionality of the form factors) and the associated time-dependent rate for various DM masses are shown in Fig.~\ref{fig:beanplots}. The anisotropy is clear, leading to a large modulation amplitude of $\simeq 50\%$ for $m_\chi = 2 \ \MeV$ for an orientation of the crystal axis with respect to the $z$-axis given by $\beta = 90^\circ$ (see Ref.~\cite{Blanco:2021hlm} for details). The 4-fold symmetry of the form factor is due to the presence of 4 molecules in the unit cell with different orientations; the anisotropy would be even larger in a crystal with fewer independent molecular orientations.
 
Note that the form factor peaks at a typical scale $q^* \simeq 6 \ \keV$, of order the inverse carbon-carbon bond length, consistent with the arguments at the beginning of this section. Since the minimum-allowed $q$ in a scattering event is $q_{\rm min} = \Delta E/v$, we can define an effective velocity scale
 \be
 v^* \equiv \frac{\Delta E}{q^*}
 \ee
 such that the peak of the form factor is accessible if $v > v^*$. In t-stilbene, $v^* \simeq 200 \ {\rm km/s}$, and the form factor peaks are accessible to the bulk of the DM velocity distribution in the SHM, a fortuitous coincidence given that $v^*$ and $v_0$ are set by completely different physical mechanisms! As we will see below, this yields a total rate per unit mass for organic scintillators which is within a factor of a few of silicon. However, this same fact leads to a smaller $\sim 10\%$ daily modulation amplitude at large DM masses, due largely to structure of the secondary peaks in the form factor at $q \simeq 1.2 \ \keV$ (comparable to the inverse size of the entire molecule). Indeed, daily modulation is mainly a \emph{threshold} effect, occurring when form factor peaks are kinematically accessible for some values of $\vecq \cdot \vecv$ but not others, and thus depends sensitively on the high-velocity tail of the DM distribution. Thus, there is an inevitable trade-off between maximizing the total rate and maximizing the modulation amplitude, an effect also seen in the analogous daily modulation signal in phonons \cite{Griffin:2018bjn,Griffin:2020lgd,Coskuner:2021qxo}. The enhanced statistical power of a modulating rate, even in the presence of a nonzero (but non-modulating) background rate, may be favorable over simply maximizing the total rate if zero-background operation is not feasible.

\subsection{Solid-state excitation and collective modes}
\label{sec:dielectric}

To probe dark matter with MeV mass or below, electronic excitation energies at the eV scale and below are required, necessitating the use of solid-state systems with the required low band gaps. Historically, much of the theoretical and experimental effort has focused on \emph{conventional semiconductor} detectors -- specifically silicon and germanium -- as well as \emph{conventional superconductors} like aluminum. Recently there has been a flourishing effort to identify new materials with particular properties which are well-suited to the kinematics of sub-GeV DM. We will survey these new candidate in materials in Sec.~\ref{sec:futuretheory}, and focus here on a particular illustrative example, \emph{Dirac materials}, which has a band structure suitable for semi-analytic calculations of the dynamic structure factor.

The first calculations of DM-electron scattering in silicon and germanium used either DFT band structures and wavefunctions \cite{Essig:2011nj,Essig:2015cda} or semi-analytic models based on hydrogenic or tight-binding orbitals \cite{Graham:2012su,Lee:2015qva}. By the arguments at the beginning of this section, the typical momentum transfers probe electrons on length scales smaller than a single unit cell as long as the gap is $\mathcal{O}(\eV)$ or larger, so both of these approaches are expected to give the correct order of magnitude for the total scattering rate, though there are important differences between the spectra at small and large recoil energies, as we will discuss shortly. 

We first review the original DFT approach of Ref.~\cite{Essig:2015cda}, which amounts to the following procedure: start with Eq.~(\ref{eq:Sqw_electron}), treat the initial and final states as Bloch states, and neglect the sum over all electrons in the operator by taking $\sum_{k} e^{i \vecq \cdot \vecr_{k}} \to e^{i \vecq \cdot \vecr}$. First, using the Bloch wave parametrization of Eq.~(\ref{eq:Bloch}) and the Fourier expansion of the cell function (\ref{eq:GExpand}), we may write the single-particle wavefunctions as
\be
\psi_{i \veck}(\vecr) = \frac{1}{\sqrt{V}} \sum_{\vecG} u_i(\veck + \vecG)e^{i(\veck + \vecG) \cdot \vecr}
\label{eq:Bloch_momentumexp}
\ee
where $i$ is a band index, $\veck$ is a crystal momentum, and $\vecG$ is a reciprocal lattice vector. The normalization of the wavefunction is enforced by $\sum_{\vecG}| u_i(\veck + \vecG)|^2 = 1$. The analogue to the atomic form factor is given by single-particle wavefunction overlaps~\cite{Essig:2015cda}
\begin{align}
|f_{i \veck \to i' \veck'}|^2 & = \sum_{\vecG'} \frac{(2\pi)^3 \delta^3(\vecq - (\veck' - \veck + \vecG'))}{V} \left | \sum_{\vecG} u^*_{i'}(\veck' + \vecG + \vecG')u_i(\veck + \vecG)\right|^2 \\
& \equiv \sum_{\vecG'} \frac{(2\pi)^3 \delta^3(\vecq - (\veck' - \veck + \vecG'))}{V} |f_{[i\veck, i'\veck', \vecG']}|^2,
\end{align}
where the delta function makes explicit that crystal momentum is conserved only up to a reciprocal lattice vector. Note that we will use $i$ to label valence bands and $i'$ to label conduction bands, assuming the target is at zero temperature. Similarly, summing over initial and final states and bands (including an explicit factor of 2 for spin degeneracy) and assuming an isotropic DM velocity distribution yields the analogue of the ionization form factor 
\be
|f_{\rm crystal}(q, \omega)|^{2} = \frac{2\pi^2}{\alpha m_e^2 \Omega \, \omega} \sum_{i, i'} \int_{\rm BZ} \frac{\Omega d^3 k}{(2\pi)^3} \frac{\Omega d^3 k'}{(2\pi)^3} \omega \delta(\omega - E_{i'\veck'} + E_{i \veck})\sum_{\vecG'} q \delta(q - |\veck' - \veck + \vecG'|) |f_{[i\veck, i'\veck', \vecG']}|^2
\label{eq:fcrystal}
\ee
referred to as the \emph{crystal form factor}. Here, the integral is taken over the 1BZ and $\Omega$ is again the unit cell volume. By analogy to Eq.~(\ref{eq:dRdEAtomic}), the spectrum is 
\be
\frac{dR}{d \ln \omega} = \frac{1}{\Omega \rho_{T}} \frac{\rho_\chi}{m_\chi} \sigmabar_e \alpha \times \left [ \frac{m_e^2}{\mu_{\chi e}^2} \int d \, \ln q \, \left(\frac{\omega}{q} \eta(\vmin) |F_{\rm DM}(q)|^2 |f_{\rm crystal}(q, \omega)|^2 \right) \right ].
\label{eq:DFTspectrum}
\ee
Note that now $\omega$ is total energy deposited, while in the atomic case we gave the differential rate with respect to the outgoing energy of the ionized electron, $E_{\rm er}$, which differs from the total deposited energy by the binding energy of the initial-state orbital. In Eq.~(\ref{eq:DFTspectrum}), the factor in square brackets is a dimensionless $\mathcal{O}(1)$ number when $q \sim p_0$ and $\omega \sim \mathcal{O}(5 \ \eV)$, since $\omega /q \simeq 10^{-3} \simeq v$ while $\eta(v_{\rm min}) \sim 1/v$. For $m_\chi = 10 \ \MeV$ and $\sigmabar_e = 10^{-37}$ cm$^{2}$ at the freeze-out cross section, this gives an expected rate of $\mathcal{O}(1)$ events/min/kg, a sizable rate compared to WIMP experiments!

\begin{figure}[t!]
\begin{center}
\includegraphics[width=0.45\textwidth]{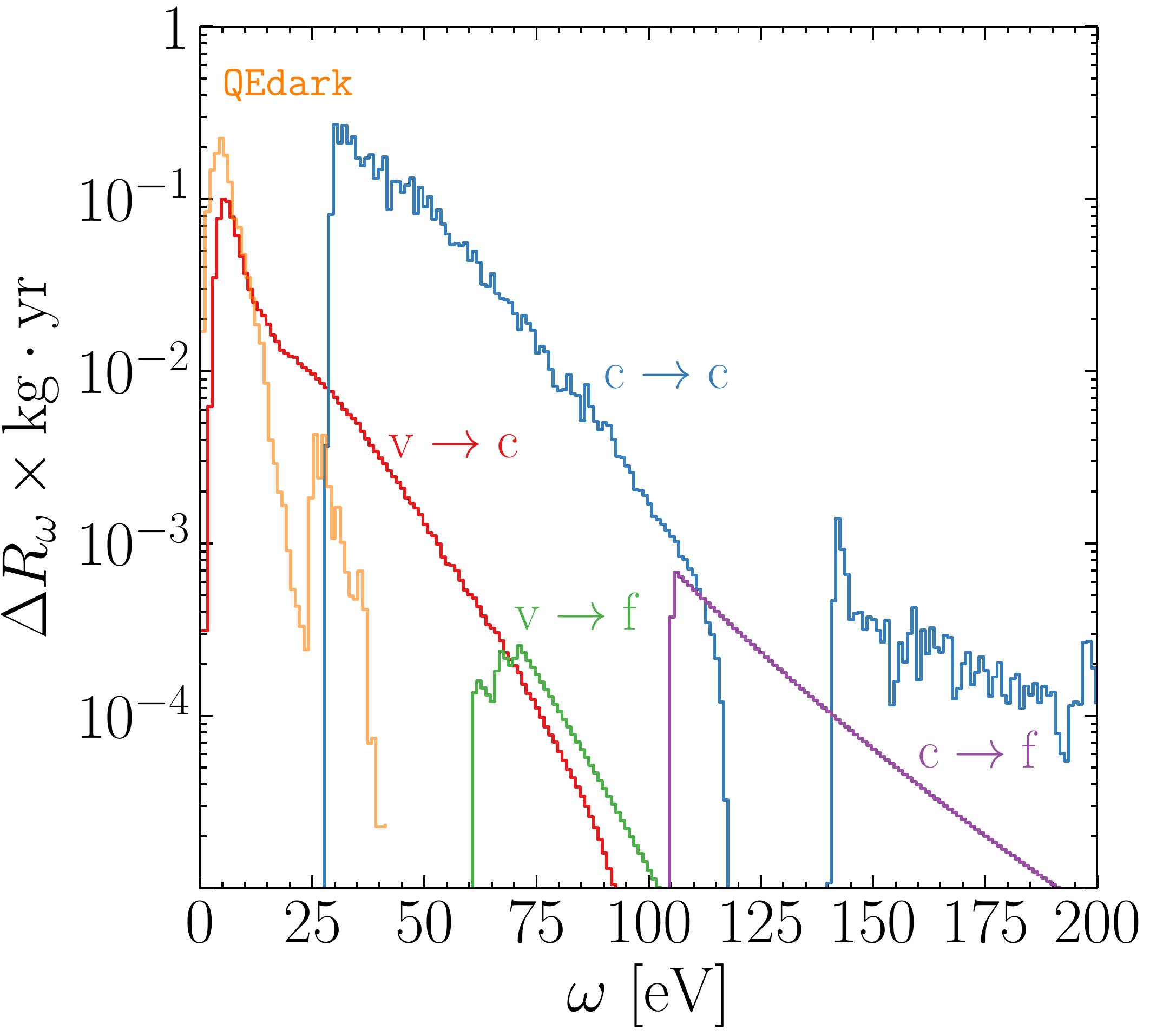}~~
\caption{ \label{fig:all_electron}
Contributions to the electron scattering spectrum in germanium for $m_\chi = 1 \ \GeV$ and $\overline{\sigma}_e = 10^{-40} \ {\rm cm}^2$, using all-electron reconstruction, including valence to conduction ($v \to c$), valence to free ($v \to f$), core to conduction ($c \to c$), and core to free ($c \to f$). The $c \to c$ transition involving the core 3d electron shell dominates the rate above 25 eV, an effect which is underestimated in the \texttt{QEdark} code \cite{Essig:2015cda} which uses only a Bloch wave basis. The $y$-axis is a binned event rate in 1 eV bins. Reproduced from Ref.~\cite{Griffin:2021znd}.
}
\end{center}
\end{figure}

One limitation of the DFT approach is that it is computationally expensive to include high-momentum components of the wavefunctions, which are needed for scattering of heavier DM and to describe excitations of core or semi-core electrons. For instance, in germanium, the 3d electron shell generates an important contribution to the spectrum above about 30 eV \cite{Lee:2015qva}. The importance of including these components was emphasized in Ref.~\cite{Liang:2018bdb}, which used the projector augmented wave (PAW) method for an ``all-electron'' reconstruction of the wavefunctions, but this treatment was still limited to DM masses below about 100 MeV. Recently, Ref.~\cite{Griffin:2021znd} extended the PAW approach and also included core electrons by a semi-analytic method similar to that of Refs.~\cite{Graham:2012su,Lee:2015qva}. In particular, Ref.~\cite{Griffin:2021znd}  showed that the core electron wavefunctions are accurately modeled in a basis of Wannier orbitals
\be
\psi_{i \veck}(\vecr) = \frac{1}{\sqrt{N}} \sum_{i = 1}^N e^{i \veck \cdot \vecR_i} \phi(\vecr - \vecR_i)
\ee
where the sum runs over all lattice sites with positions $\vecR_i$. The Wannier functions $\phi$ are typically taken to be linear combinations of atomic orbitals, similar to those used in Secs.~\ref{sec:atomic} and~\ref{sec:scintillation}. Since $q_{\rm min} \gg |\vecR_i - \vecR_j|^{-1}$ for the high-momentum transfers of interest, the sum is mostly incoherent and the crystal form factor can be approximated with an atomic form factor. The conclusion of Ref.~\cite{Griffin:2021znd} is that rates can be much larger for excitation energies $\gtrsim 25$ eV and that for Ge in particular, scattering of 3d electrons dominates for DM mass above 30 MeV and a heavy mediator, as shown in Fig.~\ref{fig:all_electron}.

\subsubsection{Dynamic structure factor for electron scattering}

To make contact with the dynamic structure factor language of Sec.~\ref{sec:DMstructurefactor}, and to see the appearance of many-body effects, recall that we determined a general relationship between the dynamic structure factor and the dielectric function $\epsilon(\vecq,\omega)$, Eq.~(\ref{eq:Sepsilon}). We can see how the single-particle form factor Eq.~(\ref{eq:fcrystal}) arises from Eq.~(\ref{eq:Sepsilon}) if we use a particular approximation for the dielectric function called the random phase approximation (RPA) \cite{PhysRev.92.609}. Within RPA, we treat the medium as a degenerate gas of weakly-coupled quasiparticle excitations and use perturbation theory to compute the loop to absorb and emit single-electron excitations, yielding the \emph{Lindhard dielectric function},
\be
\epsilon(\vecq,\omega) = 1 - \frac{4 \pi \alpha}{V q^{2}} \sum_{\veck,\veck',i, i'}  | \langle \veck', i' |e^{i \vecq \cdot \vecr} | \veck, i \rangle |^{2}   \lim_{\eta \to 0} \frac{ f^{0}(E_{i'\veck'}) - f^{0}(E_{i \veck}) }{ E_{i'\veck'} - E_{i \veck} - \omega - i \eta } ,
	\label{eq:lindhard}
\ee
where $| \veck, i \rangle$ denotes Bloch states. This approximation to the dielectric function is essentially computing the polarization in a noninteracting gas.\footnote{A potentially confusing piece of nomenclature, standard in the condensed matter literature, is that ``noninteracting'' refers to the fact that the electrons are treated as not \emph{directly} interacting particle-by-particle when calculating the leading-order response function to an external probe. However, one obtains an approximation to the dielectric function of an interacting gas by self-consistently treating each individual electron as feeling the average potential accounting for all the other electrons~\cite{Arovas20}. In field theory language, the ground state of the system is corrected by the photon vacuum polarization diagram, which is calculated by neglecting electron-electron interactions.}
Note that  $\epsilon(\vecq, \omega)$ acquires an imaginary part when the denominator becomes singular:
\be
{\rm Im} \, \epsilon(\vecq, \omega) = \frac{4 \pi^2 \alpha}{V q^{2}} \sum_{\veck,\veck',i, i'}  | \langle \veck',i' |e^{i \vecq \cdot \vecr} | \veck,i \rangle |^{2}  \delta(\omega + E_{i \veck} - E_{i'\veck'} ),
\label{eq:lindhard_imeps}
\ee
using the Dirac identity ${\rm Im}(\lim_{\eta \to 0} \frac{1}{x - i \eta}) = \pi \delta(x)$. Then the dynamic structure factor can be written as 
\be
  S(\vecq, \omega) =    \frac{q^2}{2 \pi \alpha}  {\rm Im} \left(-\frac{1}{\epsilon(\vecq, \omega)}\right) =  \frac{2 \pi}{V |\epsilon(\vecq,\omega)|^2 } \sum_{\veck,\veck',i, i'}  | \langle \veck',i' |e^{i \vecq \cdot \vecr} | \veck,i \rangle |^{2}  \delta(\omega + E_{i \veck} - E_{i'\veck'} ).
\ee

$S(\vecq,\omega)$ has the expected form if we were to start with Eq.~(\ref{eq:Sqw_electron}) and treat the initial and final states as Bloch states, which led to the rate in Eqs.~(\ref{eq:fcrystal}-\ref{eq:DFTspectrum}), but here an additional factor of $1/|\epsilon(\vecq, \omega)|^{2}$ appears. The difference arises from the fact that the true initial and final states are many-body states, and the actual result in terms of single-particle states must account for the total response of the medium, which screens any external perturbation by $\epsilon(\vecq,\omega)$. In other words, the factor of $1/|\epsilon(\vecq, \omega)|^{2}$ accounts for in-medium screening effects and is equivalent to resumming an infinite series of insertions of the polarization loop, where we see explicitly that the structure factor includes terms to all orders in $\alpha$.\footnote{These ``in-medium'' effects may be though of as corrections to the propagator of the mediator for DM-electron interactions \cite{Hochberg:2015fth}, or $q$-dependent modifications to the DM-electron coupling \cite{Trickle:2019nya}, but here we include them directly in the dynamic structure factor.}
We can also write $S(\vecq,\omega)$ explicitly in terms of the same wavefunction overlaps as before using Eq.~(\ref{eq:Bloch_momentumexp}) for the initial and final states \cite{Knapen:2021run}:
\begin{align}
S(\vecq, \omega) &= \frac{2\pi}{|\epsilon(\vecq, \omega)|^2} \sum_{i, i'} \int \frac{2 d^3 \veck}{(2\pi)^3}\frac{d^3 \veck'}{(2\pi)^3}  |f_{[i\veck, i'\veck', \vecG']}|^2 \sum_{\vecG} (2\pi)^3 \delta^3(\vecq + \veck - \veck' -\vecG) \delta(\omega - E_{i'\veck'} + E_{i \veck}) \\
 &= \frac{8 \pi^{2}\alpha^{2}m_{e}^2}{q^{5} \Omega } \frac{ |f_{\rm crystal}(q,\omega)|^2}{|\epsilon(\vecq, \omega)|^2}, \label{eq:Sqw_to_fcrystal}
\end{align}
where the second line is obtained by comparing with Eq.~(\ref{eq:fcrystal}). For typical momentum transfers $q \gg q_{\rm coh}$, $ |\epsilon(\vecq,\omega)|^2  \approx 1$ and screening will lead to $O(1)$ effects, though this is needed to obtain precise predictions of the scattering rate beyond an order-of-magnitude estimate.

While the interpretation of the dielectric function in terms of single-particle Bloch states is only true within RPA, the relationship between the dielectric function and the scattering rate is more general. For spin-independent DM-electron scattering, which may arise from exchange of a scalar or vector mediator, the transition rate for a fixed DM velocity $\vecv$ is \cite{Hochberg:2021pkt,Knapen:2021run}
\be
    \Gamma(\vecv) =   \frac{\pi \bar \sigma_{e}}{\mu_{\chi e}^2} \int\frac{d^3 \vecq}{(2\pi)^3}\,
       |F_{\rm DM}(q)|^{2} \, \left[\frac{q^2}{2 \pi \alpha} \,
        {\rm Im}\left(-\frac{1}{\epsilon(\vecq, \omega_{\vecq})}\right)\right],
\ee
which is proportional to the ELF $\mathcal{W}(\vecq, \omega)$ (Eq.~(\ref{eq:ELFdef})) and the DM-electron scattering potential $V(q)$ (or equivalently, the DM form factor $F_{\rm DM}(q)$).\footnote{See Refs.~\cite{Trickle:2020oki,Catena:2021qsr} for a discussion of interactions beyond spin-independent couplings to electron density, performed in the single-particle Bloch wave basis. See also a dedicated study of magnon excitation by spin-dependent scattering in Ref.~\cite{Trickle:2019ovy}.} Assuming an isotropic velocity dispersion, the spectrum per unit mass is
\be
    \label{eq:spectrum}
    \frac{dR}{d \omega} = \frac{1}{\rho_T} \frac{\rho_\chi}{m_\chi}   \frac{\pi \bar \sigma_{e}}{\mu_{\chi e}^2}
        \int  \frac{dq \, q^3}{(2 \pi)^3 \alpha } \, |F_{\rm DM}(q)|^{2}  \eta\bigl(v_{\mathrm{min}}(q,\omega)\bigr)  \mathcal{W}(q,\omega).
\ee
Using Eq.~(\ref{eq:Sqw_to_fcrystal}) for the Lindhard dielectric, we see that Eq.~(\ref{eq:spectrum}) exactly matches previous results~(\ref{eq:DFTspectrum}) using DFT wavefunctions except for the factor of $1/\epsilon^{2}$. This makes it clear that the objects of interest are not the single-particle wavefunctions themselves, but rather the particular combination which appears in the dynamic structure factor.

The advantages of the dielectric function formalism for spin-independent scattering are that there are many  simple analytic models and detailed condensed matter calculations for $\epsilon(\vecq, \omega)$, and that for many common (and some less-common) materials, $\epsilon(\vecq, \omega)$ has been directly measured. 
 The \emph{free-electron gas (FEG) dielectric function} (somewhat confusingly, also commonly referred to as the Lindhard function) is a particularly useful analytic model which captures important scales in the problem. In this model, the electrons are assumed to fill up an isotropic Fermi sphere with no band gap or lattice structure, which gives \cite{dressel2002electrodynamics}
\begin{multline}
    \label{eq:e-rpa-fw}
    \epsilon_{\rm FEG} (\vecq,\omega) =
    1 +  \lim_{\eta \to 0} \frac{3\omega_p^2}{q^2v_F^2}\Biggl\{
        \frac12 + \frac{k_F}{4q}\left(
            1 - \left(
                \frac{q}{2k_F} - \frac{\omega + i\eta}{qv_F}
            \right)^2
        \right)\operatorname{Log}\left(
            \frac{\frac{q}{2k_F}-\frac{\omega + i\eta}{qv_F} + 1}
                 {\frac{q}{2k_F}-\frac{\omega + i\eta}{qv_F} - 1}
        \right) \\
        +
        \frac{k_F}{4q}\left(
            1 - \left(
                \frac{q}{2k_F} + \frac{\omega + i\eta}{qv_F}
            \right)^2
        \right)\operatorname{Log}\left(
            \frac{\frac{q}{2k_F}+\frac{\omega + i\eta}{qv_F} + 1}
                 {\frac{q}{2k_F}+\frac{\omega + i\eta}{qv_F} - 1}
        \right)
    \Biggr\}
    .
\end{multline}
The free parameter in this model is the electron density $n_e$, in terms of which we have
\be
k_F = (3\pi^2 n_e)^{1/3}, \qquad v_F = \frac{k_F}{m_e}, \qquad \omega_p = \sqrt{\frac{4\pi \alpha n_e}{m_e}}.
\label{eq:FEGparams}
\ee
Typical values for the FEG model parameters are
\be
k_F \simeq 5 \ \keV \simeq p_0, \qquad v_F \simeq 10^{-2}, \qquad \omega_p \simeq 15 \ \eV.
\ee
In general, the ELF in an isotropic medium must satisfy two important consistency conditions, arising from its origin as a causal correlation function:
\be
\label{eq:Wsumrule}
\mathcal{W}(\vecq, -\omega) = -\mathcal{W}(\vecq, \omega), \qquad \int_0^\infty d\omega \, \omega \mathcal{W}(\vecq, \omega) = \frac{\pi}{2}\omega_p^2.
\ee
The integral constraint is known as the \emph{f-sum rule}, and both are satisfied by the FEG dielectric function. One limitation of the FEG is that it doesn't account for finite electron damping time or the plasmon width; this can be accounted for in the \emph{Mermin dielectric function} with damping parameter $\Gamma_p$~\cite{PhysRevB.1.2362}:
\be
	\epsilon_{\rm Mermin}(\vecq, \omega) \equiv 1 + \frac{ (1 + i \Gamma_{p}/\omega) \left[ \epsilon_{\rm FEG}(\vecq,\omega + i \Gamma_{p}) - 1 \right] }{ 1 + i \Gamma_{p}/\omega \left[ \frac{\epsilon_{\rm FEG}(\vecq,\omega + i \Gamma_{p}) - 1}{\epsilon_{\rm FEG}(\vecq,0) - 1} \right]}
	\label{eq:epsMermin}
\ee
In particular, one can also take a linear combination of Mermin models with different $n_e$, fitted to experimental measurements of energy loss or dielectric functions and appropriately weighted to satisfy the $f$-sum rule~\cite{10.1002/sia.6227,PhysRevA.58.357,VOS2019242}. This is called the Mermin oscillator model. 

\begin{figure*}[t!]
\begin{center}
\includegraphics[width=0.48\textwidth]{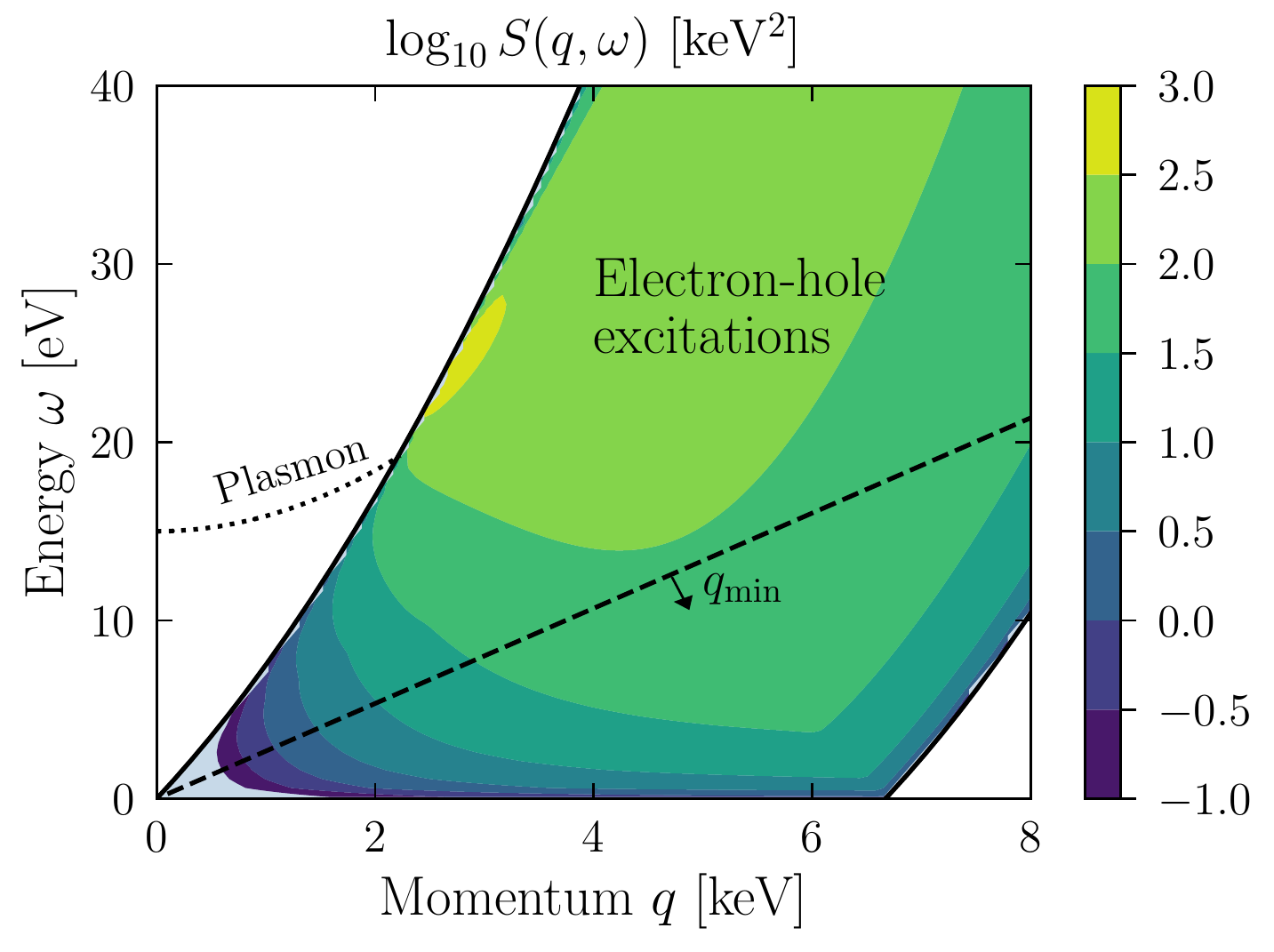}
\includegraphics[width=0.47\textwidth]{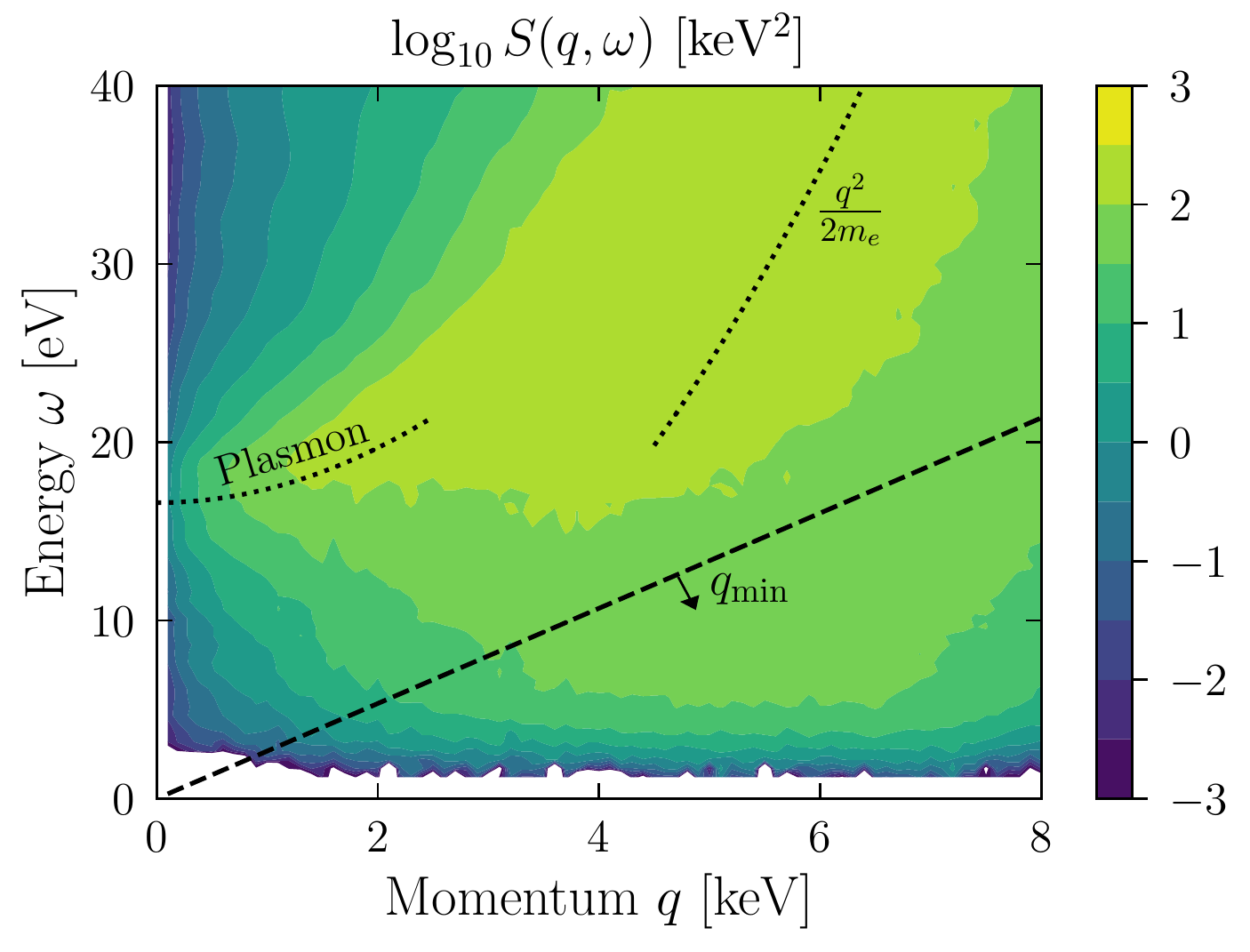}
\caption{ \label{fig:Si_sqw} (\emph{left}) Dynamic structure factor for a free electron gas with $\omega_{p} = 15$ eV and $k_{F} = 3.3$ keV, obtained by taking the $\Gamma_{p} \to 0$ limit of Eq.~(\ref{eq:e-rpa-fw}). The plasmon resonance shows up as an infinitely narrow resonance indicated by the dotted line, while the rest of the support is interpreted as electron-hole excitations. (\emph{right}) Dynamic structure factor for ionization in a Si semiconductor, based on the calculation in Ref.~\cite{Knapen:2021run}. At low $q$, it is peaked at the plasmon resonance, while at high $q$ the peak converges to the free-electron dispersion $q^{2}/(2m_{e})$, similar to the structure factor for hydrogen. In both panels, the dashed line is the minimum $q$ for DM scattering, Eq.~(\ref{eq:qminDMe}), such that the kinematically-accessible region is below and to the right of the dashed line. Thus, the peaks of the structure factor are not accessible to halo DM.
}
\end{center}
\end{figure*}

In Fig.~\ref{fig:Si_sqw}, we show the dynamic structure factor calculated with the FEG dielectric function in the limit $\Gamma_{p} \to 0$ and with a DFT calculation for Si from Ref.~\cite{Knapen:2021run}. In both cases, there is a strong resonance at low $q$ and $\omega \simeq 10-20$ eV associated with the plasmon excitation, with is infinitely narrow in the left panel but has a finite width of a few eV in the right panel. At high $q$, there is broad support for the dynamic structure factor, interpreted here as electron-hole excitations, which are peaked about the free-electron dispersion $q^{2}/(2 m_{e})$. An important difference between the two structure factors is that the dynamic structure factor goes to zero as $\omega$ approaches the band gap of the material. However, for $\omega$ somewhat above the band gap and $q > q_{\rm min}$ for DM scattering, we see that the two structure factors are qualitatively quite similar. Indeed, as shown in Refs.~\cite{Hochberg:2021pkt,Knapen:2021bwg}, the FEG model describes the DM-electron spectrum for $\omega \in [5, 15] \ \eV$ in Si and Ge semiconductors quite well. For higher energies, it is also expected that the FEG model fails to capture the localized core electrons and the associated high-momentum tails. This can be improved on by generalizing the FEG ELF to the Mermin oscillator model fitted to experimental data, which gives good agreement with DFT calculations at higher energies and momenta as well, see Refs.~\cite{Knapen:2021run,Knapen:2021bwg}. Note there is also good agreement among independent DFT calculations accounting for screening effects~\cite{Knapen:2021run,Liang:2021zkg,Griffin:2021znd}.

The upshot of this discussion is that the FEG dielectric function provides a useful analytic model to understand the dynamic structure factor in conventional semiconductors and metals, and we can use it here to explore the features and general behavior of the structure factor in several limits:
\begin{itemize}
\item For $q \gg k_F$, the FEG ELF is maximized at $\omega \simeq \frac{q^2 v_F}{2k_F}$. For $v_F = k_F/m_e$, this relation is simply the dispersion for elastic scattering, $\omega = q^2/(2m_e)$, reflecting the fact that electrons are mostly independent particles at large $q$. The width of the peak is quite broad, since the Fermi velocity of electrons at the Fermi surface is sizeable, but at $q \gg k_{F}$, the scattering rate converges to that of free-electron scattering.
\item For lower $ q \lesssim k_{F}$, screening and many-body effects effects start to play an important role. Taking $q \ll k_F$ and $\omega \ll q v_F$, we have
\be
\epsilon_{\rm FEG} (\vecq,\omega) \approx \frac{\lambda_{\rm TF}^2}{2q^2} + i \frac{3\pi \omega_p^2 \omega}{2 q^3 v_F^3} \qquad (q \ll k_F, \ \omega \ll q v_F)
  \ee
  where $\lambda_{\rm TF} = \omega_p v_F/\sqrt{3}$ is the inverse \emph{Thomas-Fermi screening length}. The fact that $|\epsilon|^2 \gg 1$ in the low-momentum regime is a manifestation of long-range charge screening. For typical values of the FEG parameters, ${\rm Re}\, \epsilon \gg {\rm Im} \, \epsilon$, so we may approximate the ELF as
  \be
 \mathcal{W}(\vecq, \omega) \approx \frac{{\rm Im} \, \epsilon(\vecq, \omega)}{({\rm Re}\, \epsilon(\vecq, \omega))^2} \approx \frac{6 \pi \omega_p^2}{\lambda_{\rm TF}^4 v_F ^3} \omega q \qquad (q \ll k_F, \ \omega \ll q v_F)
 \label{eq:FEGELF}
 \ee
which vanishes as $q \to 0$.
\item At small $q < k_{F}$ but larger energy $\omega \gtrsim q v_{F}$, collective modes emerge from the ELF. The FEG model gives rise to poles:
\be
\lim_{q \to 0} \mathcal{W}(\vecq, \omega) = \frac{\pi \omega_p}{2}\left(\delta(\omega - \omega_p) - \delta(\omega + \omega_p)\right)
\ee
which describes a \emph{plasmon} mode of energy $\omega_p$ where all of the valence electrons in the material are oscillating in phase at the classical plasma frequency. At finite $q$, the plasmon pole has a dispersion relation given by $\omega^{2} \approx \omega_{p}^{2} + 3q^{2}v_{F}^2/5$ and extends up to a cutoff momentum $q_c \simeq \omega_p/v_F$, beyond which \emph{Landau damping} occurs and the plasmon can efficiently decay into electron-hole pairs. In a realistic material, various decay paths for the plasmon are also available for $q < q_{c}$, so that we should use the Mermin dielectric function, Eq.~(\ref{eq:epsMermin}), with finite $\Gamma_{p}$ to represent the plasmon width, which yields at $q=0$
\be
\lim_{q \to 0} \mathcal{W}(\vecq, \omega) = \frac{\omega_p^2 \omega \Gamma_{p}}{(\omega_p^2 - \omega^2)^2 + \omega^2 \Gamma_{p}^2}.
\ee
Using the relation between the dynamic structure factor and the ELF, Eq.~(\ref{eq:Sepsilon}), we see that $S(\vecq, \omega)$ scales as $q^2 \omega$ at small $q$ and $\omega$ as claimed in Eq.~(\ref{eq:plasmonsimple}). This Breit-Wigner lineshape is the same as what would be obtained from treating the valence electrons as a damped harmonic oscillator (the \emph{Fr\"{o}hlich model}). The measured plasmon lineshapes in both semiconductors like Si and Ge \cite{kundmann1988study} and metals like Al \cite{gibbons1976line,sun2016calculations} are extremely close to this simple form. 
\end{itemize}

From the identification of the peaks of the ELF (and thus dynamic structure factor), we learn that conventional metals and semiconductors are not optimal in terms of matching the DM scattering phase space, which requires $\omega < q v_{\rm max}$. At $q > k_{F}$, the peak at $\omega \sim q^{2}/(2m_{e})$ is not accessible, since in this regime $\omega \gtrsim q k_{F}/m_{e} \sim q \alpha$. At low $q < k_{F}$, most of the weight in the ELF is carried by the plasmon which is located at $\omega > q v_{F}$. However, the DM velocity is typically much slower than the Fermi velocity in conventional materials, so that the DM cannot access the plasmon.\footnote{A fast subcomponent of DM, arising for example from supernova remnants \cite{Li:2020wyl}, may excite the plasmon directly \cite{Kurinsky:2020dpb}. DM-nucleus scattering can also excite the plasmon through a $2 \to 3$ process~\cite{Kurinsky:2020dpb,Kozaczuk:2020uzb}, as will be discussed in more detail in Sec.~\ref{sec:Migdal}.}  Since the integrated weight of the ELF is constrained by the $f$-sum rule, it is in this sense that DM-electron scattering in conventional materials is ``inefficient'' as was the case for multi-phonon excitations in superfluid helium. However, semiconductors are still better than a typical atomic target as in Fig.~\ref{fig:hydrogen_sqw}, due to the lower gaps. Projections in terms of the DM-electron cross section are shown in Fig.~\ref{fig:DMelectron_all} for example semiconductors of Si and Ge.

\subsubsection{Sub-MeV DM scattering}
 
The small-$q$ regime is especially important for sub-MeV DM, which carries maximum momentum $p_\chi < 1 \ \keV \ll k_F$. This regime is accessible in superconductors, which have a gap $\Delta \lesssim \ \mathcal{O}({\rm meV})$ and hence $q$ can be as small as $\mathcal{O}(\eV)$ for DM scattering. Ref.~\cite{Hochberg:2015pha} first pointed out the suitability of superconductors for sub-MeV DM, using the FEG model to compute scattering for $\omega \gg \Delta$. Ref.~\cite{Hochberg:2015fth} noted the reach is suppressed for a dark photon mediator because of screening effects; recently, Refs.~\cite{Hochberg:2021pkt,Knapen:2021run}, using the dielectric formalism, clarified that screening is present for all scalar and vector mediators (not just dark photons), and furthermore that the reach is considerably better than previously estimated because \cite{Hochberg:2015fth} effectively used a loss function which did not satisfy the causality property due to an incorrect choice of branch cut.\footnote{As noted in \cite{Hochberg:2021pkt}, there may also be additional contributions to the ELF from the plasmon tail, which would further improve the reach, but dedicated measurements are required in the small-$q$ regime to confirm the presence of the tail; Ref.~\cite{sun2016calculations} shows the tail in aluminum extending down to 0.1 eV.} The projected reach for different energy thresholds is given by the Al lines in Fig.~\ref{fig:DMelectron_all}.

The screening effects in a superconductor can be traced back to the behavior of the ELF in the low $q$ limit, Eq.~(\ref{eq:FEGELF}). This behavior gets cutoff in a semiconductor, since in the limit of low $q, \omega$ we should reproduce the static dielectric function which is typically $\mathcal{O}(1-10)$. We can see the finite behavior of the dielectric function in this case from the Lindhard formula~(\ref{eq:lindhard}). In the limit $q\to0$, the squared matrix element must then vanish like $q^2$ as $q \to 0$, since we are considering orthogonal states. At the same time, in a semiconductor or insulator, the Fermi level lies in the gap and the lowest-energy excitations are interband transitions between bands, so that energy differences $E_{i' \veck'} - E_{i \veck} = E_{i' \veck + \vecq} - E_{i \veck}$ should be approximately independent of $q$. The $q^{2}$ from the matrix overlap then cancels the Coulomb prefactor and yields a \emph{constant} $\epsilon$. An alternate parameterization of the dielectric function for semiconductors at low $q$ can be found in Ref.~\cite{PhysRevB.47.9892}.  Thus, large screening effects at small $q$ are absent in materials with a Fermi level lying between the valence and conduction bands, which motivates considering narrow-gap semiconductors as targets to enhance the sub-MeV DM scattering rate.

An example of such a narrow-gap semiconductor is a gapped Dirac material, so named because the valence and conduction bands have dispersion relations which approximate the relativistic dispersion, $E_{\pm}(\veck) = \sqrt{v_F^2 \veck^2 + \Delta^2}$, with the gap $2\Delta$ playing the role of (twice) the relativistic particle mass. The point in the BZ with $\veck = 0$ is known as the \emph{Dirac point}. Ref.~\cite{Hochberg:2017wce} first determined the scattering rate in Dirac materials by computing the single-particle matrix elements $\langle \veck + \vecq; + | e^{i \vecq \cdot \vecr} | \veck; - \rangle$ between the uppermost valence ($-$) and lowest conduction ($+$) bands, imposing a momentum cutoff at $|\veck| = \Lambda$ where the dispersion deviates from the relativistic form, and dividing by $|\epsilon(\vecq, \omega)|^2$ to account for screening. This is equivalent to simply computing the ELF directly, which we will do here as it is somewhat more transparent. Assuming that the $-$ and $+$ bands are the only bands which contribute to the dielectric function, $\epsilon(\vecq, \omega)$ may be computed in exact analogy with the 1-loop vacuum polarization in quantum electrodynamics, with appropriate factors of $v_F$ instead of $c$. The momentum cutoff corresponds to the familiar UV cutoff in 1-loop diagrams, and the result is
\begin{align}
\epsilon(\vecq, \omega) = 1 + &\frac{e^2}{4\pi^2 v_F}\int_0^1 dx\, \left \{ x(1-x) \ln \left |\frac{(2 v_F \Lambda)^2}{\Delta^2 - x(1-x)(\omega^2 - v_F^2 \vecq^2)}\right| \right \} \nonumber \\
& + i\frac{e^2}{24\pi \kappa v_F}\sqrt{1 - \frac{4\Delta^2}{\omega^2 - v_F^2 \vecq^2}}\left (1 + \frac{2\Delta^2}{\omega^2 - v_F^2 \vecq^2}\right)\Theta(\omega^2 - v_F^2 \vecq^2 - 4\Delta^2).
\label{eq:epsGap}
\end{align}
However, as noted in Ref.~\cite{Coskuner:2019odd}, this calculation for the real part of $\epsilon$ is unreliable because DFT calculations show that ${\rm Re} \, \epsilon$ receives contributions from the \emph{entire} BZ, not just the region around the Dirac point with $|\veck| < \Lambda$. Thus, we may approximate ${\rm Re} \, \epsilon(\vecq, \omega) \approx \epsilon(\mathbf{0}, 0) = \kappa \gg 1$ as a background static dielectric constant which should be measured or determined from DFT. The imaginary part is unaffected by this complication, so assuming ${\rm Im} \, \epsilon \ll \kappa$ (which is in fact required for the perturbation theory calculation we used to determine $\epsilon$) we may compute the ELF as \cite{Hochberg:2021pkt}
\be
\mathcal{W}(\vecq, \omega) \approx \frac{{\rm Im} \, \epsilon(\vecq, \omega)}{\kappa^2} =   \frac{e^2}{12 \kappa^2 \pi v_F}\sqrt{1 - \frac{4 \Delta^2}{\omega^2 - v_F^2 q^2}} \left(1 + \frac{2\Delta^2}{\omega^2 - v_F^2 q^2}\right) \Theta(\omega^2 - v_F^2 q^2 - 4 \Delta^2) \Theta(\omega_{\mathrm{max}} - \omega)
\label{eq:ELFDirac}
\ee
where we have replaced the momentum cutoff $\Lambda$ with an energy cutoff $\omega_{\rm max}$. In contrast to the FEG single-particle ELF (\ref{eq:FEGELF}), the Dirac ELF is unsuppressed by powers of $q$ and can thus lead to a larger scattering rate for small DM masses. In addition, scattering is forbidden for $v < v_F$, a peculiarity of the kinematics of linear dispersion. The projected reach for an example Dirac material of ZrTe$_{5}$ is shown in Fig.~\ref{fig:DMelectron_all}. Dirac materials may also feature anisotropic dispersion relations, with parametrically different Fermi velocities $v_{F,x}, v_{F,y}, v_{F,z}$ along different crystal axes, which allows for a large $\mathcal{O}(1)$ daily modulation signal for sub-MeV DM \cite{Geilhufe:2019ndy,Coskuner:2019odd} (due partly to the requirement $v > v_F$), greatly improving the detection prospects if such a detector can be realized in the laboratory. Finally, we note that plasmons are also expected to exist in Dirac materials \cite{kharzeev2015universality,hofmann2015plasmon,Jenkins_2016}; indeed, the ELF in Eq.~(\ref{eq:ELFDirac}) does not satisfy the $f$-sum rule and therefore cannot represent the entire excitation spectrum because it only contains single-particle contributions. The effect of these contributions to the ELF on the DM-electron scattering rate remains to be fully evaluated.

\section{The Migdal Effect}
\label{sec:Migdal}

The main narrative emphasized the dynamic structure factor including only phonon excitations (Sec.~\ref{sec:DMN}) or only electronic excitations (Sec.~\ref{sec:DMN}).  This is a good leading approximation, but there are regimes of phase space or situations where it is valuable to consider a {\emph{mixed}} response in a dynamic structure factor. For instance, for DM-electron scattering below the gap, phonon modes are the only available excitations, and thus DM-electron scattering may lead to phonon production \cite{Griffin:2018bjn,Trickle:2019nya}. At energies just above the band gap, the response function may be dominated by simultaneous production of an electron and phonon. This can be thought of as a $2 \to 3$ process, where an off-shell electron emits a phonon; in particular, it is distinct from possible subsequent interactions of the electron in the material, which may scatter off other electrons to produce phonons. This $2 \to 3$ process is known to be important in the limit of zero momentum transfer for indirect gap semiconductors, since the emission of phonon allows for conservation of crystal momentum. However, the energy of the phonon would generally be much smaller than that of the electronic excitation itself. 

For DM-nucleus interactions, a similar $2 \to 3$ process is possible, with a nuclear recoil and electronic excitation being produced simultaneously. In this case, the possibility of a mixed electronic and nuclear response provides a more interesting prospect for experimental detection, since the electronic excitation could have comparable energy to the energy in phonons or nuclear recoil. Furthermore, at the present time the detection thresholds for electronic excitations are also much lower than for nuclear recoils (see Sec.~\ref{sec:Detection}), which means that it is possible to indirectly observe nuclear recoils of much lower mass DM with electronic detection approaches. This idea of producing a charge excitation from a nuclear interaction was originally studied and proposed in the context of atomic targets, where it is known as the Migdal effect~\cite{Migdal1939,Migdal1941,Migdal:1977bq}. Historically, it played a role in explaining observations of radioactive decays of heavy elements, which cause the sudden recoil of the nucleus~\cite{PhysRevC.11.1746,LAW1977339,PhysRev.90.11,PhysRevLett.108.243201}.  Discussions of the Migdal effect in DM-nucleus collisions first appeared in Refs.~\cite{Vergados:2004bm,Bernabei:2007jz}, but it is only in recent years that interest has revived due to the interest in sub-GeV dark matter and the attainment of lower electronic thresholds~\cite{Ibe:2017yqa}. The majority of theoretical studies have treated the target material as consisting of individual atomic targets~\cite{Ibe:2017yqa,Dolan:2017xbu,Bell:2019egg,Baxter:2019pnz,Essig:2019xkx,Liu:2020pat,GrillidiCortona:2020owp,Bell:2021zkr}, similar to the original derivation by Migdal, while recently the effect has also been generalized to condensed matter targets~\cite{Kozaczuk:2020uzb,Knapen:2020aky,Liang:2020ryg}.  In principle, if the response function for nuclear recoils is obtained from experimental data, it would contain such contributions, and we could similarly determine the size of the Migdal effect for DM direct detection. However, similar to the other cases discussed in this review, DM scattering kinematics are quite different from where the Migdal effect was originally studied, and so far there are no direct measurements in the desired regimes. 

\begin{figure*}[t!]
\begin{center}
\includegraphics[width=0.49\textwidth]{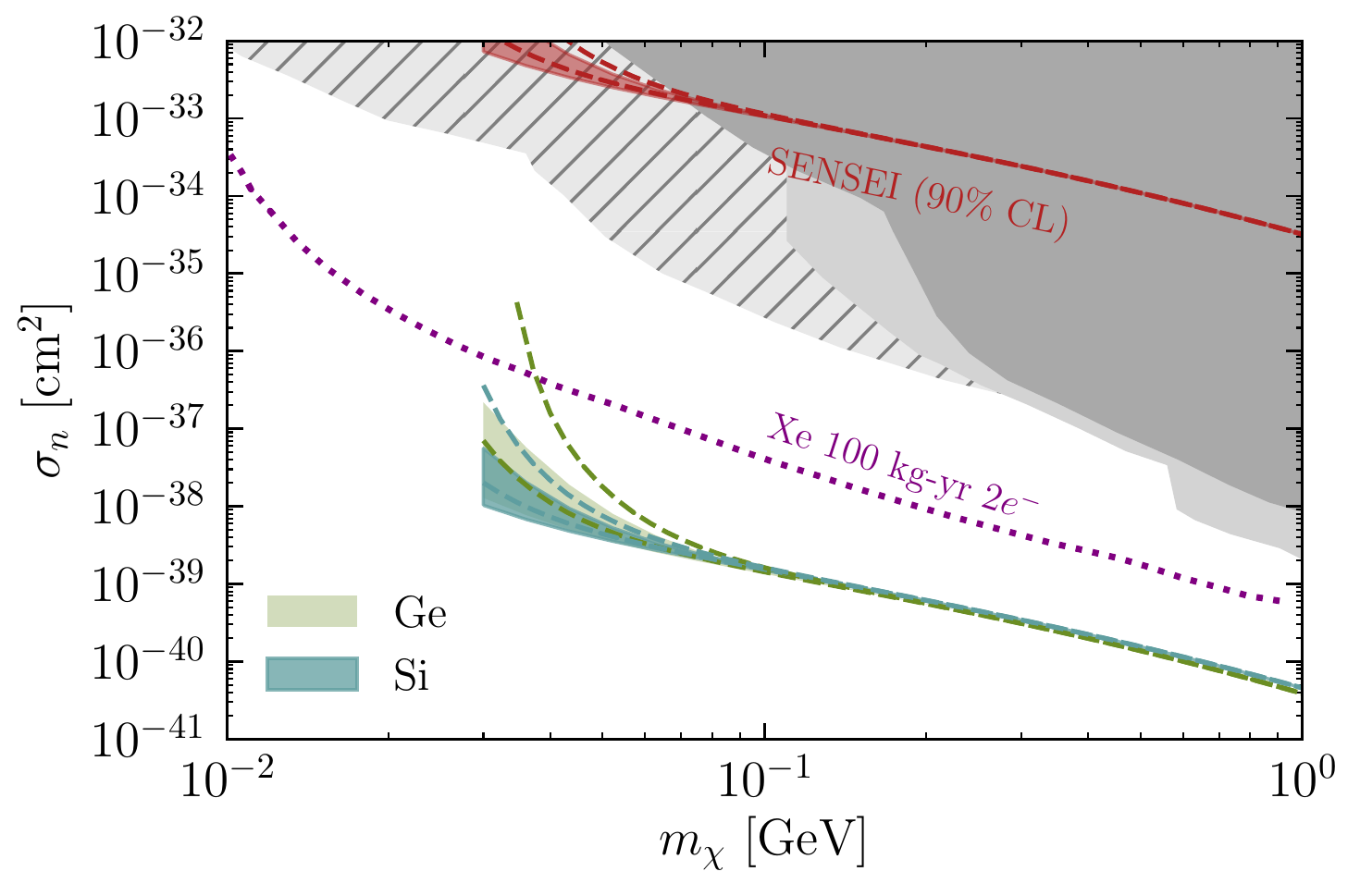}
\includegraphics[width=0.45\textwidth]{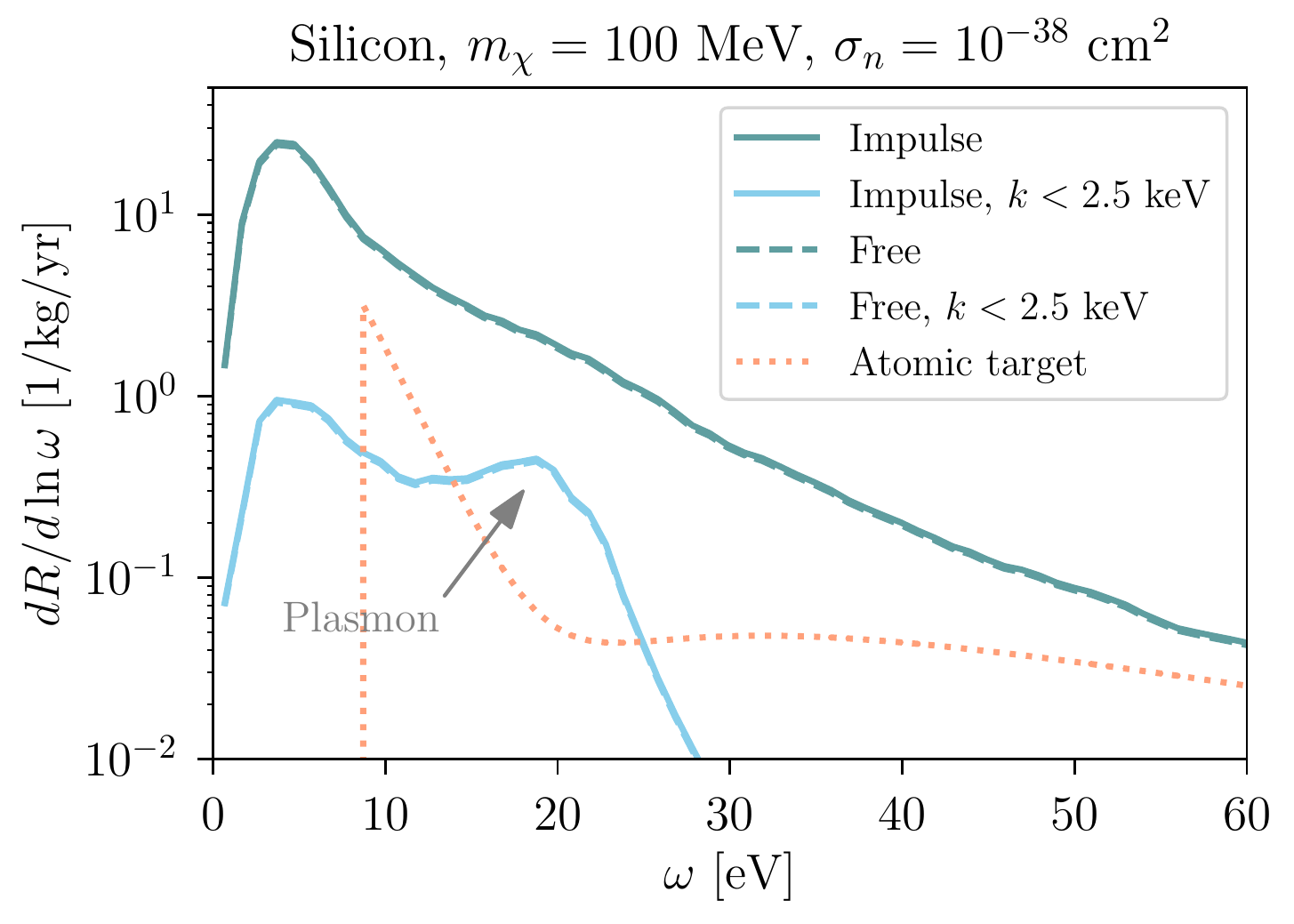}
\caption{ \label{fig:Migdal} (\emph{left}) The dark gray shaded regions are combined nuclear recoil bounds from XENON1T~\cite{XENON:2019zpr}, LUX~\cite{LUX:2018akb}, CRESST III~\cite{Abdelhameed:2019hmk} and CDEX~\cite{Liu:2019kzq}. The light gray region is a XENON1T limit using the Migdal effect~\cite{XENON:2019zpr} (see also LUX limits~\cite{LUX:2018akb}), and the hatched region is a recasted XENON limit in terms of the Migdal effect by Ref.~\cite{Essig:2019xkx}.   Si and Ge lines are expected 90\% CL sensitivity to DM--nucleon cross section $\sigma_n$ due to the Migdal effect, assuming a heavy mediator, 1 kg-year of exposure, and minimum 2 electrons observed. The shaded bands are an estimate of the theoretical uncertainty due to the impulse approximation, see Ref.~\cite{Knapen:2020aky} for details. The red line is a 90\% CL limit obtained using the recent upper limit on the 2-electron rate from SENSEI~\cite{Barak:2020fql}. For comparison, the dotted line is a  projection for the atomic Migdal effect in Xenon (dotted line) from Ref.~\cite{Essig:2019xkx}. 
(\emph{right}) Comparison of the spectrum of ionizations from the Migdal effect in atomic Si~\cite{Ibe:2017yqa} with the spectrum of electron excitations due to the Migdal effect in a Si semiconductor. Both plots are adapted or reproduced from Ref.~\cite{Knapen:2020aky}.
}
\end{center}
\end{figure*}

In this section, we will review the Migdal effect, as applied to direct detection of DM-nucleus interactions. As shown in Fig.~\ref{fig:Migdal}, initial applications of the Migdal effect have already given the leading constraints of nuclear recoils to sub-GeV dark matter compared with traditional searches for nuclear recoils (dark gray region). We will begin with a treatment of the Migdal effect in atomic targets to reflect the approach by Migdal and in the majority of studies. The argument of Migdal, however, does not apply when treating solid state targets. We therefore give several alternate ways to understand the Migdal effect, and then discuss its generalization to solid state materials.

\subsection{Atomic Migdal effect}

For a DM-nucleon interaction given by Eq.~(\ref{eq:Hnuc_contact}), recall that we obtained a nuclear recoil structure factor given by
\be
	S(\vecq, \omega) = \frac{ 2 \pi N_{\rm nuc}}{V} A^{2} \, \delta \! \left(\omega - \frac{q^2}{2 m_{N}} \right),
\ee
where we can take $F_{N}(q) \to 1$ in Eq.~(\ref{eq:Sqw_elasticNR}) for sub-GeV DM.
For the atomic Migdal effect, the dynamic structure factor contains an additional contribution that can be written as
\be
	\frac{dS^{\rm Migdal}(\vecq, \omega)}{d\omega_e} \approx \frac{ 2 \pi N_{\rm nuc}}{V} A^{2} \, \delta \! \left(\omega - \frac{q^2}{2 m_{N}} - \omega_e \right) \frac{dP}{d\omega_e}
\ee
where  $dP/d\omega_{e}$ is a differential probability for energy $\omega_{e}$ to be deposited into an electron excitation. This factorized form of the response function is approximate, because it assumes that any momentum transferred to the electron excitation is much smaller than the total momentum transfer $\vecq$. However, this will easily be valid for nuclear recoils with energy 10 meV or above, since $q = \sqrt{2 m_{N} E_{R}}$ will be much larger than the typical momentum of electron excitations, $p_0 = \alpha m_{e}$. Then $dP/d\omega_{e}$ depends only on the nuclear recoil energy $E_R = q^2/(2 m_N)$. We will clarify the validity of this factorized form further below.

Migdal gave an elegant argument to obtain $dP/d\omega_e$ by treating the nuclear recoil as ``sudden'' (fast compared to response of electronic orbitals). For a sudden collision, the original electron wavefunction has not been perturbed at all. However, after the collision, the asymptotic eigenstates are given by atomic wavefunctions with respect to the moving nucleus. Migdal thus obtains the ionization probability by finding the overlap of the original state with excited or ionized states about the moving nucleus. In particular, if the initial state is given by the ground state $|i \rangle$, then after the collision the wavefunction is given by applying a boost operation to obtain $e^{i m_{e} \vecv_{N} \cdot \sum_{\beta} \vecr_{\beta}} |i \rangle$, with $\vecv_{N} = \vecq/m_{N}$ and $\vecr_{\beta}$ the electron position operators. Then $dP/d\omega_e$ can be written as
\be
 	\frac{dP}{d\omega_e} =  \sum_{f} | \langle f | e^{i m_{e} \vecv_{N} \cdot \sum_{\beta} \vecr_{\beta}} |i \rangle |^{2} \delta(E_{f} - E_{i} - \omega_{e}).
\ee
This resembles the general form for the dynamic structure factor~(\ref{eq:Sqw_electron}), but the interpretation is quite different, and in particular the sum over electron coordinates appears inside rather than outside the phase factor. Writing the electron state as a Slater determinant of electron orbitals, and considering just single electron ionizations, the wavefunction overlap can be simplified to $\langle k_{e} \ell' m' | e^{i m_{e} \vecv_{N} \cdot \vecr} | n \ell m \rangle$. For an atomic target where it is assumed that ionizations can be detected, the outgoing states are unbound states with energy $E_{\rm er} = k_{e}^2/(2 m_{e})$, and normalized such that $\langle k_{e}' \ell' m' | k_{e}\ell m \rangle = 2 \pi \delta(k_e' - k_e) \delta_{\ell \ell'} \delta_{m m'}$ (note that this differs from the chosen normalization for atomic ionization in Sec.~\ref{sec:atomic}). This leads to the following form of the ionization probability, where we have included a factor of 2 to sum over spin and assumed degenerate spin states for simplicity:
\begin{align}
	\frac{dP}{dE_{\rm er}} &= 2 \sum_{n,\ell,m, \ell', m'} \int \frac{d k_e}{ 2\pi} |\langle k_{e} \ell' m' | e^{i m_{e}\vecv_{N} \cdot \vecr} | n \ell m \rangle|^2 \delta(E_{\rm er} - E_{n\ell} - \omega_{e}) \\
	& =  \frac{k_e}{2 \pi E_{\rm er}} \sum_{n,\ell,m, \ell', m'} |\langle k_{e} \ell' m' | e^{i m_{e} \vecv_{N} \cdot \vecr} | n \ell m \rangle|^2.
\end{align}
Here we have written the differential probability in terms of outgoing energy $E_{\rm er}$, rather than the total deposited energy $E_{\rm er} + |E_{n\ell}|$, in order to match the convention in some of the literature and to indicate this is an ionization probability. This form can be simplified further if the wavefunctions are written as $R_{n\ell}(r) Y_{\ell}^m$ and $\widetilde{R}_{k_e \ell'}(r) Y_{\ell'}^{m'}$  for the initial and final orbitals, respectively, giving the form of the differential probability in Ref.~\cite{Essig:2019xkx}. 

For sub-GeV dark matter, the nuclear recoil velocity is slow,  $v_{N} \ll \alpha$, so that $m_{e} v_{N} \ll p_0$. As a result, a very good approximation is to take the leading term in the boost operator which gives rise to transitions, with
\be
	\frac{dP}{dE_{\rm er}} \approx \frac{ 2 m_e^2 E_R}{m_N} \times  \frac{k_e}{2 \pi E_{\rm er}}  \sum_{n,\ell,m, \ell', m'} \left| \hat \vecv_{N} \cdot \langle k_{e} \ell' m' | \vecr | n \ell m \rangle \right|^2
	\label{eq:atomicMigdal_dipoleform}
\ee 
where we have pulled out the magnitude of $v_{N}$ from the matrix element squared and written $v_N^2$ in terms of $E_R$. From this form of the differential ionization probability, we see that the leading contribution comes from dipole transition matrix elements, and that overall the rate scales with $E_{R}$, at least in the rate of small enough recoil energy. Writing the matrix element as a dipole transition allows one to determine the Migdal rate in terms of an experimentally measured photoabsorption cross section~\cite{Liu:2020pat}. This form of the matrix element is therefore reminiscent of the probability for bremsstrahlung from a recoiling charged source, with the radiated field converted or absorbed into an electronic excitation. We will give an alternate derivation of the Migdal effect in terms of this interpretation shortly. 

The wavefunction overlap computed by Migdal made two key assumptions: a ``sudden'' collision, and that the asymptotic eigenstates of the electrons are obtained by boosting the wavefunctions to the frame of the recoiling nucleus.  The first assumption is not necessarily true for all collisions, and one may question whether it applies for all DM interactions, in particular long-range interactions where the momentum transfer may be slow. The second assumption breaks down for solid state targets, where there is a preferred frame of reference given by the center of mass of the crystal, and the valence electron eigenstates are defined with respect to that preferred frame. Given these limitations, we give two alternate ways to derive the Migdal effect.

The hydrogen atom provides a useful toy model where we calculate the ionization rate from the Migdal effect without any semiclassical or sudden approximation. Together with the DM, this forms a three-body system which can be solved for with a quantum mechanical treatment~\cite{Baur_1983}. First, recall that the two-body Hamiltonian for the hydrogen atom can be written in terms of a center-of-mass coordinates $\vecr_{CM} = (m_{N} \vecr_N + m_{e} \vecr_{e})/(m_{N} + m_{e})$ and relative coordinates $\vecr \equiv \vecr_{e} - \vecr_{N}$ with conjugate momenta $\vecp_{CM}$ and $\vecp$, respectively. Including also the DM kinetic energy and DM-nucleus interaction potential, the total Hamiltonian is given by
\be
	H = \frac{\vecp_{CM}^2}{2 (m_N + m_e)} + \frac{\vecp^2}{2 \mu_{e N} } - \frac{\alpha_{em}}{| \vecr |} + \frac{ \vecp_\chi^2}{2 m_\chi} + \mathcal{V}_{\chi N}(\vecr_\chi - \vecr_{N}).
\ee
Treating $\mathcal{V}_{\chi N}$ as a perturbation, the eigenstates are Fock states $| \vecp_{CM} \rangle \otimes | n\ell m \rangle \otimes | \vecp_{\chi} \rangle$ for bound states or $| \vecp_{CM} \rangle \otimes | k_e \ell m \rangle \otimes | \vecp_{\chi} \rangle$ for continuum states, where the DM and center-of-mass pieces are just plane wave states. We can treat the initial state as having $\vecp_{CM} = 0$ in the atomic ground state, and final state with $\vecp_{CM}  \approx \vecq_{N} = m_{N} \vecv_{N}$ in an excited state. 

To calculate the matrix element for the DM-nucleus interaction, we rewrite the interaction potential so that it is written in terms of CM and relative coordinates:
\be
	\langle \Psi_{f} | \mathcal{V}_{\chi N}(\vecr_\chi - \vecr_{N}) | \Psi_{i} \rangle 
	= \langle \Psi_{f} | \mathcal{V}_{\chi N}\left(\vecr_\chi - \vecr_{CM} + \frac{m_e}{m_{N} + m_{e}} \vecr \right) | \Psi_{i} \rangle.
\ee
Writing the interaction potential $\mathcal{V}_{\chi N}$ in terms of its Fourier transform,
\be
	\mathcal{V}_{\chi N}(\vecR) = \int \frac{d^3 \vecq}{(2\pi)^3}  \mathcal{V}_{\chi N}(\vecq) e^{i \vecq \cdot \vecR},
\ee
where $\vecR = \vecr_\chi - \vecr_{CM} + \frac{m_e}{m_{N} + m_{e}} \vecr$, we can now separately evaluate the matrix element for each piece of the Fock state. This gives
\be
	M_{if} \approx -i \frac{(2 \pi)^{3}}{V^2} \delta\left(\vecp^{\chi}_{f} - \vecp^{\chi}_{i} - \vecq_{N} \right)  \mathcal{V}_{\chi N}(\vecq_{N})  \times \langle k_{e} \ell m | e^{i m_{e} \vecv_{N} \cdot \vecr} | 1 0 0 \rangle
\ee
where in the last line we have used $m_{e} \vecr /(m_{N}+ m_{e}) \approx m_{e} \vecr/m_{N}$.

The first factor of this matrix element will precisely give the DM-nucleus scattering rate, while the second factor matches the wavefunction overlap derived by Migdal. Applying this argument to multi-electron atoms becomes more cumbersome as one must then use a generalized coordinate system of relative coordinates, which in general introduces additional corrections at higher order in $m_e/m_N$. However, at leading order, which gives the dipole matrix element in Eq.~(\ref{eq:atomicMigdal_dipoleform}), the result remains the same. The same ionization probability can thus be obtained without the sudden approximation or boosting argument in the dipole limit for a general atom.

It remains the case that the change of coordinates to relative coordinates applies only to atomic targets. Our final approach to the atomic Migdal effect relies on a semiclassical model, but one which generalizes more readily to a condensed matter treatment. Our discussion will closely follow Ref.~\cite{Knapen:2020aky}, although the same idea was applied in earlier studies of the Migdal effect for nuclear decays~\cite{LAW1977339,PhysRev.90.11}.  In this approach, the nucleus motion is treated with a classical trajectory $\vecr_{N}(t) = \theta(t) \vecv_{N} t$, similar to the original sudden approximation by Migdal. However, rather than considering the eigenstates with respect to the moving nucleus, we treat the sudden relative motion of the nucleus as a perturbation to the electron Hamiltonian, by splitting up the Hamiltonian as:
\be
	 H = H_{0} + H_{1}(t) = H_{0} -  \sum_{\beta} \frac{Z_{N} \alpha}{| \vecr_\beta - \vecr_{N}(t)|} +  \sum_{i} \frac{Z_{N} \alpha}{| \vecr_\beta |}
\ee
where $H_{0}$ is the time-independent Hamiltonian with the nucleus at the origin and $\vecr_\beta$ are the electron coordinates. The perturbation $H_{1}(t)$ therefore involves the difference of the full time-dependent nucleus potential and the equilibrium nucleus potential, where $Z_N$ is the charge of the nucleus.

At small $t$, the potential can be expanded about small $\vecr_{N}(t)$, yielding a dipole potential for a recoiling nucleus with dipole moment $Z_N \vecr_N(t)$:
\begin{align}
	H_1(t) \approx - \sum_{\beta} \frac{Z_N \alpha \, \hat \vecr_\beta \cdot \vecr_{N}(t)}{\vecr_\beta^2}.
\end{align}
Assuming that ionization primarily occurs at early times when the perturbation is small (since the recoiling nucleus moves very slowly), to compute the probability we include a damping factor $e^{- \eta t}$ in $H_{1}(t)$, where $\eta$ is a small positive number. Then the transition probability
between initial and final electron eigenstates $|i \rangle, |f \rangle$ is given in time-dependent perturbation theory as \cite{landau2013quantum}
\begin{align}
    P_{i \to f} = \left| \frac{1}{\omega_{fi}} \int_0^\infty dt\, e^{i (\omega_{fi} + i \eta) t}  \langle f | \frac{dH_1(t)}{dt} | i \rangle \right|^2 = \Big| \langle f | \frac{1}{\omega_{fi}^2}\sum_{\beta} \frac{Z_{N} \alpha \, \hat \vecr_\beta \cdot \vecv_N}{\vecr_\beta^2} | i \rangle \Big|^2
    \label{eq:transitionprob}
\end{align}
with $\omega_{fi} = E_f - E_i$, and where we took $\eta \to 0$ in the last equality. To see the connection with the wavefunction overlaps derived previously, we will utilize the relationship between the dipole transition matrix element and the dipole potential itself. First, we use the standard relationship for the dipole transition matrix element, generalized to a many-electron system,
\begin{align}
	\langle f | \vecr_{\beta} | i \rangle  =\frac{-i}{m_e \omega}\langle f| \vecp_\beta | i \rangle, 
\end{align}
which comes from writing $\vecp_{\beta} = i  m_{e} [H_{0}, \vecr_{\beta}]$. The matrix element of $ \vecp_\beta$ between eigenstates can further be rewritten as $ - [\vecp_\beta, H_{0}] /\omega$. Altogether,
\begin{align}
	\langle f |  \sum_\beta \vecr_\beta | i \rangle  &= \frac{ i}{m_e \omega^{2}} \langle f | \sum_\beta [\vecp_\beta, H_{0}] | i \rangle =  \frac{1}{m_e \omega^2}  \langle f|\sum_\beta \frac{ Z_N \alpha \, \hat \vecr_\beta}{\vecr_\beta^2} | i \rangle 
	\label{eq:dipoleforce_identity}
\end{align}
The second equality follows if we assume that $H_{0}$ consists of electron kinetic terms, Coulomb interactions between electrons, and the Coulomb potential of the nucleus. Because $\sum_\beta [\vecp_\beta, H_{0}]$ is proportional to the total force on all of the electrons, the interactions between electrons drop out and we are left only with the force due to the nucleus. Using Eq.~(\ref{eq:dipoleforce_identity}) in Eq.~(\ref{eq:transitionprob}), it follows that the transition probability can be written as $P_{i \to f} = | m_{e} \langle f | \vecv_{N} \cdot \sum_{\beta} \vecr_{\beta} | i \rangle |^{2}$. This is the same wavefunction overlap derived by Migdal, in the dipole limit where $v_{N} \ll \alpha$. Thus we arrive at an alternate interpretation of the Migdal effect, which is that the recoiling nucleus induces a dipole potential and therefore dipole transitions of the electronic states. In this sense, the Migdal effect is analogous to bremsstrahlung, except instead of producing transverse radiation, the dipole field induces ionizations. Note that radiation of photons from the nucleus does occur as well, and this process has been studied in Refs.~\cite{Kouvaris:2016afs,McCabe:2017rln,Kobayashi:2018jky,Bell:2019egg,XENON:2019zpr,LUX:2018akb}. However, the rate is much smaller than the Migdal effect and it generally leads to weaker direct detection constraints, and so we do not discuss it further. 

This final interpretation of the Migdal effect has the advantage that the same idea can be applied in a crystal target by considering the perturbation due to a single recoiling nucleus in a crystal. Because in a crystal target, the electron states can be modeled as Bloch states with crystal momentum $\vecp_{e}$, it is useful to rewrite Eq.~(\ref{eq:transitionprob}) as a differential probability in Fourier space:
\be
    \frac{dP}{d\omega_{e}}  \approx \left(\frac{4 \pi Z_N \alpha}{\omega_{e}^2} \right)^2 \sum_{i,f}    \Bigg|  \int\!\! \frac{d^3 \veck}{(2\pi)^3} \frac{\vecv_N \cdot \veck}{k^2} \, \langle f | e^{i \veck \cdot \vecr} | i \rangle \Bigg|^2 \,  \delta \left(\omega_{fi} - \omega  \right) .
    \label{eq:atomic_fourier}
\ee
As before, we have assumed single electron ionizations and that the wavefunction overlap can be computed with single electron orbitals.

\subsection{Migdal effect in crystal targets}

The interpretation and result of Eq.~(\ref{eq:atomic_fourier}) provides a way to generalize the Migdal effect to crystal targets, which was first discussed in Ref.~\cite{Knapen:2020aky}. In the atomic case, we assumed initial and final states are many-body states that could be build up from single-particle orbitals, and that the wavefunction overlap was computed with those orbitals. For the case of a crystal, we will treat the excitations of the valence electrons with the single-particle picture discussed in Sec.~\ref{sec:CMexcitations}, with wavefunctions given by Bloch states, Eq.~(\ref{eq:Bloch}). In this single-particle picture, the interaction between the nucleus and a valence electron is screened due to two effects: by the bound inner-shell electrons, and by all the other valence electrons in the material. The first effect can be accounted for by modeling the nucleus charge as a momentum-dependent charge $Z_{\rm ion}(k)$, which asymptotes to the nucleus plus inner-shell electron charge as $k \to 0$; this can be obtained from calculations of ionic form factors. The effect of all the other valence electrons can be accounted for by a factor of $1/\epsilon(\veck, \omega)$ in the potential, with $\epsilon(\veck, \omega)$ the longitudinal dielectric function.  Another key difference is that for atomic targets, we assumed only ionizations could be observed but not excitations to bound states, while in crystal targets it is possible to observe all electron excitations. However, we will still generally refer to the Migdal effect in crystals by an ionization probability. With these important differences in mind, we can now directly evaluate Eq.~(\ref{eq:atomic_fourier}) with Bloch states.

We will denote initial and final Bloch states with momenta $\vecp_{e}$ and $\vecp_{e}'$, suppressing the band index for simplicity. Because of the periodicity of the Bloch functions $u(\vecr)$, the wavefunction overlap can be split up as an integral over unit cell volume $\Omega$ and a sum over lattice vectors $\vecR_{n}$:
\begin{align}
	\frac{1}{V} \int d^{3} \vecr \, u^{*}_{\vecp_{e}'}(\vecr) u_{\vecp_{e}}(\vecr) e^{i (\vecp_{e} - \vecp_{e}' + \veck) \cdot \vecr } &= 
	\frac{1}{N} \sum_{\vecR_{n}} e^{i (\vecp_{e} - \vecp_{e}' + \veck) \cdot \vecR_{n} } \times \frac{1}{\Omega} \int_{\Omega} d^{3} \vecr \, u^{*}_{\vecp_{e}'}(\vecr) u_{\vecp_{e}}(\vecr) e^{i (\vecp_{e} - \vecp_{e}' + \veck) \cdot \vecr }   \\
 &= \sum_{\vecG} \delta_{\vecp_{e} - \vecp_{e}' + \veck, \vecG} \, [ \vecp_{e}' | e^{i \veck \cdot \vecr} | \vecp_{e} ]_{\Omega} \end{align}
where we used that $\sum_{\vecR_{n}} e^{i (\vecp_{e} - \vecp_{e}' + \veck) \cdot \vecR_{n} } = N \sum_{\vecG} \delta_{\vecp_{e} - \vecp_{e}' + \veck, \vecG}$ with $N$ the number of unit cells and $\vecG$ reciprocal lattice vectors. In the last line, $[ \vecp_{e}' | e^{i \veck \cdot \vecr} | \vecp_{e} ]_{\Omega}$ is shorthand for the wavefunction overlap computed over the unit cell. Taking $\veck$ to be small, such that we can neglect the sum over $\vecG$ for simplicity, the matrix element simplifies to $  \delta_{ \vecp_{e}',\vecp_{e} + \veck} [  \vecp_{e}'| e^{i \veck \cdot \vecr} | \vecp_{e} ]_{\Omega} $ with conservation of crystal momentum.

To evaluate Eq.~(\ref{eq:atomic_fourier}), we can sum over initial occupied states $\vecp_{e}$ and available final states $\vecp_{e}'$, leaving the occupation numbers factors implicit. Discrete sums and delta functions can be interchanged with continuous ones as $\sum_{\veck} \leftrightarrow \frac{V d^{3} \veck}{(2\pi)^{3}} $ and $\delta_{\veck} \leftrightarrow (2\pi)^{3}/V \times \delta(\veck)$. Then the ionization probability in a crystal is given by
\begin{align}\frac{dP}{d\omega_{e}}  &\approx \left(\frac{4 \pi \alpha}{\omega_{e}^2} \right)^2 \sum_{\vecp_{e}}    \int\!\! \frac{d^3 \veck}{(2\pi)^3} Z_{\rm ion}(k)^{2} \, \frac{|\vecv_N \cdot \veck|^{2}}{k^4} \, \frac{ \left| [  \vecp_{e} + \veck| e^{i \veck \cdot \vecr} | \vecp_{e} ]_{\Omega} \right|^{2} }{V |\epsilon(\veck, \omega)|^{2} } \,  \delta \left(E_{\vecp_{e} + \veck} - E_{\vecp_{e}}- \omega  \right)  \nonumber \\
    &\approx  \frac{4 \alpha}{\omega_{e}^{4}}  \int\!\! \frac{d^3 \veck}{(2\pi)^3} Z_{\rm ion}(k)^{2}  \frac{|\vecv_N \cdot \veck|^{2}}{k^2} {\rm Im} \left( \frac{-1}{\epsilon(\veck, \omega)} \right).
    \label{eq:crystal_fourier}
\end{align}
Instead of first integrating over $k$ and then squaring the amplitude, the momentum conservation delta function leads to an integral over the amplitude squared.  In the second line, we rewrote the ionization probability in terms of $ {\rm Im}(-1/\epsilon(\veck, \omega))$ using the relationship in Eq.~(\ref{eq:lindhard_imeps}). The form of the ionization probability in the second line can be interpreted in terms of the \emph{longitudinal} energy loss rate of the ion in a material, which depends on the energy loss function ${\rm Im}(-1/\epsilon(\veck, \omega))$. Again, the result is analogous to bremsstrahlung to transverse radiation (photons), but here we are considering the longitudinal (Coulomb) field which leads to electron excitations~\cite{Kozaczuk:2020uzb}. Comparing to the discussion in Sec.~\ref{sec:dielectric}, we also see that plasmon emission is possible, since it appears as a resonance in the loss function.

While the starting point here was a semiclassical approach to the atomic Migdal effect, Refs.~\cite{Knapen:2020aky,Kozaczuk:2020uzb,Liang:2020ryg} calculated the ionization probability more generally by considering the $2 \to 3$ process of DM + N $\to$ DM + N + $e^{-}$ with second-order perturbation theory in quantum mechanics. One arrives at the same result of Eq.~(\ref{eq:crystal_fourier}) in the ``soft'' limit, when $k \ll q_{N}$ and $|\vecq_{N} \cdot \veck | \ll m_{N} \omega_{e}$~\cite{Knapen:2020aky}. This is similar to the limit of ``soft'' bremsstrahlung radiation. The full quantum mechanical result  gives a more general version of the Migdal effect in crystals, although in practice using Eq.~(\ref{eq:crystal_fourier}) is a very good approximation for the DM mass range studied in Ref.~\cite{Knapen:2020aky}, 70 MeV -- 1 GeV. In particular, this mass range is dictated by where we can approximately treat the nucleus as a free particle, and avoiding dealing directly with multiphonon states. The treatment of Ref.~\cite{Knapen:2020aky} accounts for the fact that the initial ion is in the ground state of an approximately harmonic potential, but relies on an impulse approximation to treat the recoiling ion wavefunction as a plane wave. The impulse approximation is valid as long as the DM collision occurs on short time scales. Then the ion remains near the minimum of the potential during the collision, we can approximate the ion as a plane wave when its energy is much greater than the harmonic frequency $\omega_{0}$. The physics is very similar to that of Sec.~\ref{sec:SHOToy}: at energies well above $\omega_{0}$, the nuclear response is highly peaked at the free recoil energy, with some spread due to the multiphonon response. At $m_{\chi} \gtrsim 70$, this spread in the response is negligible. 

Compared to atomic targets of the same element, the rate for the Migdal effect in semiconductors is found to be much larger due to the strong $1/\omega_{e}^{4}$ dependence and lower band gap~\cite{Knapen:2020aky,Liang:2020ryg}. This makes crystal targets with low charge threshold particularly attractive in searching for DM-nucleus interactions. Previous attempts to estimate the Migdal effect in semiconductors tended to underestimate the ionization probability at $\omega_{e}$. For instance, Ref.~\cite{Liang:2019nnx} first attempted to treat the Migdal effect in crystals by using atom-centered Wannier functions in a tight-binding model, but noted the limitations when Wannier functions have too large of a spread, such as for Si or Ge semiconductors. In Ref.~\cite{Liang:2020ryg}, the same authors instead pursued the approach in terms of bremsstrahlung given in Eq.~(\ref{eq:crystal_fourier}), finding much larger rates at low $\omega_{e}$. 

Because the ionization probability depends on the same energy loss function as for DM-electron scattering, we expect similar benefits in going from atomic to condensed matter systems as for DM-electron scattering, although the Migdal ionization and DM-electron scattering rates are weighted towards somewhat different regions in $\omega, \veck$. As shown in \cite{Baxter:2019pnz,Essig:2019xkx}, for the atomic Migdal effect, the relationship between the Migdal and electron ionization rates is schematically
\be
\label{eq:MigdalCompare}
\frac{dR_M/d\vecq}{dR_e/d\vecq} >  Z^2  \left(\frac{m_e}{m_N}\right)^2 (\vecq r_a)^2,
\ee
where $r_a \simeq \mathcal{O}(a_0)$ is an effective atomic radius. The Migdal spectrum is therefore dominated by larger momentum transfers, and the total rate can exceed the rate for electron scattering in the dark photon model for large-$Z$ atoms, while the spectrum is peaked at lower energies, consistent with the behavior seen in crystals. While so far there are only a few studies of the Migdal effect beyond atomic targets, they make a strong case for further study of the Migdal effect in condensed matter systems and considering such higher order effects in response functions more generally.  Aside from Si and Ge, shown in Fig.~\ref{fig:Migdal}, calculations of the Migdal effect can be found for Diamond and Si in Ref.~\cite{Liang:2020ryg} and for a limited set of other target materials in Ref.~\cite{Knapen:2021bwg}.

\section{Experimental Techniques}
\label{sec:Detection}

Since the first theoretical proposals for sub-GeV DM detection, there have been numerous experimental results, both from existing experiments originally designed to look for heavier DM and more recently from dedicated experiments designed specifically to search for light DM. Here we summarize the main experimental techniques which can be harnessed for sub-GeV DM detection, focusing on the end-stage signal in the detector rather than the primary interaction between DM and the detector constituents. Indeed, the relationship between the primary interaction and the detector signal is highly nontrivial, and dedicated calculations and/or measurements of the charge multiplicity per electronic energy deposited or the phonon lifetime are still needed. Very broadly speaking, the signals appear as charge, light, or heat, often with one signal transduced into another during the detector operation.\footnote{Other signals, like magnetic fields, are more relevant for ultralight axion dark matter, but are occasionally useful for DM scattering, and we will briefly mention those signals that have been proposed in the literature.} It is worth pointing out that, because the deposited energy (eV scale or below) is so small, and the event rate so low, there is often only one quantum of excitation (electron, phonon, or photon for charge, heat, and light detection, respectively) present in the detector at a time. In this sense, light DM detectors are part of the emerging field of \emph{quantum sensing,} though most detection schemes thus far proposed have not exploited the quantum entanglement that is the usual hallmark of quantum detectors. In this section we focus primarily on experiments that are currently operational, leaving a discussion of future experiments for Sec.~\ref{sec:futureexp}.

\subsection{Charge}

DM interactions may produce mobile charge carriers in a detector, which may then be collected in a charge sensor: for simplicity, we define ``charge detectors'' to operate with the assistance of a static electric field or voltage sensor to move the charges to a desired location and/or count units of charge by measuring the static Coulomb potential. The charges may be produced as a result of direct ionization from DM-electron scattering, or as secondary excitations from a primary DM-nuclear scattering event as in the Migdal effect. In a semiconductor detector, mobile charges are the ones which have been excited from the valence band to the conduction band, while in a noble liquid detector, mobile charges must be ionized from their parent atom. In both cases, electric fields can be applied to drift the charges across the detector before they recombine and leave the detector in a charge-neutral state. In a metal (and in a superconductor in particular), there is a large density of mobile charge at the Fermi surface even when the detector is in its ground state, so charge detection as we have defined it above is not feasible; rather, quasiparticles (even if charged) must be detected as heat, which we discuss below.

Consider a charge-coupled device (CCD) detector, of which DAMIC and SENSEI are two examples, using devices made from silicon. The CCD is a thin sheet of silicon segmented into $\sim 10^7$ pixels of area $A_{\rm pixel} = (15 \ \mu{\rm m})^2$, and when an electron is excited to the valence band in the bulk of a pixel, it is drifted $\sim  700 \ \mu$m across an $\mathcal{O}(50)$ V potential to the surface. Various voltage pulses can move charge across the surface, first to a ``serial register'' which collects all the charge in a row of pixels, then to a ``sense node'' which measures the total charge collected by comparing the voltage in the sense node (proportional to the number of electrons) to a floating reference voltage. In traditional CCDs used by DAMIC at SNOLAB \cite{DAMIC:2019dcn}, this charge is measured once, leading to a typical noise of about $2e^-$ per pixel. With ``Skipper CCDs'' used by SENSEI, one can perform $N \gg 1 $ non-destructive measurements of the same pixel, leading to a noise which scales as $\sigma_e \propto 1/\sqrt{N}$ and which can be made arbitrarily small in principle \cite{Tiffenberg:2017aac}. In practice, readout time considerations favor $N \simeq 10^3$ yielding a noise of less than 0.1 $e^-$/pixel. This makes it possible to truly count integer numbers of charges in each pixel. The energy threshold for exciting electrons in a silicon CCD is equal to the band gap of silicon, 1.2 eV. The ``single-electron bin'' (\textit{i.e.} the number of pixels with exactly one electron) tends to be polluted by a \emph{dark rate} consisting mostly of charge leakage across the band gap \cite{SENSEI:2021hcn}, with an important secondary contribution from Cherenkov photons generated by high-energy charged particles such as cosmic ray muons \cite{Du:2020ldo}. Therefore, the strongest sensitivity to DM comes from the two-electron bin, which is also where the rate is expected to peak from scattering through a heavy mediator. While nuclear recoil searches
with CCDs are currently obtained with higher thresholds~\cite{DAMIC:2020cut}, the inclusion of the Migdal effect
could be used to extend the sensitivity of such searches to light dark
matter, especially with the inclusion of Skipper amplifiers. The current state-of-the-art is a SENSEI detector with a total mass of about 2 g, which has demonstrated background-free operation for 3 or more $e^-$/pixel in a 24-day run \cite{Barak:2020fql} (superseding previous test runs at the surface \cite{Crisler:2018gci} and underground \cite{Abramoff:2019dfb}). The next phase of the DAMIC experiment, DAMIC-M, will scale to a 1 kg array of Skipper
CCDs~\cite{Settimo:2020cbq}; the Oscura experiment is expected to merge the efforts of DAMIC and SENSEI to scale up to a total CCD mass of 10 kg \cite{Aguilar-Arevalo:2022kqd}.

An alternative charge detection strategy in semiconductors is to exploit the conversion between charge and heat when an electron is drifted through the conduction band at a high voltage. The \emph{Neganov-Trofimov-Luke effect}, the emission of phonons from a high-velocity electron, acts as a charge amplifier and yields a phonon signal proportional to the number of conduction-band electrons $n_{\rm eh}$ and the applied voltage $V$:
\be
E_{\rm ph} = E_r + n_{\rm eh} e V,
\ee
where $e$ is the electron charge and $E_r$ is the energy deposited into primary phonons, typically much less than the second term which is linear in $n_{\rm eh}$. In the SuperCDMS detector, a 150 V potential difference applied across a silicon chip of gram-scale mass yields quantized peaks at integer multiples of 150 eV of phonon energy detected in the calorimetric phonon readout \cite{Romani:2017iwi,Agnese:2018col,Amaral:2020ryn}; the same technique has been used in germanium by the EDELWEISS collaboration \cite{EDELWEISS:2017uga,Arnaud:2020svb}. Leakage currents are still a main background in this approach, especially at high voltages where electrons can tunnel across the interface (\emph{Schottky barrier}) between the electrode and the substrate; at lower voltages, it is possible to some extent to reject these backgrounds since they yield non-quantized phonon energies. Charge readout via calorimeters has the advantages of being able to operate at cryogenic (mK) temperatures, as opposed to the 100 K required to obtain sufficient charge mobility to read out a CCD, and of utilizing a thicker (cm-scale) sample of silicon which allows a kg-scale mass to be placed in a compact cryostat. On the other hand, the event localization is much coarser than in a CCD due to the lack of pixellated segmentation, which makes background rejection somewhat more difficult. Similar to the plans for CCDs, work is ongoing to demonstrate scalability to kg-scale masses, either by multiplexing or scaling up individual detector masses \cite{SuperCDMS:2016wui,Griffin:2020lgd}.

The large-volume noble liquid detectors like LUX, XENONnT, and Panda-X (xenon) and DarkSide (argon) which are sensitive to sub-GeV DM are \emph{dual-phase time projection chambers} (TPCs). ``Dual phase'' refers to the fact that a small layer of the gaseous phase sits above the bulk of the detector which is in the liquid phase, and ``time-projection chamber'' refers to the method by which ionized charges are detected. A large electric field is applied vertically across the detector (about 125 V/cm in XENON), drifting the charges to the liquid-gas interface, at which point they are extracted and amplified, producing scintillation light (known as S2 in the literature) proportional to the number of electrons. These detectors therefore operate as transducers from a charge signal to a light signal. If an event produces a substantial nuclear recoil, primary scintillation photons (S1) are also generated by the recoiling nucleus, and the timing difference between S1 and S2 allows the event to be localized in the vertical direction (hence the name time-projection chamber). However, sub-GeV DM always produces an S2-only signal because the maximum kinetic energy which can be transferred to the nucleus is below the S1 threshold. A well-known background to few-electron events are ionized electrons produced in a high-energy event which become trapped at the liquid-gas interface for long times; when they are later released, they appear uncorrelated with an S1 signal, thus mimicking an S2-only signal \cite{Sorensen:2017kpl}. However, there are as of yet no quantitatively satisfactory models for this background, and given that the strongest sensitivity to light DM comes from small numbers of ionized electrons  \cite{XENON:2019gfn}, modeling and reducing this background is an active area of research \cite{Bernstein:2020cpc,Kopec:2021ccm}.

\subsection{Heat}
\label{sec:Heat}

Heat detectors, or \emph{calorimeters}, are excellent multi-purpose dark matter detectors because they are sensitive to the total energy deposited in the detector, regardless of the character of the excitation. The second law of thermodynamics and the large entropy of the phonon system in a solid-state detector guarantees that any excited state of the detector, whether arising from DM-nuclear or DM-electron scattering, will eventually decay or annihilate to (possibly a large number of) phonons. The present best limits on DM-nuclear scattering for sub-GeV DM come from WIMP DM detectors like CRESST-III which have been able to lower the energy detection threshold to the $\mathcal{O}(30) \ \eV$ scale \cite{Abdelhameed:2019hmk}. The detector is calcium tungstate (CaWO$_4$), which has been used for calorimetric particle detection applications for decades, and the detection principle is that a small heat deposit somewhere in the 24 g bulk crystal will produce a large number of meV-scale phonons. Crucially, if the detector is operated at cryogenic temperatures of 15 mK ($\simeq 1 \ \mu\eV$), thermal phonons have $\mu$eV energies, so the signal phonons are \emph{non-thermal} and will propagate and scatter ballistically throughout the crystal \cite{Lang:2009ge}. The thermalization time is slow enough that there is a large probability of the phonons encountering a \emph{transition edge sensor} (TES) placed on the surface of the crystal. A TES is a thin film of superconducting metal biased to a temperature just below its critical temperature $T_C$. Absorption of a few phonons can then heat the film above $T_C$, causing a massive increase in resistance and a measurable change in current or voltage \cite{Irwin2005}. The energy threshold is in one-to-one correspondence with the number of ``noise triggers'' from the TES readout; the 30 eV threshold in CRESST-III corresponds to one noise trigger per kg-day of exposure. This general principle, where a large-mass absorber provides a large number of DM scattering targets while a small-mass thermometer can efficiently convert small amounts of heat to a measurable signal, informs the design of many calorimetric detectors. The lowest threshold among currently-operating experiments is 16.3 eV, achieved by the SuperCDMS cryogenic phonon detector (CPD), a silicon crystal with a mass of 10.6 g coupled to a tungsten TES with energy resolution of 3.86 eV \cite{Alkhatib:2020slm,Fink:2020jts}.\footnote{Recently, a tungsten TES tuned to 65 mK was able to achieve an energy resolution of 2.65 eV \cite{Ren:2020gaq}, with a full DM search analysis expected soon.}

An alternative strategy is to sacrifice the large-mass absorber and use the small-mass thermometer itself as the absorber, exploiting the incredibly sensitive energy resolution of the smaller mass. One such implementation is \emph{superconducting nanowire single-photon detectors}, originally designed for detecting optical photons through absorption but which are also sensitive to DM-electron scattering. These devices consist of meandering arrays of nm-thick wires, with a total area of about $400 \times 400 \ \mu$m$^2$ and a mass of $\sim$ ng. At an energy threshold of 0.8 eV, a prototype device with WSi wires has been shown to have \emph{zero} dark rate with $10^4$ seconds of exposure at temperatures of 300 mK \cite{Hochberg:2019cyy}. Prototype devices with NbN wires have been shown to have an energy threshold of 250 meV \cite{Marsili:2012wr}, and thus further improvement in the energy threshold is likely in the near future.

The peculiar properties of superfluid helium also make it an excellent calorimetric detector \cite{Hertel:2018aal}. First, the excitation energy to the next available $n = 2$ electronic state is 19.77 eV, so any energy deposit below this must appear purely in phonons. Similar to solid-state detectors, the quasiparticle excitations are also very long-lived, allowing ballistic propagation out to the surface of the sample. Indeed, $^4$He remains a liquid down to absolute zero, allowing cryogenic operation at the lowest temperatures available. Finally, the phenomenon of \emph{quantum evaporation} converts quasiparticle excitations with energy above about 0.6 meV to an ejected helium atom at the interface of the liquid with vacuum, and when this atom is eventually adsorbed on a nearby calorimeter surface, it gains an energy equal to the adsorption energy. Typical surfaces have an adsorption energy of 10 meV, but in principle it is possible to increase this to 42.9 meV in fluorographene \cite{Reatto_2012}; since typical quasiparticle excitations have meV energies (see Sec.~\ref{sec:helium}), this corresponds to an effective amplification by a factor 10--50. The HeRALD experiment aims to exploit these phenomenon to obtain the first sensitivity to sub-MeV DM-nuclear scattering \cite{Hertel:2018aal}.

\subsection{Light}
\label{sec:light}

The primary signal in conventional scintillation detectors is optical or near-UV photons with eV-scale energies. A common way to detect such signals is with a \emph{photomultiplier tube} (PMT), also with a long history of use in WIMP DM experiments: when a photon is incident on a PMT, it collides with a metal plate and ejects an electron via the photoelectric effect. That electron is accelerated toward another plate, where due to the gain in kinetic energy, more electrons are ejected, yielding a cascade consisting of a macroscopic current after a gain of $\sim 10^7$. This is effectively a transduction of a light signal to a charge signal, and is almost universally the way single photons of this energy are detected in practice. The first limits on sub-GeV DM-electron scattering in organic scintillators were set using a PMT readout \cite{Blanco:2019lrf}, but were limited by the dark rate of $\sim 30$ Hz, achieved at a stabilized temperature of $\sim 5^\circ$C. Even after subtracting the intrinsic dark rate of the PMT by decoupling the PMT and the scintillator, a 3.8 Hz residual single-photoelectron dark rate remained. The possibility of using a skipper CCD as a photon readout for a solid organic scintillator, where absorption of a $\sim 5$ eV scintillation photon will generate either one or two electron/hole pairs in a single pixel, is currently being investigated \cite{Blanco:2021hlm}, motivated by the much lower dark rates possible with a CCD. NaI and CsI are two standard scintillators in use for numerous particle detection applications; their band gaps of 5.9 and 6.4 eV respectively are comparable to organic materials, but the response is isotropic. GaAs \cite{Derenzo:2016fse,Derenzo:2018plr,Vasiukov:2019hwn} is another promising scintillator detector. The lower band gap of 1.52 eV results in infrared rather than optical photons, which will not typically trigger a conventional PMT but may be detected with TESs in a method analogous to heat detection described above. Indeed, most of the charge and heat sensors described above may also function as light detectors, since a photon will generically produce charge pairs above the band gap and phonons below the band gap. 

Light detection may also be useful for nuclear scattering in a solid. When a nucleus is scattered out of its equilibrium lattice position, it leaves behind a defect, whose electronic states may scatter light of a particular wavelength where the crystal is transparent. This defect is known as a \emph{color center} and has been proposed as a detection signal for DM-nucleus scattering \cite{Budnik:2017sbu}. Once created, at an energy cost of $\sim 10$ eV, color centers are typically stable on very long time scales (effectively infinite at room temperature), and can be interrogated at will by irradiating the sample with a laser and looking for scattered light. Because the light produced by the color center is due to electronic transitions, it is typically in the optical or near-UV, and may be detected with a PMT or its solid-state analogue, the silicon photomultiplier (SiPM). The latter is simply a silicon absorber biased at a large voltage: an absorbed photon with energy exceeding the 1.2 eV band gap produces a few electron/hole pairs, which create a cascade of free charges through collisions with valence band electrons, known as \emph{impact ionization}. The persistent nature of the defects is both an advantage and a disadvantage, since ordinary radiobackgrounds such as neutrons will also create defects, and these are indistinguishable from DM-created defects on an event-by-event basis. Indeed, in this setup the crystal must be scanned prior to the DM search in order to identify any existing defects, and/or undergo an annealing process at $\sim 1000^\circ$C to cure these defects.

\subsection{Other signals}

A qualitatively different signal of DM scattering arises from a detector placed in a metastable state, such that a small energy deposit from DM causes a runaway transition to the ground state, greatly amplifying the signal. Indeed, this is the operating principle of the \emph{bubble chamber}, where a superheated liquid can undergo localized transitions to the gas phase after a nuclear scattering event, forming macroscopic bubbles which can be photographed. Bubble chambers have been in use since the earliest days of particle physics, and have recently been revived for few-GeV DM-nuclear scattering in the PICO experiment \cite{Amole:2019fdf}, but the bubble nucleation threshold is too high for sub-GeV DM. However, an analogous magnetic system may be constructed using a crystal of single-molecule magnets \cite{Bunting:2017net}. The molecules in these crystals essentially act as individual isolated nano-magnets, and may be prepared in a metastable state where an order-1 fraction of the spins are anti-aligned with an external magnetic field. A localized heat deposit from DM scattering (or other particle interactions) will flip some spins, releasing the Zeeman energy in the transition to the ground state, causing more heating an a runaway spin-flip process called \emph{magnetic deflagration}. The detectable signal is a growing magnetic field which may be read out with a precision magnetometer such as a SQUID. The deflagration process may be arrested simply by turning off the external field, and dead time from background events may be reduced by using a ``powder'' of small grains rather than a large single crystal. It was recently demonstrated that one such candidate, Mn12-acetate, can function as a particle detector with a MeV energy threshold \cite{Chen:2020jia}, so considerable R\&D is still required to reach the meV-eV energy thresholds which would allow sub-GeV DM sensitivity.

\subsection{The neutrino floor}
\label{sec:neutrino}

At sufficiently small DM cross sections, neutrino-electron scattering and secondary ionization from low-energy neutrino-nucleus scattering begin to compete with the DM rate. In standard WIMP detection, the background from solar neutrinos scattering off detector nuclei is known as the ``neutrino floor'' because the sensitivity to DM degrades rapidly in the presence of a background with similar spectral shape to the signal, requiring either much larger exposures or directional detection capabilities to overcome this background. However, the situation is somewhat more favorable for sub-GeV DM because the kinematics of the signal and background are quite different. Neutrinos are produced via several nuclear fusion processes in the sun, yielding a spectrum which is peaked in the few MeV range, along with a component from the $pp$ process that extends to very low energies and is the dominant component below 0.5 MeV. Because the neutrinos are highly relativistic, the typical momentum is always much larger than $p_0$, yielding an electron spectrum from neutrino-electron scattering which is approximately flat below electron recoil energies of 10 keV and easily distinguishable from the DM-electron spectrum at few-eV energies \cite{Wyenberg:2018eyv}. On the other hand, coherent neutrino-nucleus scattering may produce a few ionized electrons as secondary excitations from the recoiling nucleus, yielding a spectrum which peaks at low energies and mimics the DM-electron spectrum \cite{Wyenberg:2018eyv,Essig:2016crl}. That said, the expected rate in the few-eV electron energy range (equivalently, the few-electron bins in semiconductor detectors) is about 1 event/kg-yr for coherent neutrino-nucleus scattering and $10^{-4}$ events/kg-yr for neutrino-electron scattering, well below the event rates for DM cross sections at the thermal targets described in Sec.~\ref{sec:DMReview}. Thus, for the near future, sub-GeV DM experiments with $\mathcal{O}$(kg-yr) exposures will be relatively unaffected by the neutrino background.

\section{Conclusion and Outlook}
\label{sec:Conclusion}

Despite being only a decade old, the field of sub-GeV DM searches with CM systems has made enormous progress and opened up new directions and collaborations, including dedicated experiments. In the next decade, experiments based on conventional semiconductors and liquid nobles will likely begin to probe the thermal-target cross sections for DM heavier than 1 MeV. If a positive signal is found, more experiments with a variety of different targets will be required to confirm the DM interpretation; if no signal is seen above background, it will be imperative to push the cross section limits to the neutrino floor, which may require new types of detectors to avoid impractically-large exposure requirements, at least for existing approaches. In either case, new detectors with sub-eV thresholds will be required to probe light DM down to the keV-scale warm DM limit. In this section, we provide our perspective on the theoretical and experimental frontiers in sub-GeV DM detection over the next decade. 

\subsection{Theory: towards a theoretically-optimal detector}
\label{sec:futuretheory}
Much of the early progress in DM searches with CM systems focused on evaluating the suitability of particular well-studied systems like noble liquids and conventional semiconductors for light DM detection. As we have discussed in this review, these systems are sub-optimal in a quantifiable sense because the DM kinematics are mismatched to the target response since the peaks of the dynamic structure factor are inaccessible. Recently, a number of groups have begun the process of identifying figures of merit which govern the DM scattering rate in terms of CM properties like the band structure, dielectric function, and Born effective charges, and leveraging the ``materials-by-design'' program such as the Materials Project \cite{Jain2013} to search large databases of CM compounds for an optimal target. As an example, Ref.~\cite{Hochberg:2017wce} identified small, anisotropic Fermi velocities and a meV-scale gap as optimal features for DM-electron scattering in Dirac materials, using ZrTe$_5$ as an example candidate material. Using these figures of merit, Refs.~\cite{Geilhufe:2018gry,Geilhufe:2019ndy} identified several additional candidate compounds, and Ref.~\cite{Inzani:2020szg} used the fact that spin-orbit interactions are a controllable way to open a gap in DFT calculations, identifying several additional families of targets. To bring the plasmon peak of the ELF within the kinematic regime accessible to DM, Ref.~\cite{Hochberg:2021pkt} proposed \emph{heavy-fermion materials} where strong electron interactions yield a renormalized electron mass that is much larger than $m_e$, reducing $v_F$ by the same factor and yielding $v > v_F$ which is not possible in conventional materials. In another approach, Ref.~\cite{Lasenby:2021wsc} used sum rules similar to Eq.~(\ref{eq:Wsumrule}) to formulate conditions on optimal dielectric functions for DM-electron scattering.

Similarly, for single-optical phonon production in polar crystals through a dark photon mediator, Ref.~\cite{Griffin:2019mvc} identified the ``quality factor''
\be
Q = \frac{1}{\overline{\epsilon}_\infty^2 \overline{\omega}_{\rm LO}} \prod_{j=1}^n \left(\frac{|Z_j^*|^2}{A_j}\right)^{2/n},
\ee
with the DM scattering rate scaling as $Q$ at large DM masses; here $j$ runs over the ions in the unit cell, $Z_j^*$ are the Born effective charges, $A_j$  are the ion mass numbers, $\overline{\epsilon}_\infty$ is the directionally-averaged high-frequency dielectric constant, and $\overline{\omega}_{\rm LO}$ is the directionally-averaged longitudinal optical phonon energy. Al$_2$O$_3$ (sapphire), previously identified as an excellent candidate polar material, turns out to have $Q = 130 \times 10^{-7}$, which can be surpassed by SiO$_2$ $(Q = 200 \times 10^{-7})$ and LiF $(Q = 270 \times 10^{-7})$. Accounting for the anisotropy of the phonon dispersion relation, Ref.~\cite{Coskuner:2021qxo} identified hexagonal boron nitride (h-BN) as having an exceptionally large daily modulation, but at the cost of a reduced total rate compared to Al$_2$O$_3$, illustrating some of the tradeoffs inherent in this target optimization.

While these candidate detector materials range from commonly-available compounds to newly-synthesized ones, requiring a commensurate range of R\&D efforts, the importance of discovering DM will likely motivate identification of theoretically-optimal materials for a given candidate DM interaction and spur additional collaborations between the DM and CM communities.

\subsection{Experiment: towards low thresholds and dark rates}
\label{sec:futureexp}

Searches for DM-electron scattering with DM lighter than 500 keV require a material with a band gap smaller than the 0.67 eV gap in germanium. Although many such searches have been proposed, none have yet been experimentally realized. While a CCD-type detector made from a narrow-gap semiconductor is certainly a theoretical possibility, the practical difficulties of obtaining a working CCD from any materials but silicon (and to a lesser extent, germanium \cite{Leitz:2019tb,Leitz:2020va}) make this approach unlikely to succeed in the short term. An alternative approach is direct charge amplification from impact ionization, where a charge drifted at high voltage can collisionally excite other charges from the valence to the conduction band. If the voltage, and hence the gain, is small and approximately linear, such a device is called an \emph{avalanche photodiode (APD)}; if the voltage is large and gain is exponential, the device is a \emph{single-photon avalanche detector (SPAD)}. The names and practical use of these devices come from the silicon particle detection community, where the typical application is a SiPM as discussed in Sec.~\ref{sec:light}. The advantage of direct charge amplification is that the energy threshold of the detector is set by the band gap, rather than the readout: as long as one electron is in the conduction band, and the impact ionization efficiency is sufficiently high, a sufficiently large applied voltage will create a current which can be detected above a noise background. APDs have already been constructed from the semiconductor InSb \cite{baertsch1967noise,abautret2015characterization}, with a gap of about 0.2 eV, but the dark rate has not yet been characterized at the low temperatures required for DM searches. There is an active R\&D effort towards developing functional avalanche detectors for even more exotic compounds with gaps down to 20 meV \cite{KurinskyLOI}. In all detector materials, calibration measurements of the many-body response, both of the ELF itself and of the correlation between energy deposited in scattering and multiplicity of final-state electron-hole pairs, will be extremely useful in reducing systematic uncertainty on the predicted DM rate.

A parallel effort is ongoing to lower the thresholds for DM-nuclear scattering toward the single-phonon threshold. One active direction is based on achieving the theoretical sensitivity floor of a TES \cite{Hochberg:2015fth},
\be
\sigma_E \propto \sqrt{V_{\rm TES} T_{c,{\rm TES}}^3},
\ee
where $V_{\rm TES}$ is the volume of the TES and $T_{c,{\rm TES}}$ is the superconducting transition temperature at which the TES is biased. In principle, meV energy resolution is achievable by both lowering $T_c$ (with the use of titanium or tungsten instead of aluminum, for example) and $V$; it is also necessarily to use a target with a sufficiently long phonon mean free path, so that phonons can scatter multiple times and still have a high probability of being absorbed at the TES. An alternative strategy is to use microwave kinetic inductance devices (MKIDs), for which the operating principle is that a small deposit of energy will change the inductance of a resonant superconducting circuit \cite{doi:10.1146/annurev-conmatphys-020911-125022,app11062671}. Since kinetic inductance exists even at zero temperature, these devices may be operated colder than TESs, and are also somewhat easier to parallelize.

\subsection{Outlook}

The nature of DM remains perhaps the most pressing mystery in fundamental physics. The overwhelming weight of the gravitational and astrophysical evidence for DM makes it our best empirical clue to physics beyond the Standard Model, and a laboratory discovery would strengthen the case even further. The rapid development of new tools for DM searches, including detectors exploiting the properties of CM systems, has greatly improved the prospects for discovery across a much larger range of DM masses than was previously thought accessible. The continued investigation of new detector materials and readout technologies, coupled with the scalability of existing detectors to larger masses, makes it likely that the simplest thermal targets will either be confirmed or ruled out within the next decade. Regardless of whether these experiments yield discoveries or stronger exclusion limits, our knowledge of the experimentally-viable parameter space for sub-GeV DM will continue to sharpen, shedding more light on the viable models for explaining the physics of the early universe. The field of light dark matter is still new and growing, and many creative ideas are sure to continue to push the field forward -- we especially encourage graduate students in particle physics and condensed matter physics to join this adventure!

\section*{Acknowledgments}

We especially thank our collaborators for generating and developing several of the ideas explored in this work, including Peter Abbamonte, Dan Baxter, Gordon Baym, Carlos Blanco, Brian Campbell-Deem, Juan Collar, Peter Cox, Jeff Filippini, Danna Freedman, Yonit Hochberg, Katherine Inzani, Sin\'{e}ad Griffin, Adolfo Grushin, Simon Knapen, Jonathan Kozaczuk, Gordan Krnjaic, Stephen von Kugelgen, Noah Kurinsky, Ben Lehmann, Ben Lillard, Mariangela Lisanti, Bashi Mandava, Sam McDermott, Tom Melia, Matt Pyle, Jessie Shelton, Javier Tiffenberg, Chris Tully, Lucas Wagner, To Chin Yu, and Kathryn Zurek. We additionally thank Dan Baxter, Carlos Blanco, Noah Kurinsky, Bashi Mandava, Jan Sch\"{u}tte-Engel, and Lucas Wagner for feedback on the manuscript. YK is supported in part by DOE grant DE-SC0015655. TL is supported in part by DOE grant DE-SC0019195 and a UC Hellman fellowship.

\bibliographystyle{apsrev4-1}
\bibliography{DMCMBib}

\end{document}